%% file: scattering.tex
\documentclass[a4paper,11pt]{article}
\pdfoutput=1 

\usepackage{jheppub} 

\usepackage[T1]{fontenc} 
\usepackage{braket}
\usepackage{upgreek}
\usepackage{subfigure}
\usepackage{xcolor}
\usepackage{bm}
\usepackage{multirow}
\usepackage{amsmath}
\DeclareMathOperator{\Tr}{Tr}
\usepackage{amsfonts}
\usepackage{bbold}
\usepackage{subcaption}
\usepackage{graphicx}
\usepackage{caption}
\usepackage{booktabs}
\usepackage{array}
\usepackage[bottom]{footmisc}
\usepackage{microtype}

\def\arraystretch{1}

\input{definitions.tex}

\everymath{\displaystyle}

\title{\boldmath The $\pi\pi$ scattering amplitude at large $\Nc$}

\author[a,b,1]{Jorge Baeza-Ballesteros\note{corresponding author},}
\author[b]{Pilar Hern\'andez,}
\author[c,d]{and Fernando Romero-L\'opez}

\affiliation[a]{Deutsches Elektronen-Synchroton DESY, Platanenallee 6, 15738 Zeuthen}
\affiliation[b]{IFIC, CSIC-Universitat de Val\`encia, Parque Científico, Catedrático José Beltrán, 2, 46980 Paterna, Spain}
\affiliation[c]{Albert Einstein Center, ITP, University of Bern, 3012 Bern, Switzerland}
\affiliation[d]{Center for Theoretical Physics, MIT, Cambridge, MA 02139, USA}

\emailAdd{jorge.baeza.ballesteros@desy.de}
\emailAdd{m.pilar.hernandez@uv.es}
\emailAdd{fernando.romero-lopez@unibe.ch}

\preprint{DESY-25-040, MIT-CTP/5850}

\abstract{We study the scaling of meson-meson scattering amplitudes with the number of colors, $\Nc$. We use lattice calculations in a theory with $\Nf=4$ degenerate flavors, with $\Nc=3-6$ and pion mass $\Mpi\approx 560$ MeV. We focus on three different scattering channels, two of which have the same quantum numbers as some tetraquark candidates recently found at LHCb: the $T_{cs0}^0(2900)$, $T_{c\bar{s}0}^{++}(2900)$, $T_{c\bar{s}0}^0(2900)$ and $T_{cs1}^0(2900)$ states. Finite-volume energies are extracted using a large set of operators, containing two-particle operators with the form of two pions or two vector mesons, and local tetraquark operators. The resulting energy spectra is used to constrain the infinite-volume scattering amplitude by means of Lüscher's quantization condition. We consider polynomial parametrizations of the phase shift, as well as one-loop chiral perturbation theory (ChPT) predictions. We find that our lattice results follow the expected $\Nc$ scaling and are sensitive to subleading $\Nc$ corrections. In addition, we constrain the scaling of different combinations of low-energy constants from matching to large $\Nc$ ChPT. The results for the channel corresponding to a $(\pi^+ D^+_s - K^+ D^+)$ state show evidence of a virtual bound state with energy $E_\text{virtual}=1.63(10)\Mpi$ for $\Nc=3$, while this pole disappears at $\Nc>3$. This may be connected to the exotic states found in experiment. 
}

\begin{document} 
\maketitle
\flushbottom

\newpage

\section{Introduction  \label{sec:intro}}

Many features of the theory of the strong nuclear force, quantum chromodynamics (QCD), remain puzzling despite decades of advances~\cite{Achenbach:2023pba}. One example is the existence of exotic hadrons, whose properties are being measured at experiments such as LHCb~\cite{LHCb-FIGURE-2021-001-report} or BESIII~\cite{Liu:2023hhl}. This is the case, for example, of the recently-found double-charmed tetraquark, the $T_{cc}(3875)$~\cite{LHCb:2021vvq,LHCb:2021auc}, and of other light tetraquarks composed of a strange, a charm and two other light quarks, the $T_{cs}(2900)$~\cite{LHCb:2020bls,LHCb:2020pxc} and $T_{c\bar{s}}(2900)$~\cite{LHCb:2022sfr,LHCb:2022lzp} states. However, a first-principles prediction of their properties---and in most cases a confirmation of their existence---is still lacking. Most exotic states appear as unstable resonances that rapidly decay into two or more stable hadrons, and so any study of their properties requires of first-principles predictions of the scattering properties of their decay products~\cite{Mai:2025wjb}.

The non-perturbative nature of QCD at low energies requires techniques that do not rely on perturbative expansions in the strong coupling constant. One of the most successful approaches to this low-energy regime is lattice QCD, in which the theory is formulated in a discretized and finite space-time, and hadronic observables are evaluated numerically as expectation values. In recent decades, progress has been made in the study of multiparticle interactions from lattice QCD, including the development of formalisms to study two-~\cite{Luscher:1986pf,Luscher:1990ux} and three-particle~\cite{Hansen:2014eka,Hansen:2015zga,Hammer:2017uqm,Hammer:2017kms,Mai:2017bge} interactions, as well as the investigation of several scattering processes and resonances~\cite{Briceno:2017max}. The study of exotic states on the lattice, however, has been focused on very specific cases, such as the $T_{cc}(3875)$~\cite{Padmanath:2022cvl,Lyu:2023xro,Chen:2022vpo,Collins:2024sfi,Whyte:2024ihh,Dawid:2024dgy} or tetraquaks with two bottom quarks~\cite{Meinel:2022lzo,Hudspith:2023loy,Alexandrou:2024iwi,Hoffmann:2024hbz}. 

A complementary approach to study the low-energy regime of QCD is the large $\Nc$ or 't Hooft limit of QCD~\cite{tHooft:1973alw}, where the number of colors, $\Nc$, is assumed to be large.  This limit constitutes a simplification of QCD that keeps most of its non-perturbative features,  and allows to compute hadronic correlation functions as a power series in $1/\Nc$. In particular, scattering amplitudes are predicted to scale as $\sim\Nc^{-1}$, and so QCD is believed to reduce to a theory of infinitely-narrow resonances at large $\Nc$. 

In recent years, some controversy has arisen regarding the existence of exotic hadrons in the large $\Nc$ limit. The traditional view due to Witten~\cite{Witten:1979kh} and Coleman~\cite{Coleman:1980nk} concluded that no tetraquarks were present in the limit, based on the factorization of four-quark operators. However, it has been pointed out that this argument implies that tetraquarks do not mix with two mesons in the large $\Nc$ limit, but says nothing about the actual existence of the exotic states~\cite{Weinberg:2013cfa}. Instead, tetraquarks may exist in at large $\Nc$ with a width that depends on their flavor composition~\cite{Knecht:2013yqa,Cohen:2014tga}. 

The study of the large $\Nc$ scaling of meson-meson interactions and of the properties of exotic states, thus, requires to constrain subleading $1/\Nc$ effects. While these corrections cannot be determined analytically, lattice QCD provides us with a tool to constrain them with calculations at varying $\Nc$. In particular, lattice techniques have been widely used to study the $\Nc$ scaling of the hadron and glueball spectrum, matrix elements, and meson interactions~\cite{Bali:2013kia,Cordon:2014sda,Bali:2013fya,Castagnini:2015ejr,DeGrand:2016pur,DeGrand:2017gbi,DeGrand:2020utq,DeGrand:2021zjw,Athenodorou:2021qvs,Donini:2016lwz,Hernandez:2019qed,Donini:2020qfu,Perez:2020vbn,Baeza-Ballesteros:2022azb,DeGrand:2023hzz,Bonanno:2023ypf,DeGrand:2024lvp,DeGrand:2024frm}---see \rcite{Hernandez:2020tbc} for a recent review.

Another approach to study multiparticle interactions are effective field theories (EFTs). These theories describe hadron interactions in terms of some effective low-energy degrees of freedom and a set of effective couplings, which need to be determined from experiment or from matching to lattice results. Chiral Perturbation Theory (ChPT)~\cite{Weinberg:1978kz, Gasser:1984gg} is the paradigmatic EFT of mesons, and has been frequently employed to complement lattice QCD studies, for example, to constrain the chiral dependence of hadron observables. In addition, ChPT can be extended to describe meson properties and interactions in the large $\Nc$ limit~\cite{DiVecchia:1980yfw,Rosenzweig:1979ay,Witten:1980sp,Kawarabayashi:1980dp,Herrera-Siklody:1996tqr,Kaiser:2000gs}. This mainly requires to include the $\eta'$ particle into the theory, which becomes degenerate with the non-singlet pseudoscalar mesons in the limit.

Finally, other methods, such as the $S$-matrix bootstrap, also provide a complementary perspective into multihadron dynamics~\cite{Guerrieri:2024jkn}, and can be used to study the large $\Nc$ limit of the theory~\cite{Albert:2022oes,Fernandez:2022kzi,Albert:2023jtd,Albert:2023seb,Ma:2023vgc}.

In this work, we exploit the synergy between lattice QCD, the large $\Nc$ limit and chiral perturbation theory to investigate meson interactions as a function of $\Nc$. We work in a theory with $\Nf=4$ degenerate dynamical quark flavors, already used in \rrcite{Hernandez:2019qed,Donini:2020qfu,Baeza-Ballesteros:2022azb}, and focus on several scattering channels: the $SS$ channel, which is analogous to the isospin-two channel of two-flavor QCD; the $AA$ channel, which is antisymmetric in both quarks and antiquarks, and exhibits attractive interactions; and the $AS$ and $SA$ channels, which are degenerate and contain odd partial waves. Note that the first two of these channels where already investigated in \rcite{Baeza-Ballesteros:2022azb} near threshold, but we here extend that study to higher energies. Preliminary results of this work were presented as conference proceedings~\cite{Baeza-Ballesteros:2024ogp,Baeza-Ballesteros:2025obw} and as part of a doctoral thesis~\cite{Baeza-Ballesteros:2024zwx}.

Due to the nature of its interactions and its flavor quantum numbers, the $AA$ channel is a candidate to contain a tetraquark state. This possibility was studied in \rcite{Baeza-Ballesteros:2022azb} based on the inverse amplitude method (IAM)~\cite{Truong:1988zp,Dobado:1989qm,Dobado:1992ha,Hannah:1995si,Dobado:1996ps} and is also supported by recent experimental findings at LHC: in our setup all the $T_{cs0}^0(2900)$, $T_{c\bar{s}0}^{++}(2900)$ and $T_{c\bar{s}0}^0(2900)$ tetraquark states found in $B$ decays are expected to have the quantum numbers of the $AA$ channel. In addition, an analogous channel to the $AA$ channel has been recently studied using lattice QCD in a setup with a heavy charm, finding evidence of a virtual bound state~\cite{Yeo:2024chk}---see also \rrcite{Guo:2009ct,Liu:2012zya,Du:2017zvv,Albaladejo:2016lbb} for related phenomenological studies. Thus, the investigation of this channel and its $\Nc$ dependence may help to shed light on the properties of these exotic states and their fate in the large $\Nc$ limit. A similar situation may happen with the $AS$ and $SA$ channels, whose flavor content match that of the recently-found $T_{c\bar{s}1}^0(2900)$ tetraquark.

We use lattice simulations with $\Nc=3-6$ at fixed lattice spacing, $a\approx 0.075$ fm and pion mass $\Mpi\approx 560$ MeV. To determine the finite-volume energy spectrum of the channels of interest, we consider an extensive set of operators consisting of two-particle operators with the form of two pions and two vector mesons, and local tetraquarks, which also allow us to investigate the impact of each type of operator in the determination of the lattice spectrum. The results for the finite-volume energies are used to constrain the infinite-volume scattering amplitudes, both considering model-agnostic parametrizations based on an effective range expansion (ERE) and ChPT predictions at large $\Nc$. In particular, we compute here for the first one-loop predictions for the $AS$ channel in large $\Nc$ ChPT. Finally, we study the presence of a tetraquark state in the scattering processes, and compare our results for the low-energy constants (LECs) from ChPT with previous literature.
\newpage

The paper is organized as follows. In \cref{sec:largeNc}, we present results for the scattering amplitude in the channels of interest based on large $\Nc$ and ChPT predictions. \Cref{sec:lattice} discusses details on the lattice simulations and the subsequent determinations of the infinite-volume scattering amplitudes. Results for the finite-volume energies and for the pion-pion scattering amplitudes are presented in \cref{sec:resultsenergies} and \cref{sec:scatteringfits}, respectively. We finally conclude in \cref{sec:conclusions}. This paper also contains three appendices: \cref{app:technicaldetailscorrelator} presents some analytic results related to the properties of the two-pion correlation functions, \cref{app:chptamplitudes} summarizes results for the ChPT amplitudes from \rrcite{Bijnens:2011fm,Baeza-Ballesteros:2022azb} that are used in the amplitude analysis, and \cref{app:summaryenergies} summarizes the pion-pion finite-volume energies used to constrain the scattering amplitudes.

\section{Meson-meson scattering at large $\Nc$}\label{sec:largeNc}

In this section we focus on the general features of meson-meson scattering in the large $\Nc$ limit. We present results for a general theory with $\Nf$ active flavors, and focus later on the $\Nf=4$ case. For any $\Nf\geq 4$, two-particle interactions between non-singlet pseudoscalar mesons, to which we refer generically as pions, can be classified in seven different scattering channels.\footnote{This is reduced to three channels for $\Nf=2$ and six for $\Nf=3$.} These channels correspond to irreducible representations (irreps) of the SU($\Nf$) flavor group~\cite{Bijnens:2011fm}. For example, for $\Nf=4$ the decomposition into irreps reads
\begin{equation}
    15\,\otimes\,15\,=\,84\,\oplus\,45\,\oplus\,45\,\oplus\,20\,\oplus\,15\,\oplus\,15\,\oplus\,1\,,
\end{equation}
where we label each irrep by its dimension and the 15-dimensional irreps on the left-hand side correspond to the adjoint irrep, under which pions transform. In this work, we aim at constraining the scattering amplitudes of four of these scattering channels below the four-pion inelastic threshold: 
\begin{itemize}
    \item The 84-dimensional irrep containing states that are symmetric under the exchange of quarks and antiquarks, which has maximal dimensionality. It is equivalent to the isospin-two channel of two-flavor QCD, and we refer to it as the $SS$ channel.
    \item The irrep containing states that are antisymmetric under the exchange of both quarks and antiquarks, which has dimensions 20 for $\Nf=4$. This only exists for $\Nf\geq 4$, and we refer to it as the $AA$ channel. As was shown in \rcite{Baeza-Ballesteros:2022azb}, this state has attractive interactions near threshold. This, together with its flavor quantum numbers, makes it a candidate to contain a tetraquark state. 
    \item Two irreps that contain states which are symmetric under the exchange of quarks and antisymmetric under the exchange of antiquarks, and vice-versa, and have dimension 45 for $\Nf=4$. We refer to them as $SA$ and $AS$ channel, respectively. These two channels are related by a $C$ transformation, and so are degenerate in the absence of a $CP$-violating theta term.
\end{itemize}
Both the $SS$ and $AA$ channels are even under particle exchange, and so the corresponding scattering amplitudes contain only even partial waves, while the $AS$ and $SA$ channels are odd under particle exchange, an so the corresponding amplitudes contain only odd partial waves. In \rcite{Baeza-Ballesteros:2022azb} we already studied pion-pion interactions in the $SS$ and $AA$ channels near threshold. Throughout this paper, we will refer to a generic irrep as $R$.

For illustration, we present a set of representative states for $\Nf=4$ for each of the channels, which make the flavor and particle-exchange symmetries explicit,\\
\noindent\begin{equation}\label{eq:isospinstates}
\begin{array}{rl}
|p_1,p_2\rangle_{SS}&=\frac{1}{2}\left[|D_s^+(p_1)\pi^+(p_2)\rangle+|D^+(p_1)K^+(p_2)\rangle+(p_1\leftrightarrow p_2)\right]\,,\\[10pt]
|p_1,p_2\rangle_{AA}&=\frac{1}{2}\left[|D_s^+(p_1)\pi^+(p_2)\rangle-|D^+(p_1)K^+(p_2)\rangle+(p_1\leftrightarrow p_2)\right]\,,\\[10pt]
|p_1,p_2\rangle_{AS}&=\frac{1}{2}\left[|D_s^+(p_1)\pi^+(p_2)\rangle+|D^+(p_1)K^+(p_2)\rangle-(p_1\leftrightarrow p_2)\right]\,,\\[10pt]
|p_1,p_2\rangle_{SA}&=\frac{1}{2}\left[|D_s^+(p_1)\pi^+(p_2)\rangle-|D^+(p_1)K^+(p_2)\rangle-(p_1\leftrightarrow p_2)\right]\,.
\end{array}
\end{equation}
Here, $p_1$ and $p_2$ denote the momenta of the two particles.

\subsection{Large-$\Nc$ scaling of amplitudes}\label{sec:largeNcamplitudes}

The large $\Nc$ or 't Hooft limit of QCD~\cite{tHooft:1973alw} allows to characterize the scaling of hadron observables with the number of colors, $\Nc$, and flavors, $\Nf$, by means of a perturbative analysis of correlations functions at large $\Nc$. The analysis of the scattering amplitudes was already presented in \rcite{Baeza-Ballesteros:2022azb} for the $SS$ and $AA$ channels near threshold up to subleading order in $\Nc$. In this section, we briefly summarize those results, which extend trivially to arbitrary momenta, and present the corresponding results for the $AS$ and $SA$ channels.

\begin{figure}[t!]
\centering
   \subfigure[Disconnected ($D$) diagram.]%
             {\includegraphics[width=0.43\textwidth]{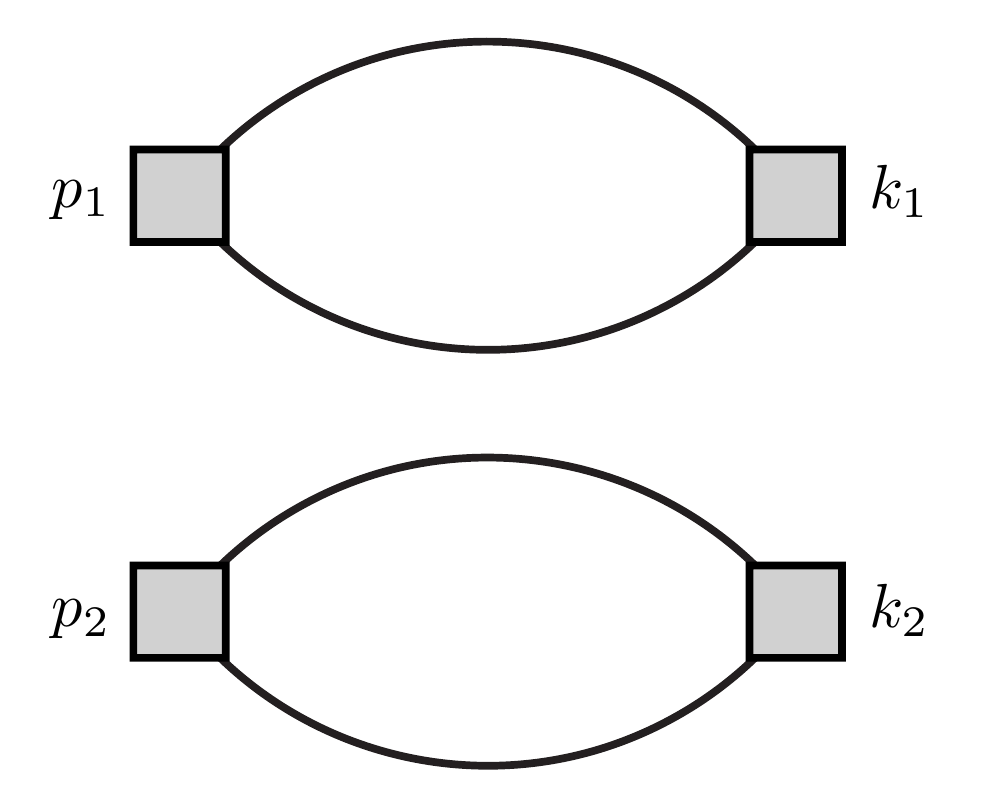} \label{fig:Ddiag}}\hspace{1cm}
   \subfigure[Connected ($C$) diagram.]%
             {\includegraphics[width=0.43\textwidth]{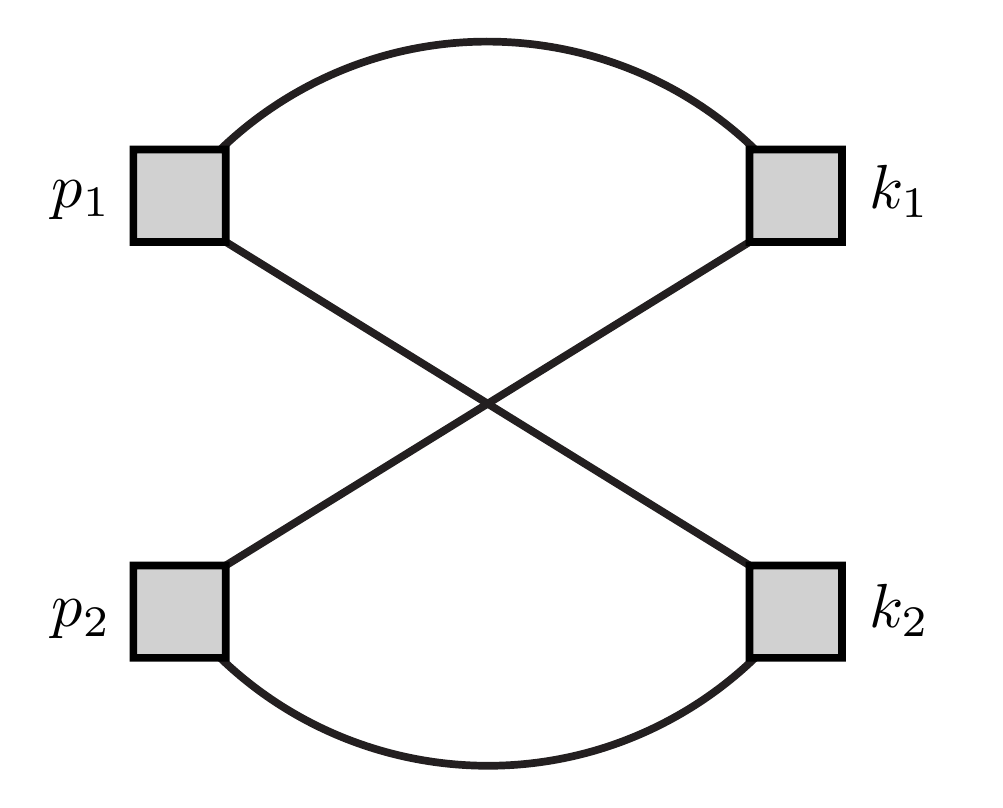} \label{fig:Cleading}} 
\caption{\label{fig:quarkcontractions} Diagrammatic representation of those Wick contractions contributing to the $SS$ and $AA$ channels. Solid lines are quark propagators, while the squares represent pion insertions, together with their corresponding momentum.}
\end{figure}

Two-pion correlation functions for all the channels considered in this work are computed as a linear combination of two quark contractions with different momentum assignments. If we let $\{\bm{k}_1,\bm{k}_2\}$ and $\{\bm{p}_1,\bm{p}_2\}$ denote the momenta of the two mesons in the initial and final state, respectively, the correlation functions are
\begin{equation}\label{eq:pionpioncorrelationfunctions}
    \begin{array}{rl}
         C_{SS}(\bm{p}_1,\bm{p}_2;\bm{k}_1,\bm{k}_2;t) = & D(\bm{p}_1,\bm{p}_2;\bm{k}_1,\bm{k}_2;t) - C(\bm{p}_1,\bm{p}_2;\bm{k}_1,\bm{k}_2;t) + (\bm{p}_1\leftrightarrow \bm{p}_2)\,,\\[5pt]
         C_{AA}(\bm{p}_1,\bm{p}_2;\bm{k}_1,\bm{k}_2;t) = & D(\bm{p}_1,\bm{p}_2;\bm{k}_1,\bm{k}_2;t) + C(\bm{p}_1,\bm{p}_2;\bm{k}_1,\bm{k}_2;t) + (\bm{p}_1\leftrightarrow \bm{p}_2)\,,\\[5pt]
         C_{AS}(\bm{p}_1,p_2;\bm{k}_1,\bm{k}_2;t) = & C_{SA}(\bm{p}_1,\bm{p}_2;\bm{k}_1,\bm{k}_2;t) =  D(\bm{p}_1,\bm{p}_2;\bm{k}_1,\bm{k}_2;t) - (\bm{p}_1\leftrightarrow \bm{p}_2)\,.
    \end{array}
\end{equation}
The disconnected, $D$, and connected, $C$, quark contractions are diagrammatically represented in \cref{fig:quarkcontractions}, and are computed in the infinite volume as
\begin{multline}\label{eq:Ddiagram}
D(\bm{p}_1,\bm{p}_2;\bm{k}_1,\bm{k}_2;t)=\int \text{d}\bm{x}_1\text{d}\bm{x}_2\text{d}\bm{y}_1\text{d}\bm{y}_2\,\text{e}^{-i(\bm{p}_1\bm{y}_1+\bm{p}_2\bm{y}_2)}\text{e}^{i(\bm{k}_1\bm{x}_1+\bm{k}_2\bm{x}_2)}\\
\times\left\langle\Tr\left[S(y_1,x_1)S^\dagger(y_1,x_1)\right]\Tr\left[S(y_2,x_2)S^\dagger(y_2,x_2)\right]\right\rangle\,,
\end{multline}
\begin{multline}\label{eq:Cdiagram}
C(\bm{p}_1,\bm{p}_2;\bm{k}_1,\bm{k}_2;t)=\int \text{d}\bm{x}_1\text{d}\bm{x}_2\text{d}\bm{y}_1\text{d}\bm{y}_2\,\text{e}^{-i(\bm{p}_1\bm{y}_1+\bm{p}_2\bm{y}_2)}\text{e}^{i(\bm{k}_1\bm{x}_1+\bm{k}_2\bm{x}_2)}\\
\times\left\langle\Tr\left[S(y_2,x_1)S^\dagger(y_1,x_1)S(y_1,x_2)S^\dagger(y_2,x_2)\right]\right\rangle\,,
\end{multline}
where $S(x_1,y_1)$ denotes the inverse Dirac operator, traces are taken over spin and color indices, expectation values are taken over all possible gauge-field configurations with the appropriate weight, and the integrals are computed over all space at fixed time, $x_i=(0,\bm{x}_i)$ and $y_i=(t,\bm{y}_i)$, for the initial and final states, respectively. We highlight the absence of the $C$ contraction in the $AS$ and $SA$ channels, which one may naively expect. This is a consequence of symmetries of the used operators---see \cref{app:technicaldetailscorrelator} for further details.

Scattering observables can be computed from the correlation function after amputating the external legs and removing the disconnected pieces, via the LSZ reduction formula. The $\Nc$ scaling can then be inferred from the amputated connected correlation function. Qualitatively, the two-pion scattering amplitude scales as, 
\begin{equation}
    \cM_2^R\sim \frac{C_R-C_\pi^2}{C_\pi^2}\,,
\end{equation}
where $C_\pi$ denotes the single-pion correlator, corresponding to one of the disconnected pieces in \cref{fig:Ddiag}. Thus, disconnected graphs in which the two pieces are not connected by the exchange of gluons  are canceled by the subtraction of $C_\pi^2$.

The large $\Nc$ scaling of scattering observables is then extracted from studying the scaling of the different quark contractions. As was discussed in \rcite{Baeza-Ballesteros:2022azb},\footnote{In this work, we include an additional $\Nc^{-1/2}$ normalization factor for each external meson insertion, as is customary in the context of the large $\Nc$ counting~\cite{Manohar:1998xv}. This factor was not taken into account in \rcite{Baeza-Ballesteros:2022azb}, but does not change the final results.} the leading contribution from $C$ originates from planar diagrams with no quark loops, while subleading corrections contain one internal loop that contributes with a power of $\Nf/\Nc$. The $D$ contraction, on the other hand, has its leading connected contribution to the amplitude coming from planar diagrams in which the two disconnected pieces are joined by the exchange of two or more gluons. Finally, the single-pion propagator has the same topology as $C$ except for two pion insertions, and so analogous large $\Nc$ scaling. All this can be summarized as follows:\\
\noindent\begin{equation}\label{eq:summarylargeNcorrelators}
    \begin{array}{rl}
         C&=\frac{1}{\Nc}\left(a+b\frac{\Nf}{\Nc}\right)+\cO(\Nc^{-3}) \,, \\[15pt]
         D&=C_\pi^2+c\frac{1}{\Nc^2}+\cO(\Nc^{-3})\,,\\[15pt]
         C_\pi&=d+e\frac{\Nf}{\Nc}+\cO(\Nc^{-2})\,.
    \end{array}
\end{equation}
Here, $a-e$ are numerical coefficients that do not depend on $\Nc$ or $\Nf$, but have some implicit dependence on the external momenta.

We can use these results to constrain the $\Nc$ and $\Nf$ scaling of the scattering amplitudes. Two-particle correlation functions in the $AA$ and $SS$ channels contain all the contributions from \cref{eq:summarylargeNcorrelators}. In particular, the scattering amplitudes scale as 
\begin{equation}\label{eq:scalingNcSSAA}
    \cM_2^{SS,AA}=\pm \frac{1}{\Nc}\left(\tilde{a}+\tilde{b}\frac{\Nf}{\Nc}\mp \tilde{c}\frac{1}{\Nc}\right)+\cO(\Nc^{-3})\,,
\end{equation}
where $\tilde{a}-\tilde{c}$ are linear combinations of $a-e$ that depend on the external momenta and, in particular, are symmetric under the exchange of momenta in the initial or final state. The $AS$ and $SA$ channels, on the other hand, have identical scattering amplitudes. What is more, only the disconnected diagram contributes. Thus, after subtracting the disconnected pieces,
\begin{equation}\label{eq:scalingNcAS}
    \cM_2^{AS,SA}=\tilde{d}\frac{1}{\Nc^2}+\cO(\Nc^{-3})\,,
\end{equation} 
where $\tilde{d}$ is another unknown coefficient that depends on the external momenta. As both channels the $AS$ and $SA$ channels are degenerate, we just focus on the former henceforth.

From these results some conclusions can be extracted. As expected in the large $\Nc$ limit, scattering between mesons is in general suppressed as $\cO(\Nc^{-1})$. However, when combined with symmetry properties of the channels under study, an additional suppression by an extra power of $\Nc$ can arise, as observed for the $AS$ channel. This means we expect interactions to be much weaker in this channel than for the other two, a prediction we verify below using lattice simulations.

\subsection{Large $\Nc$ scaling of tetraquarks and Weinberg compositeness criterium}\label{sec:tetraquarksmolecularlargeN}

A topic of great interest in the large $\Nc$ limit is the possible existence of exotic states, such as tetraquarks. The standard lore, based on the works of Witten~\cite{Witten:1979kh} and Coleman~\cite{Coleman:1980nk}, is that these states do not exist in the limit, based on the factorization property. For example, in the case of tetraquarks,  the correlation function of two local tetraquarks operators reduces to the correlation function of four non-interacting meson operators. 

This argument has been revisited in recent times following experimental findings of tetraquark states.  Weinberg~\cite{Weinberg:2013cfa} defended that, while the above conclusions are true for the disconnected part of the correlation function, the connected bit, while formally subleading, may describe different physics. Thus, tetraquarks could still arise from poles in the connected part of the correlation function, and it is not possible to rule out their existence. Moreover, Weinberg concluded that, if they existed in the large $\Nc$ limit, tetraquarks would have a decay width that scales as, $\Gamma_\text{tetra}\sim \cO(\Nc^{-1})$, like other standard resonances.

This conclusion was later revisited in \rcite{Knecht:2013yqa}, where it was shown that the particular scaling of tetraquarks could be narrower, depending on their flavor composition. In particular, open-flavor tetraquarks, such as those we study in this work, would have a narrower width, $\Gamma_\text{tetra}\sim\cO(\Nc^{-2})$. This scaling was also found in \rcite{Cohen:2014tga}, which argued against the existence of open-flavor tetraquark states in the large $\Nc$ limit. 

In parallel to their large $\Nc$ scaling, another open question regarding tetraquark states is that of their internal structure. There has been much controversy on whether they have a compact structure (and so are part of the Hilbert space of the low-energy effective theory) or if they are instead composite molecular states. For example, the $T_{c\bar{s}}(2900)$ state has been argued to be vector-meson molecule, due to its proximity to the $D^*K^*$ and $D_s^*\rho$ thresholds~\cite{Molina:2022jcd}.

A standard criteria to  differentiate between compact and molecular states is Weinberg compositeness criteria~\cite{Weinberg:1965zz,Matuschek:2020gqe,vanKolck:2022lqz}. If a bound state is close to a two-particle threshold, the behavior of the scattering amplitude close to threshold provides information about the nature of this state. In particular, Weinberg criteria relates the probability of a state being a compact particle or a molecular state to its field renormalization factor, also known as compositeness parameter,
\begin{equation}\label{eq:compositeness}
    Z=1-\left[1+\left|\frac{2r_\ell}{a_\ell}\right|\right]^{-1/2}\,,
\end{equation}
where $a_\ell$ and $r_\ell$ are the $\ell$-wave scattering length and the effective range of the interaction. A value of $Z$ close to zero is an indication of a hadron molecule, while a value close to unity can be related to a compact state.\footnote{Some authors have argued that any $Z\neq 0$ would imply the state contains a compact component, and only if $Z=0$ the state can be argued to be purely molecular~\cite{Maiani:2017kyi}.}

\subsection{Chiral Perturbation Theory predictions}

Chiral Perturbation Theory (ChPT) is an effective field theory that describes QCD at energies below the QCD scale in terms of the lightest multiplet of non-singlet pseudoscalar mesons. These are the pseudo-Nambu-Goldstone bosons arising from spontaneous chiral symmetry breaking. ChPT allows to obtain predictions for mesonic observables, such as scattering amplitudes, as a perturbative expansion, following the standard power counting~\cite{Weinberg:1978kz, Gasser:1984gg},
\begin{equation}
    \cO(m_q)\sim\cO(M_\pi^2)\sim\cO(k^2)\,.
\end{equation}
Here $m_q$ is the quark mass, $M_\pi$ is the pion mass and $k$ refers to the external-meson momenta. These predictions typically depend on a number of unknown couplings, the so-called low-energy constants (LECs), which need to be fixed from matching to experiment or from first principles, i.e., from lattice QCD. 

ChPT admits an extension to study meson interactions at large $\Nc$. The mass of the isospin-singlet pseudoscalar meson---the  $\eta'$---receives a contribution from the axial anomaly. According to the Witten-Veneziano relation~\cite{Witten:1979vv,Veneziano:1979ec},
\begin{equation}\label{eq:WittenVeneziano}
    M_{\eta'}^2=M_\pi^2+\frac{2\Nf \chi_\text{YM}}{\Fpi^2}\,,
\end{equation}
with $\chi_\text{YM}$ the topological susceptibility of the pure Yang-Mills theory and $F_\pi$ is the pion decay constant.\footnote{We use the normalization for $F_\pi$ corresponding to $\sim93$ MeV in QCD.} At large $\Nc$, $\chi_\text{YM}\sim\cO(\Nc^0)$ and $\Fpi^2\sim\cO(\Nc)$, and so the contribitution to the $\eta'$ mass from the axial anomaly is suppressed. This implies that the singlet meson becomes degenerate with non-singlet ones in the large $\Nc$ limit, and the chiral symmetry breaking pattern changes to
\begin{equation}
    \text{U}(\Nf)_\text{L}\times \text{U}(\Nf)_\text{R}\rightarrow \text{U}(\Nf)_\text{V}\,.
\end{equation}
Thus, the low-energy effective theory needs to be modified at large $\Nc$ to account for both these facts. The resulting extension of ChPT is commonly known as large $\Nc$ or U($\Nf$) ChPT~\cite{DiVecchia:1980yfw,Rosenzweig:1979ay,Witten:1980sp,Kawarabayashi:1980dp,Herrera-Siklody:1996tqr,Kaiser:2000gs}.

As in SU($\Nf$) ChPT, the large $\Nc$ theory also makes it possible to obtain prediction for meson observables in perturbation theory, but requires an extended counting scheme that includes the scaling on $\Nc$~\cite{Herrera-Siklody:1996tqr,Kaiser:2000gs},
\begin{equation}\label{eq:countinglargeNcChPT}
    \cO(m_q)\sim\cO(M_\pi^2)\sim\cO(k^2)\sim\cO(\Nc^{-1})\,.
\end{equation}
It also requires to take into account the implicit dependence of the LECs on $\Nc$~\cite{Kaiser:2000gs}. Those relevant for this work are,
\begin{equation}
    \cO(\Nc):\,\Fpi^2,L_0,L_3,L_5,L_8\,,\quad\quad \cO(\Nc^0):\,\Mpi^2,L_1,L_2,L_3,L_6\,,
\end{equation}
with subleading corrections to this leading scaling appearing with powers of $\Nf/\Nc$. Note that we use the same notation for the LECs in the two theories, although their actual value may not the same. In the decoupling limit of the $\eta'$, one can match the two theories, finding a relation between the LECs in both theories. In particular, for the relevant LECs here, one finds that all are expected to be the same at one-loop order, except for $L_6$ and $L_8$---see \rrcite{Peris:1994dh,Herrera-Siklody:1998dxd}. 

\subsubsection{SU($\Nf$) ChPT predictions}\label{sec:chptSUpredictions}

Results for two-meson scattering amplitudes for $\Nf$ degenerate flavors are known up to next-to-next-to-leading order (NNLO) in SU($\Nf$) ChPT~\cite{Bijnens:2011fm}. For the channels of interest for this work, the leading order (LO) predictions are
\begin{equation}\label{eq:chptLOpredictions}
    \cM_2^{SS,AA,\LO}=\pm\frac{\Mpi^2}{\Fpi^2}(2-\tilde{s})\,,\quad\quad \cM_2^{AS,\LO}=0\,.
\end{equation}
where $\tilde{s}=(k_1+k_2)^2/\Mpi^2$ is the usual Mandelstam variable normalized by the pion mass squared. At next-to-leading order (NLO), the predictions depend on some linear combinations of LECs and some additional terms, which we call $f_R$, containing chiral logarithms. Here, we give explicit expressions for the former ones and relegate the latter to \cref{app:chptamplitudes}. For the $SS$ and $AA$ channels, the $s$-wave projected amplitudes take the form
\begin{equation}\label{eq:chptNLOpredictionsSSAA}
    \begin{array}{rl}
         \frac{\Fpi^4}{\Mpi^4}\cM_{2,s}^{SS,\NLO} & = \phantom{+}32 L_{SS} + 32 q^2 L_{SS}'+ \frac{128}{3}q^4 L_{SS}''+f_{SS}(\Mpi^2, q^2, \mu, \Nf)\,, \\[10pt]
         \frac{\Fpi^4}{\Mpi^4}\cM_{2,s}^{AA,\NLO} & = - 32 L_{AA} - 32 q^2 L_{AA}' - \frac{128}{3}q^4 L_{AA}''+f_{AA}(\Mpi^2, q^2, \mu, \Nf)\,, \\ 
    \end{array}
\end{equation}
where $q=\sqrt{\tilde{s}/4-1}$ is the magnitude of the relative momentum normalized by the pion mass, and we indicate the dependence of the last term on $\Nf$ and on the scale, $\mu$, on which also the LECs depend, $L_R\equiv L_R(\mu)$. In terms of the LECs appearing in the Lagrangian, we have\footnote{We have changed the definition of $L_{AA}'$ and $L_{AA}''$ by a sign compared to \rcite{Baeza-Ballesteros:2022azb} to ensure all LECs terms have a common large $\Nc$ limit between both channels.}
\begin{equation}\label{eq:chptNLOpredictionsSSAALECs}
    \begin{array}{rl}
         L_{SS,AA}&=L_0\pm2L_1\pm 2L_2+L_3\mp 2L_4-L_5\pm L_6 + L_8\,,  \\[5pt]
         L_{SS,AA}'&= 4L_0\pm 4L_1\pm 6L_2+2L_3 \mp 2L_4 - L_5\,,\\[5pt]
         L_{SS,AA}'' &= 3L_0 \pm 2L_1\pm 4L_2 + L_3\,,
    \end{array}
\end{equation}
where the upper (lower) signs refer to the $SS$ ($AA$) channel.

Results for the $AS$ channel are also known and start at NLO. The $p$-wave-projected amplitude is
\begin{equation}\label{eq:chptNLOpredictionsAS}
    \frac{\Fpi^4}{\Mpi^4}\cM_{2,p}^{AS,\NLO}  = +64 q^2 L_{AS}  + 64 q^4 L_{AS}' + f_{AS}(\Mpi^2, q^2, \mu, \Nf)\,,
\end{equation}
where we have defined
\begin{equation}\label{eq:chptNLOpredictionsASLEcs}
    \begin{array}{rl}
    L_{AS} & =  -2L_1+L_2+L_4\,,\\[5pt]
    L_{AS}' & = -2L_1 + L_2\,.
    \end{array}
\end{equation}

From these results for the scattering amplitudes, one can easily check that both the LO parts and the NLO terms without LECs follow the expected $\Nc$ scaling from \cref{eq:scalingNcSSAA,eq:scalingNcAS}, and so this remains true for the LECs. For the $SS$ and $AA$ channels, it is possible to parametrize them as
\begin{equation}\label{eq:SUparammetrizationSSAALECs}
    L_{SS,AA}=\Nc L^{(0)}+L_{SS,AA}^{(1)}+\cO(\Nc^{-1})
\end{equation}
and similarly for $L_{SS,AA}'$ and $L_{SS,AA}''$. This makes explicit the common large $\Nc$ limit of both channels. For the $AS$ channel we have, on the other hand,
\begin{equation}\label{eq:NcpredictionsLECsAS}
    L_{AS}=L_{AS}^{(1)}+\cO(\Nc^{-1})
\end{equation}
and similarly for $L_{AS}'$. It is worth stressing that interactions are very weak in the $AS$ channel: they are one order further suppressed than those in the $SS$ and $AA$ channel both in the standard chiral counting and in $\Nc$.

 The scaling with $\Nf$, on the other hand, is not so clear. The chiral logs within $f_R$ depend explicitly on inverse powers of $\Nf$, which are not expected from \cref{eq:scalingNcSSAA,eq:scalingNcAS}. This arise since we are considering an inconsistent effective theory when taking the $\Nc \to \infty$ limit. In the large $\Nc$ limit, these terms are canceled by implicit terms within the LECs. 

\subsubsection{U($\Nf$) ChPT predictions}\label{sec:chptUpredictions}

The large $\Nc$ limit of meson-meson scattering amplitudes can be studied using U($\Nf$) ChPT. In the counting in~\cref{eq:countinglargeNcChPT}, NLO predictions are given by tree-level diagrams including insertions from $\cO(\Nc)$ LECs, while one-loop diagrams first appear at NNLO, together with products of $O(\Nc)$ LECs, and tree-level diagrams with $\cO(1)$ LECs and subleading terms from $\cO(\Nc)$ LECs---see \cref{app:chptamplitudes} for an explicit expression of NLO and NNLO part of the amplitude. It is at this order that the differences between the SU($\Nf$) and the U($\Nf$) theory become explicit for the observables of interest for this work. In particular, loops including internal $\eta'$ propagators start to contribute at NNLO.

The scattering amplitudes for the $AA$ and $SS$ channel were first computed up to NNLO in the U($\Nf$) theory in \rcite{Baeza-Ballesteros:2022azb}. The results for these channels up to one loop can be summarized as 
\begin{equation}\label{eq:UchptNLOpredictionsSSAA}
    \begin{array}{rl}
        \frac{\Fpi^4}{\Mpi^4}\left.\cM_{2}^{SS,\text{LO+NLO+NNLO}}\right|_{\text{U}(\Nf)} & = \frac{\Fpi^4}{\Mpi^4}\left.\cM_{2}^{SS,\text{LO+NLO}}\right|_{\text{SU}(\Nf)}\\[15pt] & +\frac{\Mpi^2}{\Fpi^2}(K_{SS}+q^2 K_{SS}')+\Delta \cM_2^{SS}(\Mpi^2, q^2, \mu, \Nf)\,, \\[20pt]
        \frac{\Fpi^4}{\Mpi^4}\left.\cM_{2}^{AA,\text{LO+NLO+NNLO}}\right|_{\text{U}(\Nf)} & =  \frac{\Fpi^4}{\Mpi^4}\left.\cM_{2}^{AA,\text{LO+NLO}}\right|_{\text{SU}(\Nf)}\\[15pt]
        & -\frac{\Mpi^2}{\Fpi^2}(K_{AA}+q^2 K_{AA}')+\Delta \cM_2^{AA}(\Mpi^2, q^2, \mu, \Nf)\,, 
    \end{array}
\end{equation}
where $\Delta \cM_{SS,AA}$ are corrections originating from loops that contain at least one internal $\eta'$ propagator, and are summarized in \cref{app:chptamplitudes}, and we have introduced some new LECs formed from the product of $\cO(\Nc)$ LECs,\footnote{There was a typo in the expression for $K_{SS,AA}$ in eq.~(2.18) of the published version of \rcite{Baeza-Ballesteros:2022azb}, which however does not affect the results of that work.}
\begin{equation}\label{eq:largeNcscalingK}
    \begin{array}{rrll}
         K_{SS} & = K_{AA} & = -256(L_5-2L_8)^2\,,  \\[5pt]
         K_{SS}' & = K_{AA}' & = -256L_5(L_5-2L_8)\,.  \\
    \end{array}
\end{equation}
We note they are equal in both channels, and scale as $\cO(\Nc^2)$.

\begin{figure}[t!]
\centering
   \subfigure[]%
             {\includegraphics[scale=0.32]{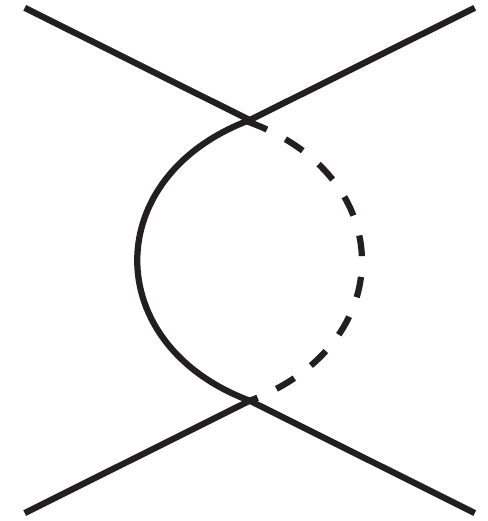}  \label{fig:loop1eta}}\hspace{3cm}
   \subfigure[]%
             {\includegraphics[scale=0.32]{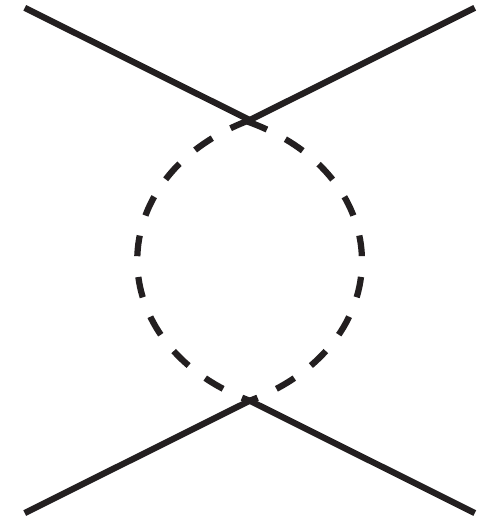}  \label{fig:loop2eta}}
\caption{One-loop Feynman diagrams required (together with their $u$-channel counterparts) to determine the two-pion scattering amplitude in the $AS$ channel at NNLO in U($\Nf$) ChPT. Solid lines depict non-singlet mesons, while dotted ones represent the $\eta'$.}\label{fig:etapdiagrams}
\end{figure}

On the other hand, the scattering amplitude for the $AS$ channel starts at NNLO in U($Nf$) ChPT. The full prediction has not been worked out before, and we present it here for the first time. In this case, the only corrections come from $t$- and $u$-channel diagrams that contain an internal $\eta'$ propagator---see \cref{fig:etapdiagrams}---and the corresponding $u$-channel counterparts. There is not, however, any new contribution coming from products of $\Nc$-leading LECs. Overall, the NNLO amplitude in the $AS$ channel is
\begin{equation}\label{eq:UchptNLOpredictionsAS}
    \frac{\Fpi^4}{\Mpi^4}\left.\cM_{2}^{AS,\text{NNLO}}\right|_{\text{U}(\Nf)}  = \frac{\Fpi^4}{\Mpi^4}\left.\cM_{2}^{AS,\NLO}\right|_{\text{SU}(\Nf)}+\Delta \cM_2^{AS}(\Mpi^2, \tilde{t}, \tilde{u}, \mu, \Nf)\,,
\end{equation}
where $\tilde{t}$ and $\tilde{u}$ and the Mandelstam variables normalized by the pion mass squared, and
\begin{equation}\label{eq:ASUchptcorrection}
    \Delta \cM_2^{AS}(\Mpi^2, \tilde{t}, \tilde{u},\mu,\Nf)= \frac{4}{\Nf^2}B_1(\tilde{t})+\frac{2}{\Nf^2}\bar{J}\left(\tilde{t}\frac{\Mpi^2}{\Metap^2}\right)-(\tilde{t}\leftrightarrow \tilde{u})\,,
\end{equation}
with $B_1$ defined in \cref{eq:chptB1} and $\bar{J}(z)$ the standard loop integral---see \rrcite{Passarino:1978jh,Scherer:2002tk}.

From the results for the scattering amplitudes, we can study the scaling with $\Nc$ and $\Nf$. Taking the large $\Nc$ limit, $\Metap\rightarrow \Mpi$, all factors of $1/\Nf$ appearing in the chiral logs cancel and the expected scaling from \cref{eq:scalingNcSSAA,eq:scalingNcAS} is recovered for those terms. This implies that the same scaling also holds for the $L_R$ terms, allowing to parametrize them as
\begin{equation}\label{eq:structureLECsAASSAS}
    \begin{array}{rl}
         \left.L_{SS,AA}\right|_{\text{U}(\Nf)} &=\Nc L^{(0)}+\Nf L_\text{c}^{(1)} \mp L_\text{a}^{(1)}+\cO(\Nc^{-1})\,,  \\ [5pt]
         \left.L_{AS}\right|_{\text{U}(\Nf)}&  = L_{AS}^{(1)}+\cO(\Nc^{-1})\,,
    \end{array}
\end{equation}
and similarly for $L_{SS,AA}'$ and $L_{SS,AA}''$, and $L_{AS}'$, respectively. 

In the case of the $SS$ and $AA$ channels, it is possible to disentangle the subleading term that is proportional to $\Nf$ from the other one by combining the results for the two channels, even when working at fixed $\Nf$. This property only holds for $\Nf\geq 4$, when the $AA$ channel is available. In particular, when studying the scattering process up to the inelastic threshold, one expects to be able combine the results for the two channels to obtain predictions about the scaling of the following combinations of LECs,
\begin{equation}
    \begin{array}{rl}
         L_0 + L_3 - L_5 + L_8 & = \Nc L^{(0)} + \Nf L_\text{c}^{(1)}+\cO(\Nc^{-1})\,,  \\ [5pt]
         4L_0 +2 L_3 - L_5 & = \Nc L^{\prime(0)} + \Nf L_\text{c}^{\prime(1)}+\cO(\Nc^{-1})\,,  \\ [5pt]
         3L_0 - L_3 & = \Nc L^{\prime\prime(0)} + \Nf L_\text{c}^{\prime\prime(1)}+\cO(\Nc^{-1})\,,  \\ [5pt]
         2 L_1 + 2 L_2 - 2 L_4 + 2 L_6 & = L_\text{a}^{(1)}+\cO(\Nc^{-1})\,,  \\ [5pt]
         4 L_1 + 6 L_2 - 2 L_4 & = L_\text{a}^{\prime(1)}+\cO(\Nc^{-1})\,,  \\ [5pt]
         2 L_1 + 4 L_2 & = L_\text{a}^{\prime\prime(1)}+\cO(\Nc^{-1})\,.
    \end{array}
\end{equation}
When working close to threshold, as we did in \rcite{Baeza-Ballesteros:2022azb}, only $L_{SS,AA}$ could be constrained. 

\subsubsection{The inverse amplitude method in ChPT}\label{sec:IAM}

Lattice simulations allow to constrain the scattering amplitude of two pions from first principles. By the matching the lattice data to the ChPT predictions presented in the previous section, one is able to constrain the values of the LECs. In particular, if lattice simulations are performed for varying values of $\Nc$, the scaling of the LECs towards the large $\Nc$ limit can be characterized. 

It has been known for a long time that naive ChPT predictions for the scattering phase shift converge poorly as the energy is increased above threshold. An alternative approach that allows to improve the convergence of the chiral expansion is the inverse amplitude method (IAM)~\cite{Truong:1988zp,Dobado:1989qm,Dobado:1992ha,Hannah:1995si,Dobado:1996ps}, which imposes perturbative unitarity and analyticity for the solution. This has been used to describe experimental data~\cite{Dobado:1992ha}, to analyze lattice results~\cite{Guo:2016zos,Guo:2018zss,Mai:2019pqr,Culver:2019qtx,Fischer:2020yvw}, and also in the context of the large $\Nc$ limit~\cite{Pelaez:2003ip,Pelaez:2005pi,Pelaez:2006nj,Guo:2013nja}.

When working up to NLO, as is the case of the SU($\Nf$) theory, the IAM $\ell$-wave scattering amplitudes are rewritten as
\begin{equation}\label{eq:IAMSU}
    \cM_{2,\ell}^\text{IAM}=\frac{\left(\cM_{2,\ell}^{\LO}\right)^2}{\cM_{2,\ell}^{\LO}-\cM_{2,\ell}^{\NLO}}\,.
\end{equation}
Expanding as a power series, one recovers the initial scattering amplitude to NLO. 
We note that the numerical values of the LECs may be different in the unitarized theory~\cite{Truong:1988zp,Dobado:1992ha,Dobado:1996ps}, coming from a redefinition of the higher orders. 

The validity of IAM fails in the subthreshold region in the neighborhood of an Adler zero (a zero of the scattering amplitude), for which it predicts a double zero. To extend the applicability of the IAM in the presence of an Adler zero, a modified IAM (mIAM) was proposed in \rcite{GomezNicola:2007qj}, in which the scattering amplitude is rewritten as
\begin{equation}\label{eq:mIAM}
    \cM_{2,\ell}^\text{mIAM}=\frac{\left(\cM_{2,\ell}^{\LO}\right)^2}{\cM_{2,\ell}^{\LO}-\cM_{2,\ell}^{\NLO}+A(s)}\,.
\end{equation}
where $A(s)$ is an analytic function of the total energy,
\begin{equation}
    A(s)=\cM_{2,\ell}^{\NLO}(s_\LO)-\frac{(s_\LO-s_\text{A})(s-s_\LO)}{s-s_\text{A}}\left[\partial_s \cM_{2,\ell}^{\LO}(s_\LO)-\partial_s\cM_{2,\ell}^{\NLO}(s_\LO)\right]\,,
\end{equation}
with $\partial_s$ the derivative with respect to $s$, $s_\LO$ the location of the Adler-zero at LO and 
\begin{equation}
    s_\text{A}=s_\LO-\cM_{2,\ell}^{\NLO}(s_\LO)/\partial_s\cM_{2,\ell}^{\LO}(s_\LO)\,,
\end{equation}
its position at LO+NLO. For the $SS$ and $AA$ chanels, one finds $s_\LO=\Mpi^2$. It is worth noting that the mIAM remains close to the IAM far from the Adler zero region.

In the case of the U($\Nf$) theory, one needs to be more cautious. Although we are working up to NNLO, order-by-order unitarity relations are imposed in a different manner that in SU($\Nf$) ChPT. The optical theorem takes the form
\begin{equation}
   \text{Im}\left.\cM_{2,\ell}^{\NNLO}\right|_{\text{U}(\Nf)} = \frac{q^{2\ell+1}}{16\pi \sqrt{s}}\left(\left.\cM_{2,\ell}^{\LO}\right|_{\text{U}(\Nf)}\right)^2\,,
\end{equation}
and an imaginary part of the amplitude first appears at NNLO. In this case, several options exist that lead to a unitarized amplitude. One alternative would be to use an expression for the IAM amplitude as the one used in heavy-baryon ChPT~\cite{GomezNicola:1999pu,GomezNicola:2000wk}. In this work, however, we have treated NLO and NNLO together for simplicity, and the IAM form in \cref{eq:IAMSU} is naively extended to large $\Nc$ ChPT by substituting \mbox{$\left.\cM_{2,\ell}^{\NLO}\right|_{\text{SU}(\Nf)}\longrightarrow\left.\cM_{2,\ell}^{\NLO}\right|_{\text{U}(\Nf)}+\left.\cM_{2,\ell}^{\NNLO}\right|_{\text{U}(\Nf)}$}.

\section{Lattice QCD methods}\label{sec:lattice}

\subsection{Ensembles and single-meson operators }

We study pion-pion scattering in the $SS$, $AA$ and $AS$ channels using lattice QCD simulations at different values of $\Nc=3-6$, working on replicas of the heaviest-pion-mass ensembles used in \rrcite{Hernandez:2019qed,Donini:2020qfu,Baeza-Ballesteros:2022azb}. We use a modified version of the HiRep code running on CPUs~\cite{DelDebbio:2008zf,DelDebbio:2009fd} to generate the configurations and measure correlation functions\footnote{For a GPU version of the HiRep code, see \rcite{Drach:2025eyg}.} . Ensembles are generated with the Iwasaki gauge action~\cite{Iwasaki:1983iya} and a fermion action with $\Nf=4$ degenerate flavors of dynamical clover-improved Wilson fermions, with periodic boundary conditions.\footnote{We note that the use of periodic boundary conditions may lead to freezing of the topological charge, specially at higher values of $\Nc$. We neglect the impact of the topological charge on scattering observables in this work, and leave its study for the future. } For the $\Nc=3$ simulations, the value of $c_\text{sw}$ is determined from the one-loop result for $\Nc=3$~\cite{Aoki:2003sj} boosted by the plaquette, and this is kept fixed for all values of $\Nc$ in accordance to its leading $\Nc$ scaling---see \rcite{Hernandez:2019qed} for a detailed explanation.

A summary of the simulation parameters is presented in \cref{tab:ensembles}, where we also indicate the values of $\Fpi$ for these ensembles as obtained in \rcite{Baeza-Ballesteros:2022azb} using a mixed-action setup with maximally twisted-mass valence fermions~\cite{Shindler:2007vp}. On the other hand, the lattice spacing was determined using the gradient flow scale $t_0/a^2$---see \rcite{Hernandez:2019qed}.

\begin{table}[b!]
\centering
\begin{tabular}{ccccccc}
\toprule
  Ensemble& $L^3 \times T$ &$\beta$ & $c_\text{sw}$ &$am$ & $a$ (fm) & $aF_\pi$  \\ \midrule 
3A11 & $24^3 \times 48$ & 1.778 &  \multirow{4}*{1.69} & $-0.4040$ & $0.075(2)$ & 0.0452(3)   \\ 
4A10 & $20^3 \times 36$ & 3.570 & & $-0.3735$ &  $0.076(2)$ & 0.0521(4) \\ 
5A10 & $20^3 \times 36$ & 5.969 & & $-0.3458$ & $0.075(2)$ &0.06154(22) \\ 
6A10 & $20^3 \times 36$ & 8.974 & & $-0.3260$ & $0.075(2)$ & 0.06874(21)   \\  \bottomrule 
\end{tabular}
\caption{Summary of the ensemble parameters used in this work. $L$ and $T$ indicate the number of points  of the lattices in the spatial and temporal directions, respectively, $\beta$ is the gauge coupling, $c_\text{sw}$ is the Sheikloleslami-Wohlert coefficient, and $am$ is the bare mass of the Dirac operator. Finally, $a$ is the lattice spacing, determined from the flow scale $t_0/a^2$~\cite{Hernandez:2019qed}, and $\Fpi$ is are the results for the pion decay constant from \rcite{Baeza-Ballesteros:2022azb}.  }
\label{tab:ensembles}
\end{table}

Single-meson masses are needed to study two-meson scattering processes, in particular to know the location of inelastic thresholds. For this reason, we begin by computing the mass of the lowest-lying mesons.
We compute correlation functions between quark bilinear operators with different $J^{CP}$ quantum numbers, focusing on non-singlet flavor states. In particular, we consider local meson operators with different Dirac structures,
\begin{equation}\label{eq:singlemesonoperator}
    O_{f_1,f_2}^M(\bm{p},\Gamma;t)=\sum_{\bm{x}} \bar{q}_{f_1}(\bm{x};t)\Gamma q_{f_2}(\bm{x};t)\text{e}^{-i\bm{px}}\,,
\end{equation}
where, $f_i$ are flavor indices, used to project to definite flavor quantum numbers, $\Gamma$ is the product of different gamma matrices, and $\bm{p}$ is the meson momentum. 

We focus on studying the masses, $M_M$, of the pseudoscalar ($\Gamma=\gamma_5$, to which we refer as $\pi$), scalar ($\Gamma=\mathbbm{1}$, called $a_0$), vector ($\Gamma=\gamma_i$, called $\rho$) and axial ($\Gamma=\gamma_5\gamma_i$, $a_1$) non-singlet mesons. These are determined from a single-state fit to the correlation functions computed with $\bm{p}=0$. For the pseudoscalar and vector mesons, we also determine the correlation functions, $C_{M,\bm{p}}(t)$, for $\bm{p}\neq 0$, from which the single-meson energies at definite momentum are extracted, $E_{M}^\text{latt}(\bm{k})$. These can be used to study the lattice dispersion relation of these states, which can be compared to the expected continuum dependence,
\begin{equation}\label{eq:freemesonenergycontinuum}
    E_{M}^\text{cont}(\bm{k})=\sqrt{M_M^2+\bm{k}^2}\,.
\end{equation}

\subsection{Two-particle operators}

To determine the two-particle finite-volume spectrum, we compute a matrix of correlation functions, 
\begin{equation}
    C_{ij}(t)=\left\langle O_i(t)O_j^\dagger(0)\right\rangle\,,
\end{equation}
for a set of interpolating operators, $\{O_i\}$, which have the correct quantum numbers of the channel of interest. In particular, one must use a sufficiently large set of operators, having significant overlap onto all the finite-volume states expected in the energy range of interest.

For this work, we consider three classes of interpolating operators: two-particle operators with the form of two pions, $\pi\pi$, or two vector mesons, $\rho\rho$, and local tetraquark operators, $T$. In all cases, we consider different values of the total momentum, $\bm{P}$.\footnote{Note, however, that we do not average over equivalent momentum frames, and we only consider one fixed momentum in each case.} The inclusion of $\rho\rho$ operators has a double motivation: vector mesons are stable in our ensembles and so their inclusion may be needed to correctly determine the finite-volume energies below the four-pion threshold, and also they may be useful to investigate tetraquark states, as some of the tetraquarks of interest have been argued to be molecular states of two vector mesons~\cite{Molina:2022jcd}---see \cref{sec:tetraquarksmolecularlargeN}. The use of $T$ operators, on the other hand, may also help in the search for tetraquark states~\cite{Stump:2024lqx}.

Two-particle operators are constructed from the product of two single-meson operators, 
\begin{equation}\label{eq:largeNmesons:pipirhorhoproduct}
\begin{array}{rl}
(\pi\pi)(\bm{p}_1,\bm{p}_2)&=O^M(\bm{p}_1,\gamma_5)O^M(\bm{p}_2,\gamma_5)\,,\\
(\rho_i\rho_j)(\bm{p}_1,\bm{p}_2)&=O^M(\bm{p}_1,\gamma_i)O^M(\bm{p}_2,\gamma_j)\,,
\end{array}
\end{equation} 
These are projected to definite isospin channel and to irreps of the the cubic group (for $\bm{P}=0$) or the relevant little group (for $\bm{P}\neq 0)$. Both projections block-diagonalize the matrix of correlation functions, simplifying the subsequent analysis. 

Isospin projection is performed for the channels of interest. Projected states are presented in the case of two pions in  \cref{eq:isospinstates} for the $\pi\pi$ case, and analogous states also hold for $\rho\rho$ operators, using the corresponding vector mesons ($\rho$, $K^*$, $D^*$, $D_s^*$). Projection to irreps of the cubic group or the relevant little group, denoted generically as $\cG$, is performed following \rrcite{Dudek:2012gj,Morningstar:2013bda}. Given some operator $O$, we project it to irrep $\Lambda$ as
\begin{equation}
    O^{\Lambda\lambda}(t)\propto \sum_{R\in\mathcal{G}}\Gamma_{\lambda\lambda}^\Lambda(R) \, U_R \, O(t) \, U_R^\dagger\,,
\end{equation}
where $R$ are the elements of $\cG$, $U_R$ is the operator that applies that transformation to $O$ and $\Gamma^\Lambda(R)$ is the representation of $R$ in irrep $\Lambda$. The index $\lambda$ labels the component within the irrep, which may have dimension larger than one. For irreps having dimension larger than one, we have averaged the results over each component of the projected operators. Note that each irrep contains contributions from different angular momenta, since rotational invariance is broken by the cubic lattice. In this work, we will restrict ourselves to those irreps that have a contribution from the lowest partial wave, and work in the lowest-partial wave approximation when matching to infinite-volume amplitudes. That is, we neglect contributions from higher partial waves, which we justify later. 

Cubic-group projection is straightforward for pion-pion operators, but becomes cumbersome for two vector mesons, due to the additional vector indices. In particular, for a fixed non-interacting energy, several projections may exist to one same irrep. For example, if we consider the $\bm{P}=(0,1,1)$ frame with $|\bm{p}_1|=0$ and $|\bm{p}_2|=\sqrt{2}$ (in units of $2\pi/L$), three different states can be constructed that transform under the $A_1$ irrep,
\begin{equation}
\begin{array}{rl}
(\rho\rho)_1&=\rho_2([0,0,0])\rho_3([0,1,1])+\rho_3([0,0,0])\rho_2([0,1,1])\,,\\[5pt]
(\rho\rho)_2&=\rho_2([0,0,0])\rho_2([0,1,1])+\rho_3([0,0,0])\rho_3([0,1,1])\,,\\[5pt]
(\rho\rho)_3&=\rho_1([0,0,0])\rho_1([0,1,1])\,,
\end{array}
\end{equation}
where the numbers in brackets are the three-momenta of each pion, in units of $2\pi/L$. In general, it is possible to project states of two vector meson to any cubic-group irrep. For this work, however, we restrict ourselves to those irreps containing pion-pion states at lowest partial wave, this is, $s$-wave for the $SS$ and $AA$ channels, and $p$-wave for the $AS$ channel. The total number of two-particle  operators used in this work for each channel and cubic-group irrep is presented in \cref{tab:quantityofoperators}.

\begin{table}[b!]
\centering
\begin{tabular}{c>{\centering\arraybackslash}p{1.cm}>{\centering\arraybackslash}p{1.cm}>{\centering\arraybackslash}p{1.cm}>{\centering\arraybackslash}p{1.cm}>{\centering\arraybackslash}p{1.cm}>{\centering\arraybackslash}p{1.cm}}
\toprule
 \multirow{2}{*}{Irrep}     &    \multicolumn{3}{c}{$SS$ and $AA$ channels} &    \multicolumn{3}{c}{$AS$ channel} \\ 
  & $\pi\pi$ & $\rho\rho$ &  $T$ & $\pi\pi$ & $\rho\rho$ &  $T$ \\ \midrule 

$A_1^+(0)$ & 5 & 6  & 5 & --- & --- & --- \\ 
$T_1^-(0)$ & --- & ---  & --- & 4 & 8 & 2 \\  
$A_1(1)$ & 5 &  9 & 7 & 5 & 9 & 2 \\  
$E(1)$ & --- &  --- & --- & 3 & 13 & 2 \\  
$A_1(2)$ & 8 & 18  & 9 & 6 & 12 & 2 \\ 
$B_1(2)$ & --- & ---  & --- & 4 & 15 & 2 \\  
$B_2(2)$ & --- & ---  & --- & 6 & 16 & 2 \\  
$A_1(3)$ & 5 & 7  & 7 & 5 & 7 & 2 \\ 
$E(1)$ & --- & ---  & --- & 5 & 12 & 2 \\  
$A_1(4)$ & 5 & 8  & 7 & 2 & 3 & 2 \\ 
$E(1)$ & --- & ---  & --- & 3 & 9 & 2 \\  \bottomrule
\end{tabular}
\caption{Number of operators of each type used in each isospin channel and cubic-group irrep. Numbers in parenthesis on the left column indicate $|\bm{P}|^2$ in units of $(2\pi/L)^2$. We note that operators in the $E$ and $T_1^+$ irreps have multiplicity two and three, respectively, which is not included in the counting.   }

\label{tab:quantityofoperators}
\end{table}

The last class of interpolating operators that we use are local tetraquark operators. These are constructed from the local product of two quark bilinears, projected to definite total momentum,
\begin{equation}
    T_{\Gamma_1\Gamma_2}(\bm{P},t)=\sum_{\bm{x}}T_{\Gamma_1\Gamma_2}(x)\text{e}^{-i\bm{Px}}\,,
\end{equation}
where the coordinate-space tetraquark operator is, 
\begin{equation}
    T_{\Gamma_1\Gamma_2}(x)=\bar{q}_{f_1}(x)\Gamma_1 q_{f_2} \bar{q}_{f_3}(x)(x)\Gamma_2 q_{f_4}(x)\,,
\end{equation}
and flavor indices are used to project to the channels of interest, in analogy to what is done for two-particle operators. 
We consider several combinations of $\{\Gamma_1,\Gamma_2\}$ for all scattering channels, having the correct spin, $J$, parity, $P$, and charge-conjugation, $C$, quantum numbers. In particular,
\begin{equation}\label{eq:Diracstructuresummary}
\begin{array}{rl}
SS\text{ and } AA\text{ channels, }J^{PC}=0^{++}:&\{\gamma_5,\gamma_5\},\{\gamma_5\gamma_0,\gamma_5\gamma_0\},\{\mathbbm{1},\mathbbm{1}\},\{\gamma_i,\gamma_j\},\{\gamma_5\gamma_i,\gamma_5\gamma_i\}\,,\\
AS\text{ channel, }J^{PC}=1^{-+}:&\{i\gamma_5,\gamma_5\gamma_i\},\{\gamma_5\gamma_0,\gamma_5\gamma_i\}\,,
\end{array}
\end{equation}
where the $i$ factor in the first combination of the $1^{-+}$ tetraquark is required to ensure hermiticity of the operator. In analogy to two-particle states, tetraquark operators are also projected to definite irreps of the flavor and cubic groups. The total number of tetraquark operators used in this work for each channel and irrep is also presented in \cref{tab:quantityofoperators}.

A complete list of the cubic-group projected operators used in this work is available as ancillary material with the arXiv submission \cite{ancillary}.

\subsection{Computation of correlation functions}\label{sec:computationcorrelationfunctions}

Correlation functions between the operators described in the previous section can be computed for the channels of interest as a combination of the connected and disconnected quark contractions, similarly to the results presented for pions in \cref{eq:pionpioncorrelationfunctions},
\begin{equation}
\begin{array}{rl}
    C_{SS}(t) & = D(\bm{p}_1,\Gamma_1',\bm{p}_2,\Gamma_2';\bm{k}_1,\Gamma_1,\bm{k}_2,\Gamma_2;t)-C(\bm{p}_1,\Gamma_1',\bm{p}_2,\Gamma_2';\bm{k}_1,\Gamma_1,\bm{k}_2,\Gamma_2;t)\\ &+(\bm{p}_1\leftrightarrow \bm{p}_2, \Gamma_1\leftrightarrow \Gamma_2)\,,\\
    C_{AA}(t) & = D(\bm{p}_1,\Gamma_1',\bm{p}_2,\Gamma_2';\bm{k}_1,\Gamma_1,\bm{k}_2,\Gamma_2;t)+C(\bm{p}_1,\Gamma_1',\bm{p}_2,\Gamma_2';\bm{k}_1,\Gamma_1,\bm{k}_2,\Gamma_2;t)\\ &+(\bm{p}_1\leftrightarrow \bm{p}_2, \Gamma_1\leftrightarrow \Gamma_2)\,,\\
    C_{AS}(t) & = D(\bm{p}_1,\Gamma_1',\bm{p}_2,\Gamma_2';\bm{k}_1,\Gamma_1,\bm{k}_2,\Gamma_2;t)-(\bm{p}_1\leftrightarrow \bm{p}_2, \Gamma_1\leftrightarrow \Gamma_2)\,.
\end{array}
\end{equation}
where $\{\bm{k}_1,\bm{k}_2\}$ and $\{\bm{p}_1,\bm{p}_2\}$ indicate the momenta of the particles in the initial and final state, respectively, with $\bm{P}=\bm{k}_1+\bm{k}_2=\bm{p}_1+\bm{p}_2$, and similarly for the Dirac structures, $\{\Gamma_1,\Gamma_2\}$ and $\{\Gamma_1',\Gamma_2'\}$. The dependence on these quantities has been omitted on the left-hand-side of the equalities. (Anti)symmetrization over momenta and Dirac structures is performed simultaneously, as indicted in the parenthesis. 

On the lattice, the connected and disconnected quark contractions take the form,
\begin{multline}\label{eq:fullcorrelationfunctionD}
D(\bm{p}_1,\Gamma_1',\bm{p}_2,\Gamma_2';\bm{k}_1,\Gamma_1,\bm{k}_2,\Gamma_2;t)=\sum_{\bm{x}_1,\bm{x}_2}\sum_{\bm{y}_1,\bm{y}_2}\text{e}^{-i(\bm{p}_1\bm{y}_1+\bm{p}_2\bm{y}_2)}\text{e}^{i(\bm{k}_1\bm{x}_1+\bm{k}_2\bm{x}_2)}\Theta(\bm{x}_1,\bm{x}_2) \Theta(\bm{y}_1,\bm{y}_2)\\
\times\langle\Tr\left[\hat{\Gamma}_1'S(y_1,x_1)\tilde{\Gamma}_1S^\dagger(y_1,x_1)\right]\Tr\left[\hat{\Gamma}_2'S(y_2,x_2)\tilde{\Gamma}_2S^\dagger(y_2,x_2)\right]\rangle\,,
\end{multline}
\begin{multline}\label{eq:fullcorrelationfunctionC}
C(\bm{p}_1,\Gamma_1',\bm{p}_2,\Gamma_2';\bm{k}_1,\Gamma_1,\bm{k}_2,\Gamma_2;t)=\sum_{\bm{x}_1,\bm{x}_2}\sum_{\bm{y}_1,\bm{y}_2}\text{e}^{-i(\bm{p}_1\bm{y}_1+\bm{p}_2\bm{y}_2)}\text{e}^{i(\bm{k}_1\bm{x}_1+\bm{k}_2\bm{x}_2)}\Theta(\bm{x}_1,\bm{x}_2) \Theta(\bm{y}_1,\bm{y}_2)\\
\times\langle\Tr\left[\hat{\Gamma}_2'S(y_2,x_1)\tilde{\Gamma}_1S^\dagger(y_1,x_1)\hat{\Gamma}_1'S(y_1,x_2)\tilde{\Gamma}_2S^\dagger(y_2,x_2)\right]\rangle\,,
\end{multline}
where we have defined $\Gamma^\text{i}=\gamma_0\Gamma^\dagger\gamma_0\gamma_5$ and $\Gamma^\text{f}=\gamma_5\Gamma$, and we use $x_i=(t_\text{src},\bm{x}_i)$ and $y_i=(t_\text{src}+t,\bm{y}_i)$, with $t_\text{src}$ the source time, of which the results are independence due to time-translation invariance. Also, we have introduced functions $\Theta$ that generalize the correlation functions to include tetraquark operators,
\begin{equation}
    \Theta(\bm{x}_1,\bm{x}_2)=\left\{
\begin{array}{ll}
1 \quad\quad\quad\quad\quad& \text{Two particles at source}\,,\\
\delta_{\bm{x}_1,\bm{x}_2}\quad\quad\quad & \text{Tetraquark  at source}\,,\\
\end{array}\right.
\end{equation}
and similarly for the final-state and $\Theta(\bm{y}_1,\bm{y}_2)$.

To evaluate the correlation functions, we use two different techniques depending on whether we have a  two-particle or a tetraquark operator at source. Sums at sink, on the other hand, are performed exactly. In the case of two particles at source we use time- and spin-diluted $\mathbb{Z}_2\times\mathbb{Z}_2$ stochastic sources, $\xi_c(\bm{x})$~\cite{Dong:1993pk,Wilcox:1999ab,Foley:2005ac}, where $c$ indicates the color index. Each particle is represented with an independent source, and the computations are repeated on 36 source times with different sources for each configuration. For mesons with non-zero momentum, we boost the stochastic sources before solving the Dirac equation, $\xi_c^{\bm{q}}(\bm{x})=\text{e}^{i\bm{qx}}\xi_c(\bm{x})$. Then, the Fourier factor of the corresponding meson is recovered from the inner product of boosted stochastic sources,
\begin{equation}
    \xi_{c_1}^{\bm{q}_1}(x_1)\xi_{c_2}^{\bm{q}_2}(x_2)=\text{e}^{-i(\bm{q}_1-\bm{q}_2)\bm{x}_1}\delta_{\bm{x}_1,\bm{x}_2}\delta_{c_1,c_2}+\cO(N^{-1/2})\,,
\end{equation}
where the resulting meson at source would have momentum $\bm{q}=\bm{q}_1-\bm{q}_2$. We have computed the minimal set of boost momenta that need to be used to construct all meson operators of interest, which reduces to only 12 different non-zero values.

Correlation functions involving tetraquark operators at source are computed using point sources localized on a sparse lattice at fixed time slice,
\begin{equation}
\Lambda_\text{S}(t)=\left\{(t,s_1+dn_1,s_2+dn_2,s_3+dn_3)\,|\,\,n_i\in\mathbb{Z},0\leq n_i < L/d,0\leq s_i < d\right\}\,,
\end{equation}
where we use a coarsening step $d/a=4$. For each configuration the offsets, $0\leq s_i < d$ are randomly generated, and the computation is repeated over two maximally separated time-slices, the location of which is also selected at random. For the computation of the correlation functions, the use of a coarse lattice amounts to substituting
\begin{equation}
    \sum_{\bm{x}_1,\bm{x}_2}\longrightarrow\sum_{\bm{x}_1,\bm{x}_2\in\Lambda_\text{S}(t)}\,,
\end{equation}
in the definitions in \cref{eq:fullcorrelationfunctionD,eq:fullcorrelationfunctionC}.

For the channels of interest, no equal-time quark propagators need to be computed, and so we found the use of stochastic and point sources sufficient for the computations, and we do not use smearing techniques. The application of variational techniques, as described in the previous section, shows to be sufficient to confidently determine the finite-volume energy spectrum in the elastic region. 

After computing the matrices of correlation functions, we use the relations $C(t)=C^\dagger(t)$ and $C(T-t)=C^*(t)$ to average the numerical results, which hold due to the hermiticity of the used operators and the periodicity in time (antiperiodicity for fermions) of the lattice ensembles. Also, since the correlation functions are in general expected to be purely real---see \cref{app:technicaldetailscorrelator}---we drop the imaginary parts of the results, which are a numerical artifact arising from the use of stochastic sources for meson-meson operators. The only exception are correlation function in the $AS$ channel that involve a single $T_{\gamma_5\gamma_0,\gamma_5\gamma_i}$ operator, which are imaginary, and so we drop the real part.

\subsection{Determination of finite-volume energies}\label{sec:energyextraction}

\subsubsection{Cancellation of thermal effects}

Finite-volume energies can be extracted from the matrix of correlation functions using a variational analysis. However, the applicability of these techniques is  limited by the presence of thermal effects, which arise due to the limited time extent of our lattices. To eliminate, or at least mitigate, thermal effects, we use a generalized version of the shift-reweighting technique first used in \rcite{Dudek:2012gj}, which also incorporates thermal backpropagating effects. When Hermitian operators are considered, the real part of the matrix of correlators takes the form
\begin{equation}\label{eq:realcorreatorthermal}
     \text{Re}\,\,C(t)= \sum_n A_n\cosh(E_nt')+\sum_{m,n} B_{nm}\cosh(\Delta E_{mn}^\text{th} t')\,,
\end{equation}
where $t'=t-T/2$, sums are performed over all states $|n\rangle$ in the Hilbert space of the theory that have a non-zero overlap onto the operators, and the real coefficients $A_n$ and $B_{mn}$ are related to the operator matrix elements between the vacuum and the $|n\rangle$ state, or between the $|n\rangle$ and $|m\rangle$ states, respectively. The first sum depends on the two-particle energies that we want to determine, while the second depends on energy differences, and represent the main thermal pollution limiting the applicability of the variational techniques. The dominant terms contributing to this pollution are related to  the energy difference between single-meson states,
\begin{equation}
    \Delta E_{mn}^\text{th}\approx E_{M}(\bm{q}_m) - E_M(\bm{q}_n)\,,
\end{equation}
for momenta $\bm{q}_1-\bm{q}_2=\bm{P}$, where $\bm{q}_i$ is the momenta associated to the single-particle state $|i\rangle$. 

To reduce the thermal effects we subtract the contribution from the dominant term in the second sum in \cref{eq:realcorreatorthermal}, which corresponds to a two single-pion state with momenta $\bm{p}_i$, with magnitudes summarized in \cref{tab:thremalmomentumsubtractions}. The energy difference is therefore \mbox{$\Delta E^\text{th}=E_{\pi}^\text{cont}(\bm{p}_1)-E_{\pi}^\text{cont}(\bm{p}_2)$}, and the corresponding  thermal effects can be eliminated by defining a shift-reweighted correlator as
\begin{equation}\label{eq:shiftreweightcosh}
   \tilde{C}(t)=\frac{1}{2}\left\{\frac{\cosh(\Delta E^\text{th} t')}{\cosh[\Delta E^\text{th} t'_+]}C(t+a)-\frac{\cosh(\Delta E^\text{th} t')}{\cosh[\Delta E^\text{th} t'_-]}C(t-a)\right\}\,.
\end{equation}
 Here $t'_\pm=t'\pm a$, and the single-pion energies are computed using the continuum dispersion relation, \cref{eq:freemesonenergycontinuum}, rather than the results obtained on the lattice for boosted pions. We note that for the rest frame as well as the $(0,1,1)$ and $(0,0,2)$ frames, this procedure basically amounts to computing the numerical derivative of the matrix of correlators.

\begin{table}[t!]
\centering
\begin{tabular}{ccc}
\toprule
  $\bm{P}/(2\pi/L)$ & $|\bm{p}_1|^2/(2\pi/L)^2$ & $|\bm{p}_2|^2/(2\pi/L)^2$  \\ \midrule 
$(0,0,0)$ & 0 &  0 \\  
$(0,0,1)$ & 1 &  0 \\   
$(0,1,1)$ & 1 &  1 \\  
$(1,1,1)$ & 2 &  1 \\  
$(0,0,2)$ & 1 &  1 \\   \bottomrule 
\end{tabular}
\caption{Magnitude of the single-particle momenta used to perform the shift-reweighting technique in \cref{eq:shiftreweightcosh} that approximately eliminates leading thermal effects.   }

\label{tab:thremalmomentumsubtractions}
\end{table}
\subsubsection{Generalized eigenvalue problem}

After reducing thermal effects, we use $\tilde{C}(t)$ to solve a Generalized Eigenvalue Problem (GEVP),
\begin{equation}
    \tilde{C}(t)v_n(t,t_0)=\lambda(t,t_0)\tilde{C}(t_0)v_n(t,t_0)\,,
\end{equation}
where the eigenvalues can be proven to obey~\cite{Luscher:1990ck,Blossier:2009kd}
\begin{equation}\label{eq:eigenvaluescorrectionexpected}
    \lambda_n(t,t_0)= \lambda_n^{(0)}(t)\left[1+\cO(\text{e}^{-\Delta_n t})\right]\,,
\end{equation}
where $n=0,...,N-1$, $\Delta_n=E_N-E_n$ and we have defined
\begin{equation}\label{eq:eigenvaluesdependence}
   \lambda_n^{(0)}(t)=B\left\{\frac{\cosh(\Delta E^\text{th} t')}{\cosh[\Delta E^\text{th} t'_+]}\cosh(E_n t'_+)-\frac{\cosh(\Delta E^\text{th} t)}{\cosh[\Delta E^\text{th} t'_-]}\cosh(E_n t'_-)\right\}\,,
\end{equation}
which encodes the dominant energy dependence of the eigenvalue, with $B$ an unknown amplitude. Strictly speaking, the form of the correction in \cref{eq:eigenvaluescorrectionexpected} only holds as long as $t_0\geq t/2$, although imposing this is not always feasible when analyzing lattice results. 

The eigenvectors, $v_n(t,t_0)$, determine the overlaps of the operators into the states $|n\rangle$, 
\begin{equation}\label{eq:eigenvectors} 
    \tilde{C}(t_0)v_n\approx \langle 0| O_i | n \rangle\text{e}^{-E_n t_0/2}\,,
\end{equation}
where we have considered the states $|n\rangle$ to be normalized to unity, i.e., $\langle m|n\rangle=\delta_{mn}$. Also, we are neglecting thermal effects here, which should be negligible at early times at which the GEVP is solved---see the discussion below. Furthermore, the eigenvectors define optimized operators that have maximal overlap onto the $| n \rangle$ states, and the eigenvalues of the GEVP correspond to the correlation function of these optimized operators.

Using the eigenvectors, one can define the overlap of the operator $O_i$ into the state $|n\rangle$, relative to other states, as
\begin{equation}\label{eq:overlaps}
    \cZ_{ni}=\frac{\langle 0 | O_i|n\rangle}{\sum_m \langle 0 | O_i|m\rangle}\,,
\end{equation}
where the computation of $\langle 0 | O_i|n\rangle$ with \cref{eq:eigenvectors} requires of the use of the associated finite-volume energies. It is also possible to define overlaps relative to other operators $O_j$. While this has the advantage that the energies are not needed, it may lead to misleading conclusions since the normalization of each operator is arbitrary.

In this work, we use the \textit{single-pivot procedure}~\cite{Blanton:2021llb,Bulava:2022vpq} to solve the GEVP, in which it is only solved at a single time $t_\text{p}>t_0$, 
\begin{equation}\label{eq:GEVPsinglepivot}
    \tilde{C}(t_\text{p})v_n(t_\text{p},t_0)=\lambda(t_\text{p},t_0)\tilde{C}(t_0)v_n(t_\text{p},t_0)\,,
\end{equation}
and only for the mean of the bootstrap samples. This GEVP can be rewritten as a normal eigenvalue problem,
\begin{equation}
    \tilde{C}^{-1/2}(t_0)\tilde{C}(t_\text{p})\tilde{C}^{-1/2}(t_0)\tilde{v}_n(t_\text{p},t_0)=\lambda(t_\text{p},t_0)\tilde{v}_n(t_\text{p},t_0)\,.
\end{equation}
where $\tilde{v}_n=\tilde{C}^{1/2}(t_0) v_n$, and solved with standard methods. The eigenvectors $\tilde{v}_n(t_\text{p},t_0)$ are used to rotate $\tilde{C}^{-1/2}(t_0)\tilde{C}(t)\tilde{C}^{-1/2}(t_0)$ for all time slices and bootstrap samples, and the principal correlation functions are taken to be the diagonal elements of the result. This procedure is found to be more stable than solving the GEVP on all samples or on all time slices. We find our results to be stable under reasonable changes of $t_0$ and $t_\text{p}$---see \cref{fig:GEVPcomparison} below---and use $(t_0,t_\text{p})=(3a,5a)$ in our analysis.

Before moving on, it is worth commenting about the applicability of the GEVP to our setup, as we are defining tetraquark operators in a different way at source and sink, this is, with a sparse definition of the momentum projection at source and an exact sum at sink. However, we are using random offsets on each configuration to define the sparse sublattice, and so averaging over multiple configurations leads to the non-sparsened result. Even if this was not the case, exact momentum projection at sink ensures that the only states that contribute to the sums in \cref{eq:realcorreatorthermal} are the same ones as those that would contribute if exact Fourier projection was also performed at source. Thus, we can apply the GEVP without further complications.

\subsubsection{Finite-volume energies from ratios of correlation functions}\label{sec:ratiofunctions}

To extract the finite-volume energies from the eigenvalues of the GEVP,  we use ratios between two- and single-particle correlation functions. This has been empirically found to cancel correlations between one- and two-particle correlation functions, and improve the finite-volume energy determinations~\cite{Umeda:2007hy,Bulava:2022vpq}. In particular, for those states that have a maximum relative overlap into a two-particle operator of the form $O_M(\bm{k}_1)O_M(\bm{k}_2)$\footnote{While the exact determination of the relative overlaps from \cref{eq:overlaps} requires to know the finite-volume energies a priori---see \cref{eq:eigenvectors}---, we find that one components of the eigenvector always dominates, and so the identification of the corresponding states only an estimate of the energy is needed, for example from computing the effective mass as some fixed time.}, we construct the following ratio
\begin{equation}\label{eq:ratiofunction}
    R_n(t)=\frac{\lambda_n(t)}{\partial_0[C_{M,\bm{k}_1}(t)C_{M,\bm{k}_2}(t)]}\,.
\end{equation}
To extract the finite-volume energies, we assume one-particle correlation functions to be dominated by a single state,
\begin{equation}
    C_{M,\bm{k}}(t)=A\cosh(E^\text{latt}_{M,\bm{k}} t')\,,
\end{equation}
and consider the generalized eigenvalues of shift-reweighted matrix of correlation functions to behave as $\lambda_n^{(0)}(t)$, given in \cref{eq:eigenvaluesdependence}.
These two expressions are then feed into \cref{eq:ratiofunction}, where we use a neutral finite-difference discretization of the time derivative in the \mbox{denominator}, to define the final fit function. Fitting this function to the lattice results, we determine an overall amplitude plus a result for the finite-volume energy, $E_n^\text{latt}$. We then perform a correction of the finite-volume energy
\begin{equation}
    E_n = E_n^\text{latt}-E_{M}^\text{latt}(\bm{k}_1)-E_{M}^\text{latt}(\bm{k}_2) + E_{M}^\text{cont}(\bm{k}_1)+E_{M}^\text{cont}(\bm{k}_2)\,,
\end{equation}
where, recall, $E_{M}^\text{cont}(\bm{k})$ is the single-meson energy defined using the continuum dispersion relation in \cref{eq:freemesonenergycontinuum}. 
This correction reduces leading discretization effects in the energy determinations~\cite{Hansen:2024cai}. 

Correlated fits to the ratios are performed over a range of times $t\in[t_\text{min},t_\text{max}]$ for each energy level independently, and the final result is extracted from where the fit results show a plateau. Typically, this is done by choosing a single fit range from visual inspection. In this work, instead, we average over multiple fit ranges, varying both $t_\text{min}$ and $t_\text{max}$,\footnote{We keep $t_\text{max}-t_\text{min}\geq 4$, to ensure the fits have at least two dof.} using weights based on the Akaike Information Criteria (AIC)~\cite{Jay:2020jkz,Neil:2023pgt},
\begin{equation}\label{eq:weightsplateaux}
    w_i \propto \text{exp}\left[-\frac{1}{2}\left(\chi_i^2-2N_i+2N_\text{par}\right)\right]\,,
\end{equation}
where $\chi_i$ is the chi-square of the fit, $N_i$ is the number of fit points used for the fit and $N_\text{par}=2$ is the number of fit parameters. 
If we let $E_{n,i}$ and $\sigma_{n,i}^2$ be the results for the energy and its variance for fit $i$, the final averaged results are computed as
\begin{equation}
\begin{array}{rl}
    \langle E_n \rangle = & \sum_i w_i E_{n,i}\,,\\[10pt]
    \sigma_n^2= & \sigma_{n,\text{stat}}^2+\sigma_{n,\text{syst}}^2\,,
    \end{array}
\end{equation}
where we have defined a statistical and systematic contribution to the variance,
\begin{equation}
    \begin{array}{rl}
       \sigma_{n,\text{stat}}^2 = & \sum_i w_i \sigma_{n,i}^2  \\
       \sigma_{n,\text{syst}}^2 = & \sum_i w_i E_{n,i}^2 - \langle E_n \rangle^2\,, 
    \end{array}
\end{equation}
where the latter is related to the choice between different fit ranges. 
In our bootstrap analysis, weights are determined from a fit of the mean, and are used for all bootstrap samples. Using the samples, the statistical part of the variance can be determined straightforwardly, but this is not the case for the systematic part. To incorporate the latter into the total variance of the bootstrap samples, we perform a rescaling
\begin{equation}
    E_{n,\alpha} = \langle E_n \rangle + \frac{\sigma_n^2}{\sigma_{n,\text{stat}}^2}(E_{n,\alpha} - \langle E_n \rangle)\,,
\end{equation}
where $\alpha$ labels the bootstrap sample and $\langle E_n \rangle$ refers to the bootstrap mean. This leads to a sample distribution with the same mean, and variance $\sigma_n^2$ that includes statistical and systematic contributions. The correlations between energy levels, on the other hand,  are kept the same, as the selection of fit range is performed independently for each level, and so the systematic error arising from the range choice does not affect the correlation between energy levels.

We note that, while the use of weights sometimes allows to automatically determine the plateau, one often needs to restrict the fit ranges taken into consideration. In particular, it is vital that $t_\text{min}$ is kept lower than times where the error blows out, and so we exclude large values of $t_{\rm min}$ from the average as the statistical errors are very large.  
Otherwise, such fits usually get an artificially low $\chi^2$ and may get large weights, leading to incorrect results. More sophisticated criteria~\cite{Neil:2022joj,Frison:2023lwb,Neil:2023pgt} may allow to take the error of the final result into account when computing the weights, but we restrict ourselves to the AIC in this work.

\begin{figure}[!t]
    \centering
    \includegraphics[width=0.75\textwidth]{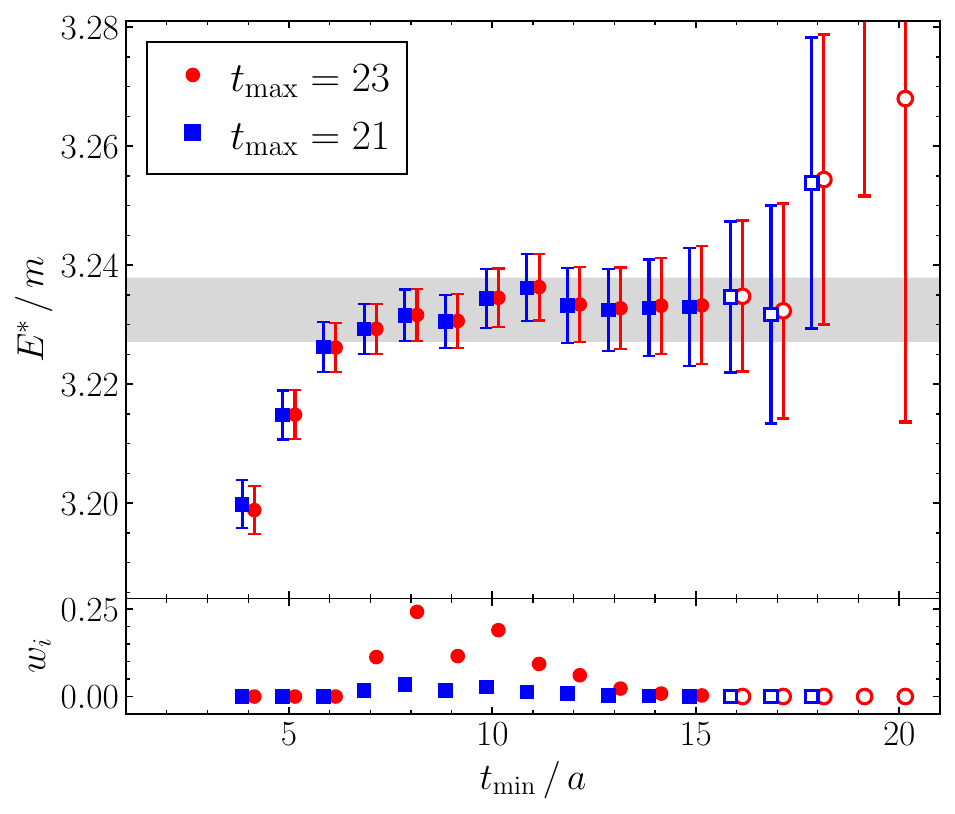}
    \caption{
        Best-fit results to \cref{eq:ratiofunction} for the ground-state energy in the CM, in the $\bm{P}=[0,0,1]$ frame of the $AA$ channel with $\Nc=3$, for different values of $t_\text{min}$ and two choices of $t_\text{max}$. The final result (gray band) is obtained by averaging the results using the weights from \cref{eq:weightsplateaux}, in the bottom panel. Empty points are manually excluded from the average.}
    \label{fig:plateaux}
\end{figure}

An example of the results of the fits is presented in \cref{fig:plateaux}, corresponding to the ground state in the $\bm{P}=(0,0,1)$ frame of the $AA$ channel for $\Nc=3$. Blue and red points in the upper panel represent the fit results for different fit ranges, with the blue band corresponding to the final averaged result. The lower panel indicates the weights associated to each fit range. Empty points are manually left out of the average.

\begin{figure}[hp!]
\centering
\includegraphics[width=\textwidth]{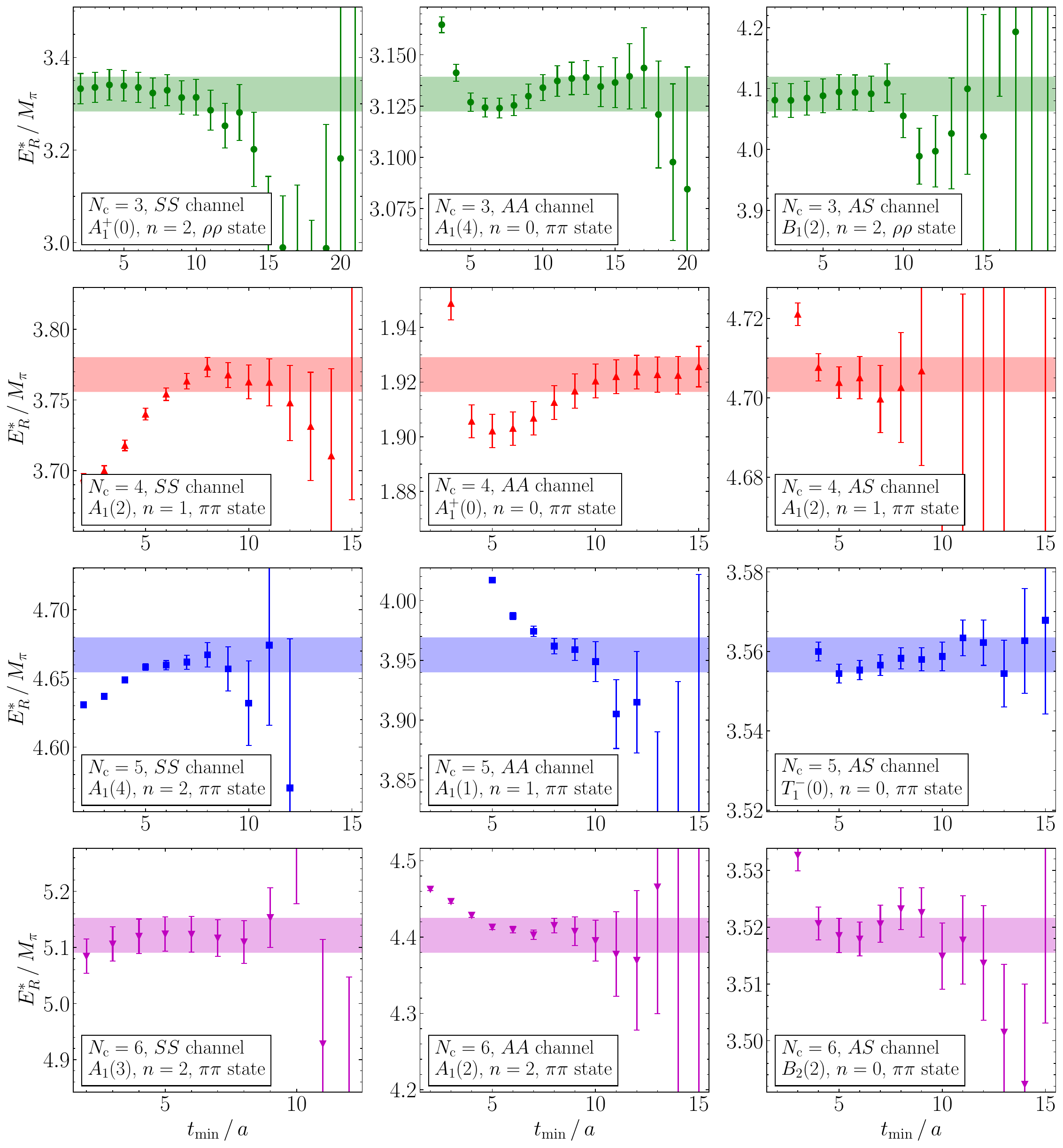} 
\caption{ Best-fit results for the finite-volume energies for a selection of states corresponding to different $\Nc$, channel, cubic-group irrep and total momentum, as indicated in each panel. Each point corresponds to the results of a fit for different $t_\text{min}$ with fixed $t_\text{max}$ as indicated. Horizontal bands are the final results obtained from averaging several fit ranges, including multiple $t_\text{max}$, as explained in \cref{sec:ratiofunctions}.}
\label{fig:plateauxallenergies}
\end{figure}

\begin{figure}[h!]
\centering
\includegraphics[width=0.93\textwidth]{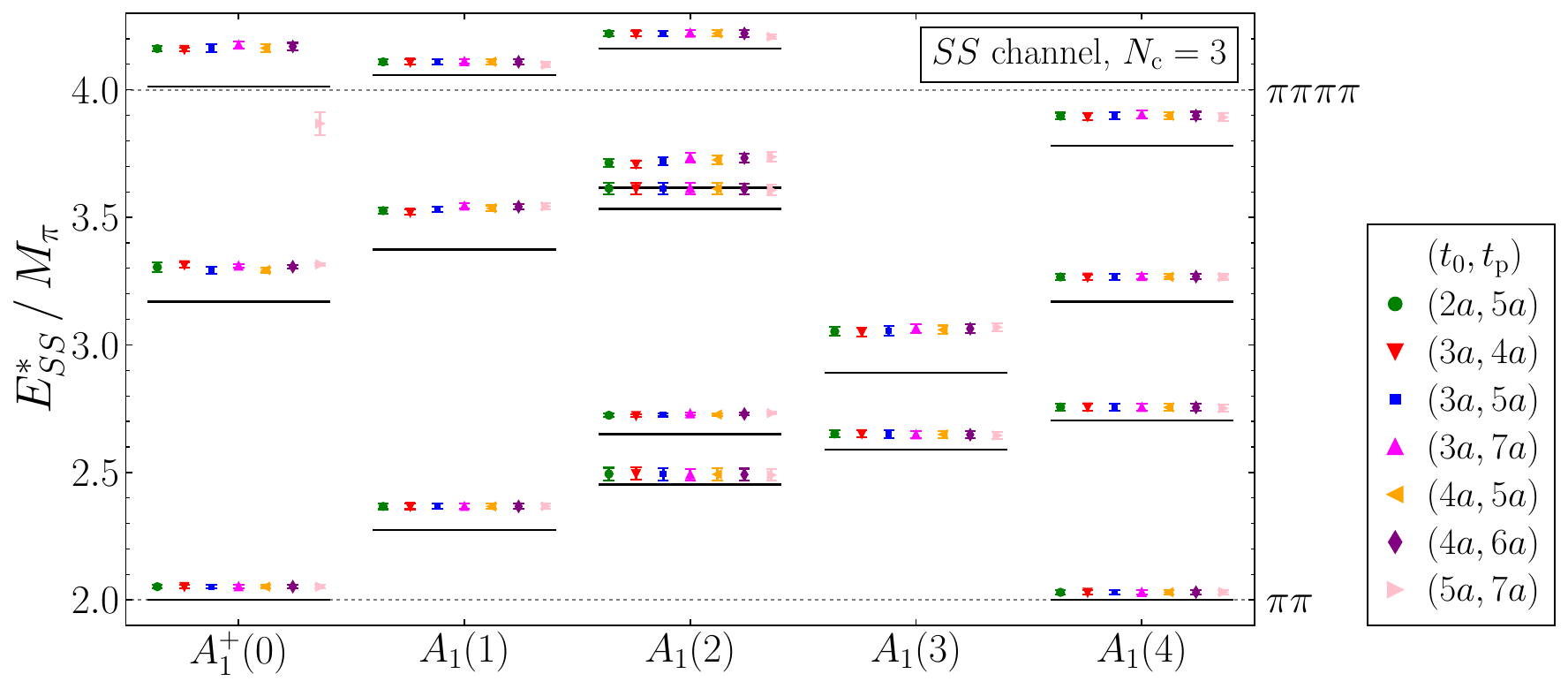}\\[0.35cm] 
\includegraphics[width=0.93\textwidth]{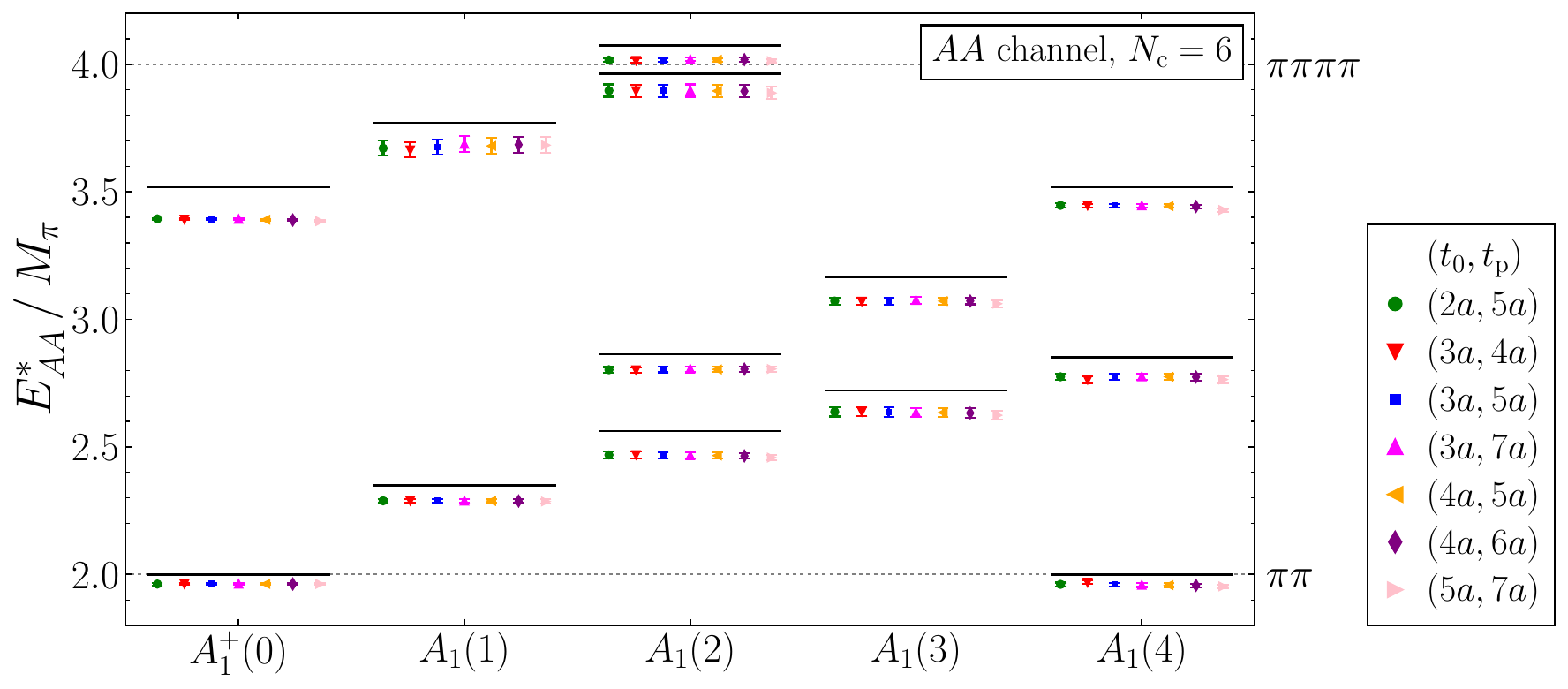} \\[0.35cm]
\includegraphics[width=0.93\textwidth]{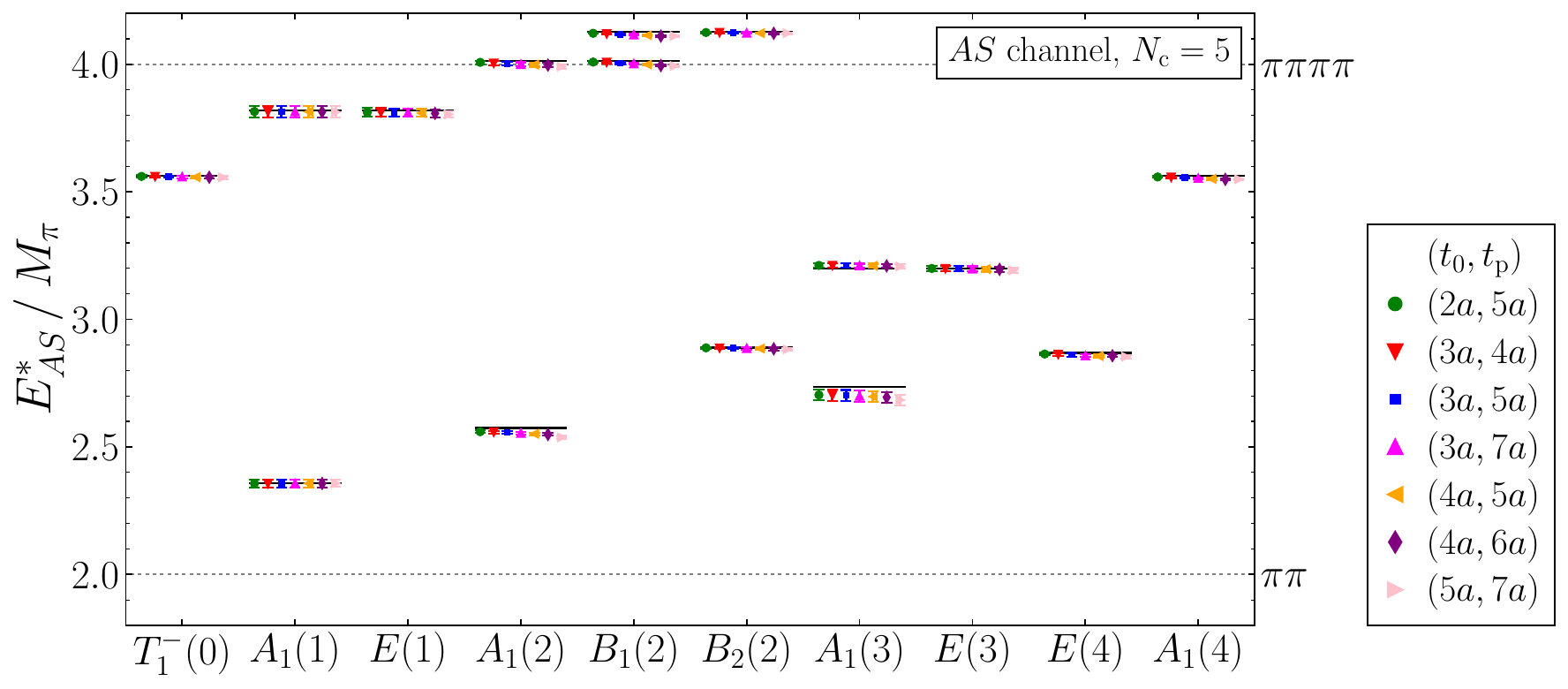} 
\caption{ Selection of results of the finite-volume energies of two-pion states for different choices of $t_0$ and $t_\text{p}$ used to solve the GEVP---see \cref{eq:GEVPsinglepivot}. Horizontal black segments indicate the non-interacting energies, and the dotted lines are the relevant two-pion inelastic thresholds. We set $(t_0,t_\text{p})=(3a,5a)$ for the analysis presented in this work. } 
\label{fig:GEVPcomparison}
\end{figure}

To conclude this section, we highlight that the procedure outlined to determine the finite-volume energies is found to sufficiently reduce excited-state contamination and allows us to confidently determine the finite-volume spectrum. As an illustration, we present in \cref{fig:plateauxallenergies} the results from different fit ranges for a selection of finite-volume states, together with the final averaged results. In all cases, a plateau can be identified, indicating a successful isolation of the lowest-lying states of each GEVP eigenvalue. We also note that our results are stable under reasonable changes of the parameters $t_0$ and $t_\text{p}$ used to solve the GEVP. This, as can be observed in \cref{fig:GEVPcomparison}, where we present a selection of results of the $\pi\pi$ energies for different combinations of these parameters. 

\subsection{The two-particle finite-volume formalism}

The two-particle energy spectra determined on the lattice can be used to constrain the infinite-volume scattering amplitude, using the so-called quantization conditions. In the case of two-particles, the formalism was first developed for two identical scalar particles~\cite{Luscher:1986pf,Luscher:1990ux}, and has since been extended to include any two particle system~\cite{Luscher:1991cf,Rummukainen:1995vs,Kim:2005gf,He:2005ey,Bernard:2008ax,Luu:2011ep,Briceno:2012yi,Briceno:2014oea,Gockeler:2012yj}. In general, it takes the form of a matrix equation,
\begin{equation}\label{eq:quantizationcondition}
\left.\text{det}\left[\cK_2^{-1}(E)+F(L,\bm{P};E)\right]\right|_{E=E_n}=0\,,
\end{equation}
where $\cK_2$ is the two-particle $K$-matrix, related to the infinite-volume scattering amplitude, and $F$ is a geometric factor that depends on the lattice geometry and the total momentum, and contains power-law finite-volume effects. The finite-volume energies are given by the solutions to this equation. We note that the matrices in the quantization condition can be projected to the different irreps of the cubic group and isospin channels, and the finite-volume energies can be determined independently for each of them.

In this work, we focus on studying single-channel pion-pion interactions, for which we identify those states associated to two pions using the relative overlap factors introduced in \cref{eq:overlaps}, and neglect interactions to states of two vector mesons. All the cubic-group irreps considered in this work are dominated by contributions from the lowest partial waves, as already mentioned: $s$ wave for the $SS$ and $AA$ channels, and $p$ wave for the $AS$ channel. For the analysis, we work in the lowest-partial wave approaximation, and neglect higher partial wave contributions. Higher-partial-wave effects are expected to be very small, as they only start at NLO in ChPT (NNLO in large $\Nc$ ChPT) and are suppressed as $p^{2\ell+1}$ close to threshold. Moreover, previous lattice studies of two-pion interactions have observed suppressed higher partial waves in the elastic region~\cite{Dudek:2010ew,Blanton:2021llb}.

In the single-channel and lowest-partial-wave approximation,  \cref{eq:quantizationcondition} can be reduced to an algebraic equation that relates each finite-volume energy to a point in the scattering phase shift. In the case of $s$-channel interactions, it takes the form,\\
\noindent\begin{equation}\label{eq:Luscheralgebraicswave}
    k\cot\delta_0=\frac{2}{\gamma L \pi^{1/2}}\cZ_{00}^{\bm{P}}\left(\frac{kL}{2\pi}\right)\,,
\end{equation}
where $k$ is the magnitude of the relative momentum in the CM, $\gamma$ is the boost factor to the CM frame and $\cZ$ is the generalized Lüscher zeta function~\cite{Luscher:1986pf,Luscher:1990ux,Rummukainen:1995vs}. In the case of $p$-wave interactions, on the other hand, the particular form of the quantization conditions depends on the total momentum and the irrep of the cubic group---see appendix A in  \rcite{Dudek:2012xn} for an explicit form for the cases of relevance to our work. 

\subsection{Fitting procedure to extract scattering amplitudes}\label{sec:fittingprocedure}

To constrain the scattering amplitudes using the quantization condition, we choose a parametrization of the scattering amplitude and the related phase-shift, eq.~({\ref{eq:kcot}}), that depends on some parameters, $\bm{c}$. The quantization condition allows one to obtain predictions of the finite-volume energies for this choice of the parameters, $E_n^\text{pred}(\bm{c})$, which can be compared to the lattice results, $E_n$, to constrain the values of the parameters that provide the best description of the phase-shift. More concretely, we minimize a chi-square function,
\begin{equation}
    \chi^2=\sum_{i,j} \left[E^\text{pred}_i(\bm{c})-E_i\right](C^{-1})_{ij}\left[E^\text{pred}_j(\bm{c})-E_j\right]\,,
\end{equation}
where $C^{-1}$ is the inverse covariance matrix of the finite-volume energies, and the indices \mbox{$i$ and $j$} label the different energy levels considered for a particular fit. In this work, we use the Nelder-Mead algorithm~\cite{Gao:2012guu} to find the best-fit parameters. Also, we use the Lüscher zeta function implementation from \rrcite{NPLQCD:2011htk,LuscherZetaGit}. 

We match our lattice results for two-pion energies to different parametrizations of the scattering amplitude. The effective-range expansion (ERE) is a model-agnostic parametrization based on the near-threshold behavior of the partial-wave-projected scattering phase shift,
\begin{equation}\label{eq:ERE}
    k^{2\ell+1}\cot \delta_\ell = \frac{1}{a_\ell}+\frac{1}{2}r_\ell k^2+...
\end{equation}
where $a_\ell$ is the scattering length\footnote{Note we use a sign convention for the scattering length that is negative for repulsive interactions and positive for attractive ones.} (although only $a_0$ has units of length) and $r_0$ is the effective range. This formula gives a description of the data near threshold, but its applicability range is limited in different cases. In particular, when the scattering amplitude presents an Adler zero below threshold, the effective range expansion has a convergence radius given by the distance to this zero. This is expected to be the case for the $SS$ and $AA$ channels. At LO in ChPT, both are predicted to have an Adler zero at $q^2=-\Mpi^2/2$ .

The convergence of the ERE can be extended in these cases by the ad-hoc inclusion of a pole in the amplitude. A widely used parametrization is the so-called modified ERE (mERE)~\cite{Yndurain:2002ud,Pelaez:2019eqa},
which for $s$-channel amplitudes takes the form,
\begin{equation}\label{eq:modifiedERE}
    \frac{k}{\Mpi}\cot\delta_0 = \frac{\Mpi E^*}{E^{*2}-2z^2}\left[B_0+B_1 \frac{k^2}{\Mpi^2}+...\right]\,,
\end{equation}
with $E^*=2(k^2+\Mpi^2)$ the total CM energy. In what follows, we treat $B_0$ and $B_1$ as fit parameters to be determined from matching to the lattice results, while we set $z=\Mpi$, corresponding to the LO ChPT prediction. Leaving $z$ as a fit parameter has been found to lead to very unstable fits. The parameters of the mERE can then be related to those of the ERE after expanding around threshold. Up to $\cO(k^2/\Mpi^2)$ and using $z=\Mpi$, one finds,
\begin{equation}\label{eq:EREandmERErelation}
\Mpi a_0=\frac{1}{B_0}\,,\quad\quad\quad\quad \Mpi^2 r_0 a_0 =2\frac{B_1}{B_0}-3\,.   
\end{equation}

An alternative parametrization uses the ChPT predictions presented in \cref{sec:chptSUpredictions,sec:chptUpredictions,app:chptamplitudes}, which depend on the values of the LECs beyond LO. In the $SS$ and $AA$ channels, we determine the scattering phase shift using the IAM, as described in \cref{sec:IAM}, which we found leads to much better convergence away from threshold. In the $AS$ channel, which vanishes at LO, we use standard ChPT predictions. Also, we use the mean values of $\Fpi$ from \rcite{Baeza-Ballesteros:2022azb}, which are summarized in \cref{tab:ensembles}, and choose a renormalization scale related to $4\pi \Fpi$, following \rrcite{Hernandez:2019qed,Baeza-Ballesteros:2022azb},
\begin{equation}
    \mu^2=\frac{3}{\Nc}(4\pi\Fpi)^2\,.
\end{equation}
In the case of the U($\Nf$) theory, we take the mass of the $\eta'$ to be related to that of pions via the Witten-Veneziano formula, 
\begin{equation}
    \frac{\Metap^2}{(4\pi\Fpi)^2}=\frac{\Mpi^2}{(4\pi\Fpi)^2}+\frac{c_0}{\Nc^2}\,,
\end{equation}
with $c_0= 6.5$, determined from \cref{eq:WittenVeneziano} and the topological susceptibility from \rcite{Ce:2016awn}.

The ensembles used in this work have a single value of $\Mpi$. This implies we cannot distinguish the effect of the $K$ LECs in \cref{eq:UchptNLOpredictionsSSAA} from those of the $L$ LECs in the U($\Nf$) theory. We opt to combine them in the fit parameters,
\begin{equation}\label{eq:LECsredefinition}
    \begin{array}{rl}
       32\tilde{L}_{SS,AA}  & = 32 L_{SS,AA}+\frac{\Mpi^2}{\Fpi^2}K_{SS,AA}\,,  \\[10pt]
        32\tilde{L}'_{SS,AA}  & = 32 L'_{SS,AA}+\frac{\Mpi^2}{\Fpi^2}K'_{SS,AA}\,,
    \end{array}
\end{equation}
Both $\tilde{L}_{SS,AA}$ and $\tilde{L}'_{SS,AA}$ have the same large-$\Nc$ scaling properties as $L_{SS,AA}$ and $L'_{SS,AA}$, and a common large $\Nc$ limit.
In general, this redefinition will affect both leading and subleading $\Nc$ terms in the LECs. However, at the order we are working in the counting of \cref{eq:countinglargeNcChPT}, one can parametrize $K_R=K^{(0)}\Nc^2+\cO(\Nc)$, and so only the leading part of $L_R$ is affected, with corrections to subleading terms originating from higher orders in the perturbative expansion. 

Finally, we make some cuts to our data for the fits. We only consider pion-pion energy levels below the four-pion threshold, and work in the single-channel scenario, neglecting possible interactions to other two-particle states, and in particular, to states of two vector mesons. In the $AA$ channel, we observe our results for the scattering phase shift to be systematically shifted for the $|\bm{P}|^2=4(2\pi/L)^2$ moving frame---see  \cref{fig:SSAAkcot}. We believe these effects are related to discretization effects present in this channel---see \rcite{Baeza-Ballesteros:2022azb}--- or contamination from higher partial waves, and opt to leave these states out from the analysis. Other choices of data cuts, such as discarding energies above some energy $E_\text{cut}<4\Mpi$, are found to have very minor impact on our final results.

\section{Results for the finite-volume energy spectrum}\label{sec:resultsenergies}

\subsection{Single-meson energies}

We present results for the single-meson masses in \cref{tab:singlemesonmasses}. We note there is a small mistuning of the pion masses between the different ensembles, specially for that with $\Nc=4$.
Also, we observe that the $\rho$ meson is stable in our ensembles, although it lies very close to the two-pion threshold in most cases. For the axial and singlet mesons, these results are just estimated, as we are neglecting interactions to two-meson states, into which they could decay. A precise determination of their mass should thus come from a study of two-meson interactions. In the case of the vector meson, which is stable in all ensembles, interactions with two-particle states only lead to exponentially-suppressed finite-volume effects~\cite{Luscher:1991cf}.

\begin{figure}[!t]
    \centering
    \includegraphics[width=0.65\textwidth]{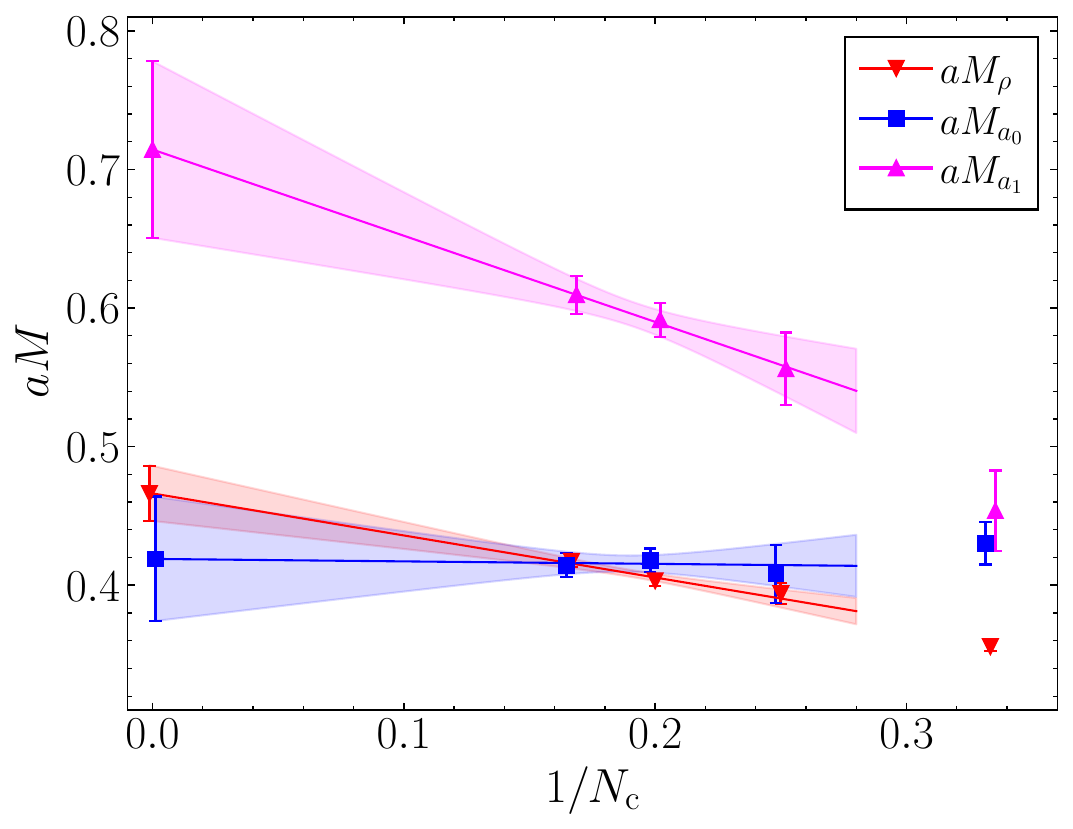}
    \caption{
        Results for the non-singlet meson masses extracted from single-meson correlation functions, together with the results at large $\Nc$, using $\Nc=4-6$ results for the linear extrapolation.  We neglect possible interactions to two-particle states in the mass determination. A small offset is added for better legibility.
        }
    \label{fig:singlemesonmasses}
\end{figure}

In \cref{fig:singlemesonmasses}, we study the dependence of the vector-, scalar- and axial-meson masses on $\Nc$, where we also represent a linear large $\Nc$ extrapolation including $\Nc=4-6$. These results are summarized in \cref{tab:singlemesonmasses}, where we also show those from a quadratic extrapolation including all $\Nc$. We find that including $\cO(\Nc^{-2})$ terms allows to describe all our data. However, we do not have enough values of $\Nc$ to get an accurate large $\Nc$ extrapolation. 

For both vector and axial mesons, the mass seems to become larger with $\Nc$. For scalar mesons, on the other hand, the $\Nc$ dependence is milder, becoming slightly lighter at \mbox{large $\Nc$}. Studies based on chiral effective models have predicted vector and scalar mesons to have the same mass in the large $\Nc$ limit~\cite{Nieves:2009ez,Nieves:2011gb}. We find that our extrapolated results are consistent with these expectations. 

The values for the meson masses at large $\Nc$ are compared in \cref{tab:singlemesonmassescomparison} against other results in the literature, obtained using quenched QCD simulations~\cite{Bali:2013kia}, which should be appropriate to study the leading $\Nc$ behavior, as well as those using volume-reduction techniques~\cite{Perez:2020vbn}.  We present the results in units of the string tension, $\sigma$, which we determine using the large-$\Nc$ result $\sqrt{8t_0\sigma}=1.078(9)$ from \rcite{Bonanno:2023ypf} together with the result for $\sqrt{t_0}$ at large $\Nc$, $\sqrt{t_0}=0.1450(39)\times\sqrt{8/9}$ fm~\cite{Hernandez:2019qed}, where the factor of $\sqrt{8/9}$ is an approximate linear correction that accounts for the different definition of the $t_0$ scale at large $\Nc$~\cite{Bonannopersonal}.\footnote{Refs.~\cite{Hernandez:2019qed, Bonanno:2023ypf} use different definitions of $t_0$, $$\langle t^2E(t)\rangle_{t=t_0}=\frac{3}{8}\frac{\Nc^2-1}{\Nc}c\quad\text{in ref.~\cite{Hernandez:2019qed}}\,, \quad\quad\quad \langle t^2E(t)\rangle_{t=t_0}=\frac{\Nc}{3}c \quad\text{in ref.~\cite{Bonanno:2023ypf}}\,,$$
 where $c=0.3$ and $E(t)$ is the field strength tensor. The correction factor of $\sqrt{8/9}$ arises from working at large $\Nc$ and approximating the dependence of $\langle t^2E(t)\rangle$ on $t$ as linear in the range of $t$ of interest~\cite{Luscher:2010iy}.} Also, the comparison is performed at a pion mass of $\Mpi/\sqrt{\sigma}=1.01(3)$, obtained averaging over our ensembles.\footnote{To obtain the results from \rcite{Bali:2013kia} at this pion mass, we use eq. (3.15) in that paper to determine the corresponding PCAC mass at large $\Nc$, and input it into eqs. (3.19) and (3.28), together with the results from table 24. In the case of \rcite{Perez:2020vbn}, we use eq. (5.1) from that work complemented by table 10.} We note that the results from \rcite{Perez:2020vbn} are extrapolated to the continuum, while those from \rcite{Bali:2013kia} are presented at fixed lattice spacing. The impact of cutoff effects on the results from \rcite{Bali:2013kia} were studied in \rrcite{Bali:2013fya,Castagnini:2015ejr}, where they were found to be of the order of $\sim 10-15\%$ for the meson masses considered here. However, no continuum-extrapolated results are presented for the chiral dependence, so we limit our comparison to \rcite{Bali:2013kia}. Overall, we find that our results for the meson masses at large $\Nc$ extracted from a quadratic extrapolation are consistent with those from \rrcite{Bali:2013kia,Perez:2020vbn}. A more accurate comparison would require calculations at larger values of $\Nc$.

\begin{table}[b!]
\centering
\begin{tabular}{ccccc}
\toprule
  $\Nc$ & $\Gamma=\gamma_5\,(aM_\pi)$ & $\Gamma=\gamma_i\, (aM_\rho)$ & $\Gamma=\mathbbm{1}\,(aM_{a_0})$  & $\Gamma=\gamma_5\gamma_i\,(aM_{a_1})$  \\ \midrule 

3 & 0.2129(7) &  0.356(4) & 0.430(15) & 0.45(3) \\  
4 & 0.2027(6) &  0.394(7) & 0.408(21) & 0.56(3) \\  
5 & 0.2132(5) &  0.403(4) & 0.408(14) & 0.591(12) \\ 
6 & 0.2169(5) &  0.417(4) & 0.405(20) & 0.610(14) \\ 
$\infty$ (linear) & --- &  0.466(20) & 0.42(5) & 0.71(6) \\ 
$\infty$ (quad.) & --- &  0.46(6) & 0.44(12) & 0.59(19) \\ 
\bottomrule 
\end{tabular}
\caption{Masses of non-singlet mesons, characterized by different choices of Dirac structure, $\Gamma$, in \cref{eq:singlemesonoperator}. The results are determined from the single particle correlator, neglecting interactions with multiparticle states. We also present the result of two extrapolations to large $\Nc$: a linear extrapolation obtained using the $\Nc=4-6$ results, and a quadratic extrapolation using all $\Nc$ data. 
}

\label{tab:singlemesonmasses}
\end{table}

\begin{table}[t!]
\centering
\begin{tabular}{cccc}
\toprule
    & $M_\rho/\sqrt{\sigma}$ & $M_{a_0}/\sqrt{\sigma}$  & $M_{a_1}/\sqrt{\sigma}$  \\ \midrule 

Linear fit &   2.22(12) & 2.01(25) & 3.4(3) \\  
Quadratic fit &   2.2(3) & 2.1(6) & 2.8(0.9) \\  
\rcite{Bali:2013kia} &   1.74(14) & 2.63(14) & 3.05(12) \\  
\rcite{Perez:2020vbn} &   1.86(14) & 2.09(14) & 3.15(12) \\\bottomrule 
\end{tabular}
\caption{Comparison of our results for the meson masses at large $\Nc$, in units of the string tension, $\sigma$, to previous results in the literature, obtained using quenched QCD simulations~\cite{Bali:2013kia} and volume-reduction techniques~\cite{Perez:2020vbn}. The comparison is performed at a pion mass $\Mpi/\sqrt{\sigma}=1.01(3)$, corresponding to an average over our ensembles. See the main text for further details on the comparison.
}

\label{tab:singlemesonmassescomparison}
\end{table}

\begin{figure}[!t]
    \centering
    \includegraphics[width=0.82\textwidth]{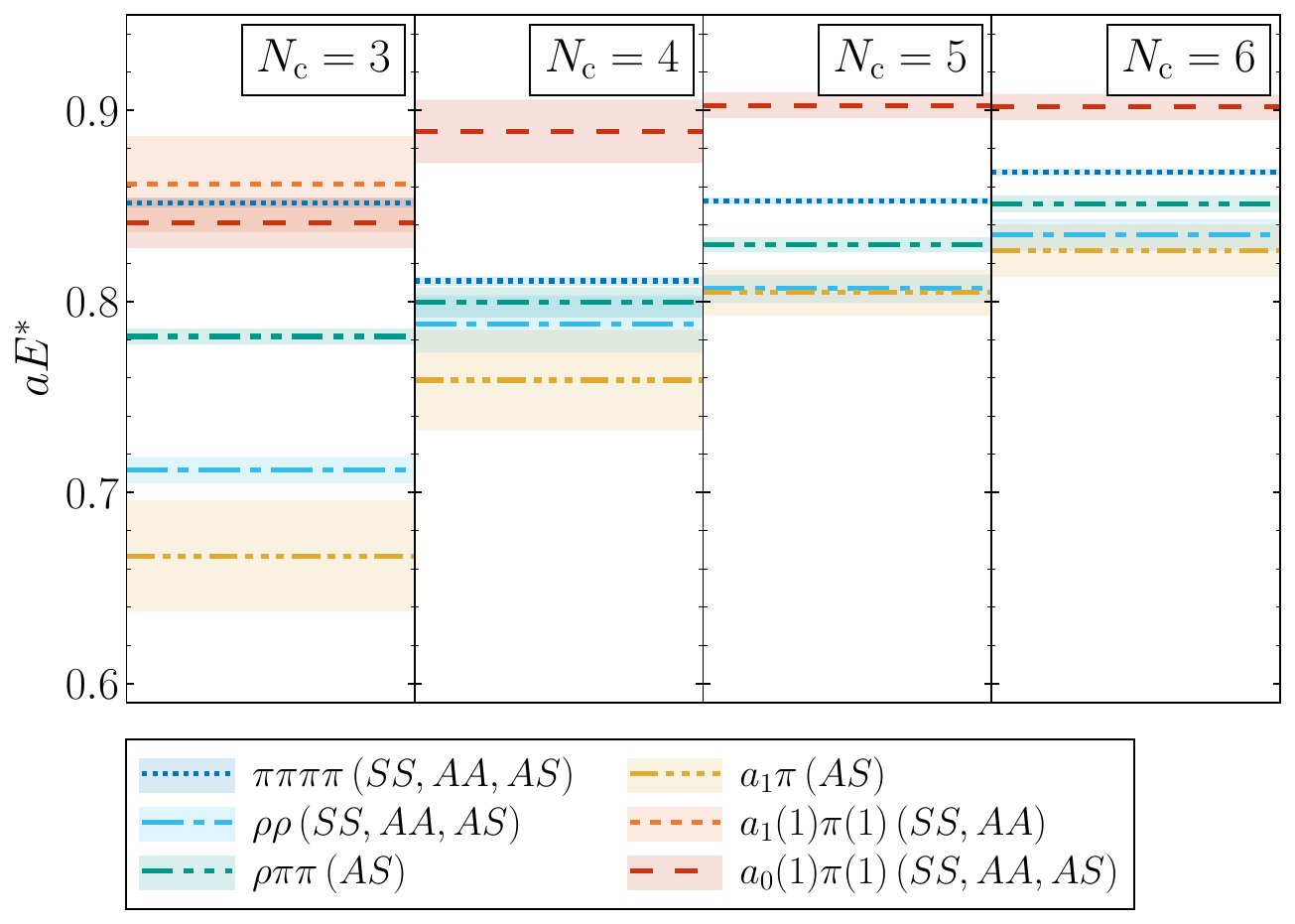}
    \caption{
        Summary of some of the relevant inelastic thresholds and non-interacting states composed of non-singlet mesons, which are present in our ensembles. We indicate in which channels they appear in the legend. 
        }
    \label{fig:inelasticthreshold}\vspace{0.2cm}

    \centering
    \includegraphics[width=0.75\textwidth]{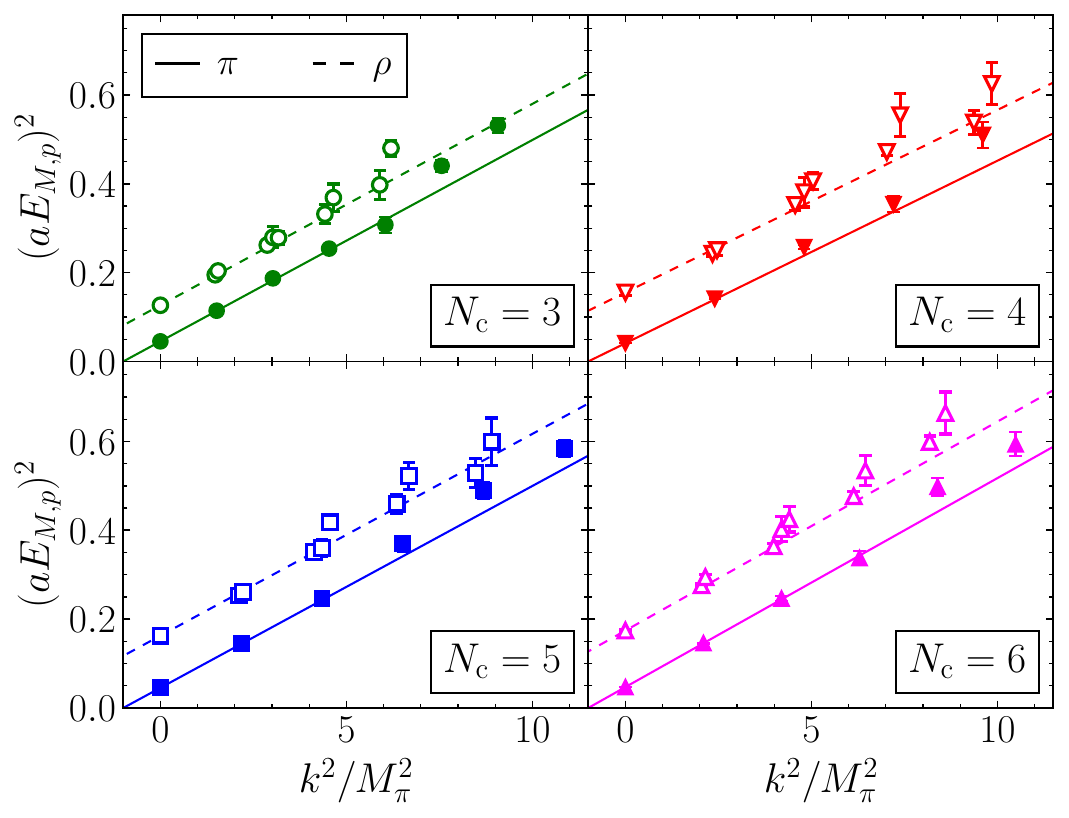}
    \caption{
        Results for the single-meson energies for different values of the momentum, together with the continuum dispersion relation. Results are presented for pions (filled points and solid line) and vector mesons (empty points and dashed line). 
        }
    \label{fig:dispersionrelation}
\end{figure}

Using the results for the meson masses, we determine the location of the most relevant inelastic thresholds that appear in the two-particle sector. These are represented in \cref{fig:inelasticthreshold}, where the legend indicates in which channels they are present. For completion, we also indicate the location of some finite-volume states in the non-interacting limit. Note that we have not studied isoscalar mesons, and the mass of the scalar singlet---known as the $\sigma$---may also lie below $2\Mpi$, meaning that the $\pi\pi\sigma$ threshold may also be present below that of four pions. However, for the analysis presented below, we neglect mixing to any such state. 

We also study the lattice dispersion relation for pions and vector mesons. The results are presented in \cref{fig:dispersionrelation}, together with the expected continuum dependence in \cref{eq:freemesonenergycontinuum}. 
We note that, in some cases, a vector meson state with some given momentum can be projected to several irreps of the cubic group, leading to independent determinations of the energy at that momentum. We observe that our results follow the continuum dispersion relation closely for energies $(aE_{M,p})^2\lesssim 0.3$. Above this threshold, lattice results start to separate from the continuum expectation due to cutoff effects.

\subsection{Results for the finite-volume energy spectrum}

\begin{figure}[h!]
\phantom{AAAA}
\centering
\includegraphics[width=\textwidth]{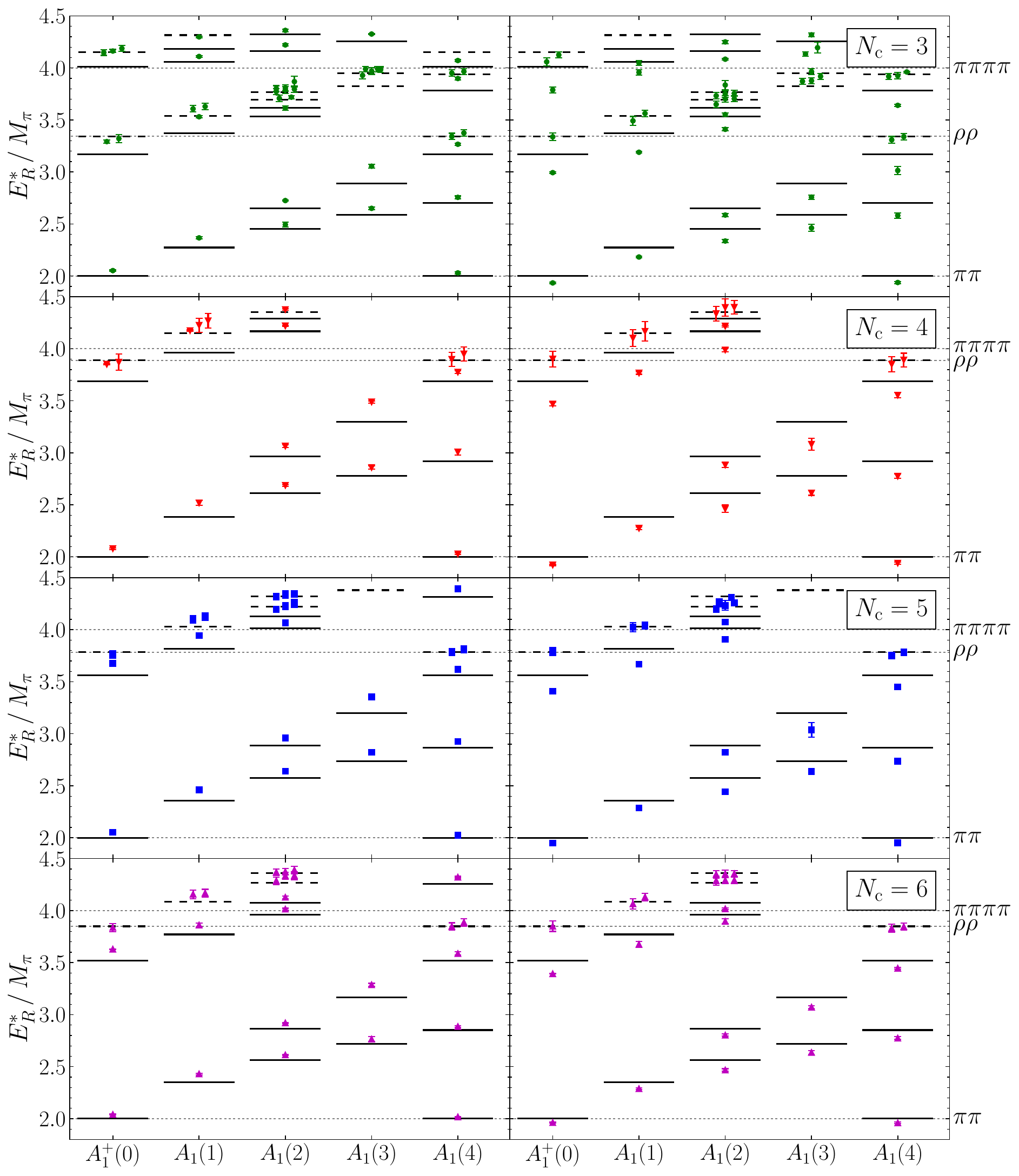} 
\caption{\label{fig:SSAAenergies} Results for the finite-volume energy spectra in the CM frame for the $SS$ (left) and $AA$ (right) channels. Each column corresponds to a different irrep of the cubic group and momentum frame, with the number in parenthesis indicating $|\bm{P}|^2$ in units of $(2\pi/L)^2$. Horizontal segments are the free energies of two pions (solid) and two vector mesons (dashed), while dotted horizontal lines are the most relevant inelastic thresholds---see \cref{fig:inelasticthreshold} for a more detailed summary.}
\end{figure}

\begin{figure}[!h]
    \centering
    \includegraphics[width=\textwidth]{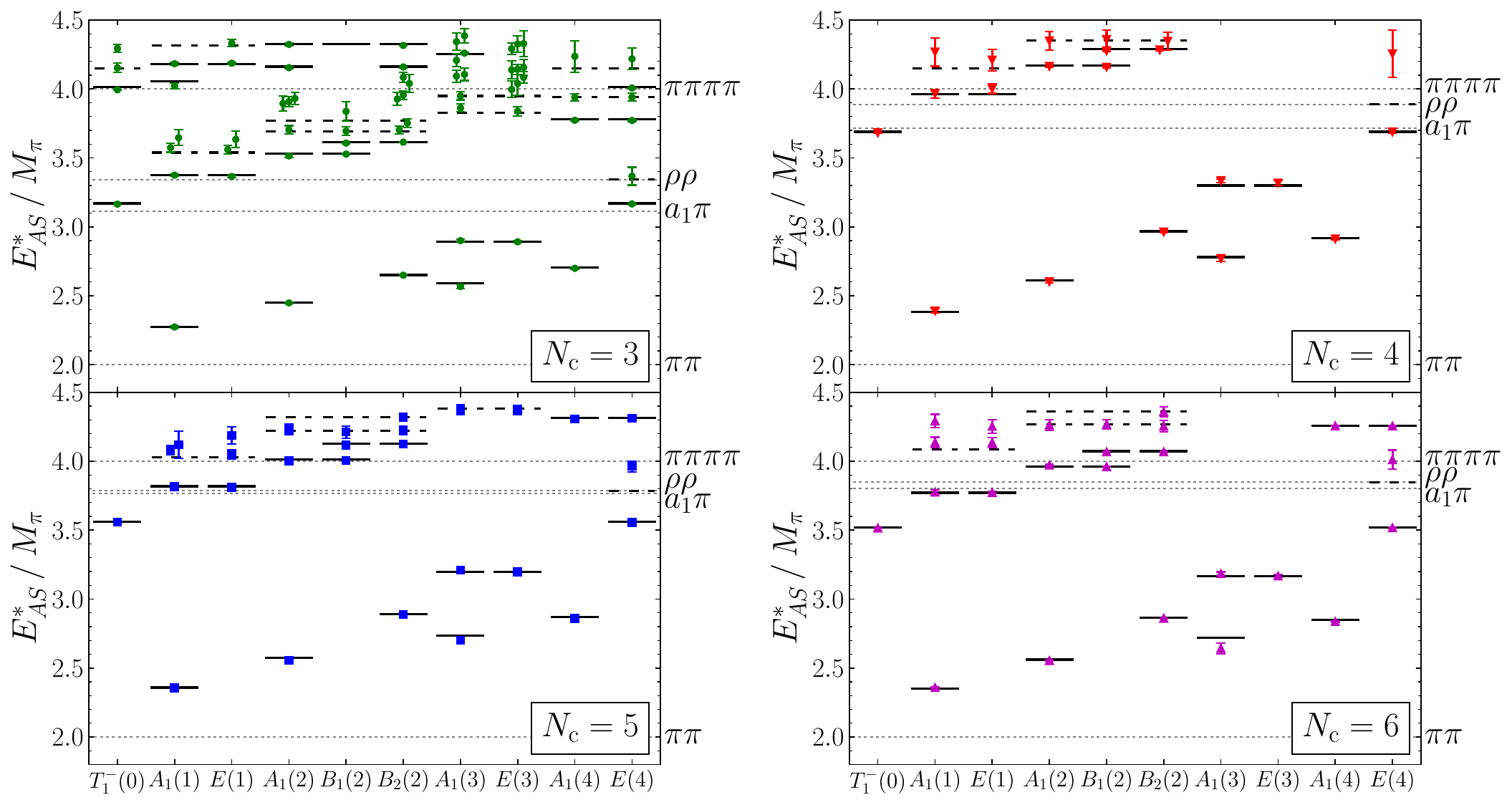}
    \caption{
        Same as \cref{fig:SSAAenergies} for the $AS$ channel.}
    \label{fig:energyAS}\vspace{-0.3cm}
\end{figure}

Using the procedure presented in \cref{sec:energyextraction}, we determine the finite-volume energy spectrum for the $SS$, $AA$ and $AS$ channels for all four ensembles considered. The results  in the CM frame are presented in \cref{fig:SSAAenergies} for the $SS$ and $AA$ channels, and in \cref{fig:energyAS} for the $AS$ channel. In all cases, we also indicate the non-interacting $\pi\pi$ and $\rho\rho$ energies (solid and dashed black lines, respectively) and the most relevant inelastic thresholds (gray dotted and dashed-dotted lines). For a detailed summary of other relevant inelastic thresholds, see \cref{fig:inelasticthreshold}. 

\begin{figure}[!t]
    \centering
    \includegraphics[width=\textwidth]{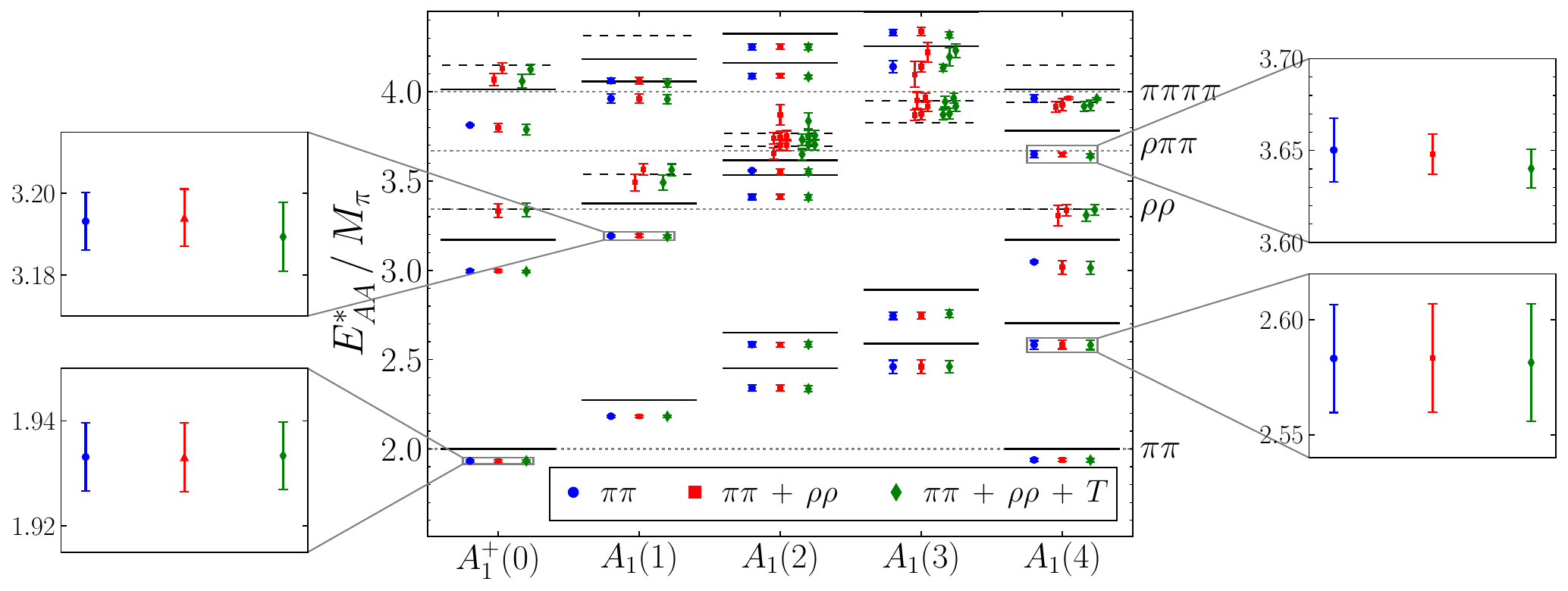}
    \caption{
        Finite-volume energy spectra of the $AA$ channel in the CM with $\Nc=3$, with and each marker type referring to a different set of operators used to compute the matrix of correlation functions.}\vspace{0.4cm}
    \label{fig:zoomedinenergies}

    \centering
    \includegraphics[width=0.86\textwidth]{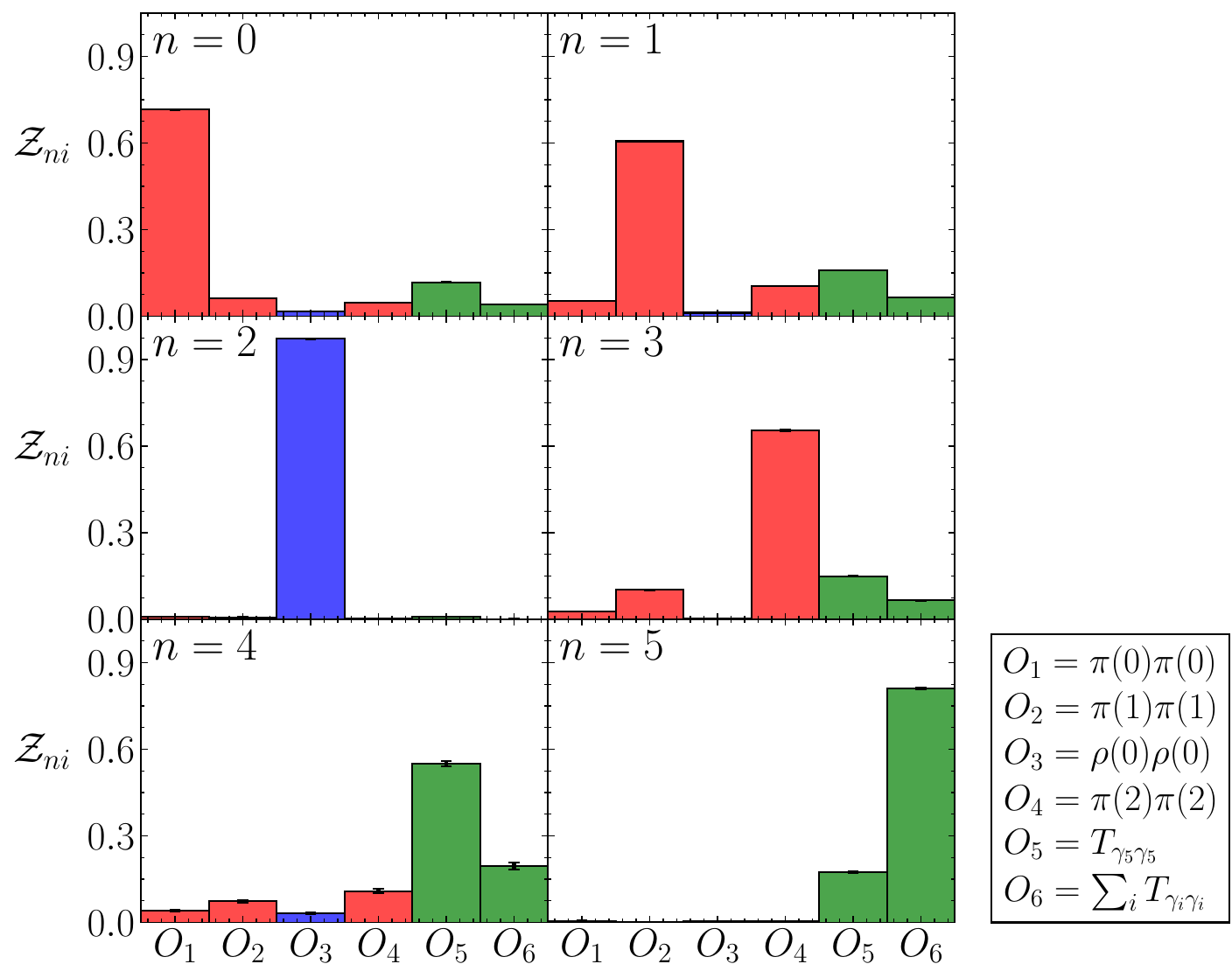}
    \caption{
         Relative overlaps of the lowest-lying finite-volume states, $|n\rangle$, with the operators used to compute the matrix of correlation functions---see \cref{eq:overlaps}---, for the rest-frame $A_1^+$ irrep in the $AA$ channel with $\Nc=3$. Different colors correspond to different types of operators: $\pi\pi$ (red), $\rho\rho$ (blue) and local tetraquarks (green). Numbers in parenthesis in the legend indicate the magnitude squared of the single-particle momenta, in units of $(2\pi/L)^2$.}
    \label{fig:histogram}
\end{figure}

We study the impact of the different operators in the determination of the finite volume energies. We have analyzed the effect of varying the operator set used to extract the finite-volume energies. Results for the energy spectra of the $AA$ channel for $\Nc=3$ are presented in \cref{fig:zoomedinenergies} for different choices of the operator set used to compute the matrix of correlation functions. We observe that the inclusion of the $\rho\rho$ operators leads to the determination of a large number of energy levels, which are associated to states of two vector mesons. However, it has minimal impact on the states associated to two pions. The inclusions of tetraquark operator in the $AA$ channel has a similar effect on $\pi\pi$ states, helping to improve the determination and slightly reduce the error for excited states. This is more clearly seen in the zoomed-in panels of \cref{fig:zoomedinenergies}.

We also investigate the relative overlaps of the different operators into the lowest-lying finite-volume states, which serve as an indication of the effect of each operator in the extraction of finite-volume energies.  An example of the relative overlaps is shown in \cref{fig:histogram} for the rest-frame $A_1^+$ irrep in the $AA$ channel with $\Nc=3$. For this example, the eigenvectors are determined from solving a GEVP in which only the correlation functions between the operators shown are considered, while we use the six lowest-lying finite-volume energies determined from the full analysis as presented above. Each panel represents the relative overlap of one operator into the six lowest lying states. Numbers in parenthesis in the naming of the operators indicate the magnitude of the momenta in units of $2\pi/L$.

We observe that, in general, operators have a dominant overlap into those states that lie closest to the free energy associated to the operator. For example, for the $\pi(1)\pi(1)$ operator, this would correspond to a state of two pions with relative momenta $|\bm{p}|=2\pi/L$. Also note that the $n=2$ state in this case is predominantly a $\rho\rho$ state. Finally, we observe that tetraquark operators have maximum overlap into the most excited states, which lie above the range of interest, and possibly correspond to more energetic two-meson states. An important remark is that adding tetraquark operators does not lead to new energy levels, but only helps cleaning up excited states in the low-lying spectrum. Similar qualitative behavior has been seen in the case of the $T_{cc}$~\cite{Vujmilovic:2024snz,Stump:2024lqx}.

\section{Results for the scattering amplitudes}\label{sec:scatteringfits}

\subsection{ Model-agnostic fits at fixed $\Nc$}\label{sec:fitsERE}

We fit our lattice results for the $SS$ and $AA$ channels to the mERE in \cref{eq:modifiedERE}. We perform fits at fixed $\Nc$ for each channel separately, fix the location of the Adler zero $z=\Mpi$ and constrain the values of $B_0$ and $B_1$.  Results are presented in \cref{tab:SSAAmEREresults}, together with the $\chi^2$ of the fits and the number of degrees of freedom (dof). We observe very good description of the data in the $SS$ channel, while the value of the $\chi^2$ is systematically higher in the case of the $AA$ channel. This may indicate the presence of unaccounted systematic effects. For example, these could be related to cutoff effects, which were found to be significant in the $AA$ channel in \rcite{Baeza-Ballesteros:2022azb}, and limit the applicability of Lüscher's formalism. A detailed investigation of their impact would require ensembles at several lattice spacings, and also the use of an adapted version of Lüscher quantization condition~\cite{Hansen:2024cai}. 

\begin{table}[b!]
\centering
\begin{tabular}{ccccc}
\toprule
 Channel     &    $\Nc$ &    $B_0$ & $B_1$ & $\chi^2\,/\,\text{dof}$\\  \midrule 

\multirow{4}{*}{$SS$} & 3 & $-2.37(4)$  & $-2.59(18)$ & $10.4/12=0.87$ \\
& 4 & $-3.11(9)$ & $-3.21(19)$ &  $5.5/8=0.69$ \\
& 5 & $-3.81(5)$ & $-3.76(20)$ & $22.8/8=2.8$ \\
& 6 & $-4.46(5)$ & $-4.09(29)$ & $9.9/8=1.24$ \\ \midrule
\multirow{4}{*}{$AA$} & 3 & $0.932(17)$  & $2.76(13)$ & $17.4/10=1.74$ \\
& 4 & $1.78(5)$ & 3.24(23) &  $16.1/7=2.31$ \\
& 5 & $2.52(5)$ & 3.57(17) & $22.0/7=3.14$ \\
& 6 & $3.33(7)$ & 3.86(19) & $19.3/7=2.75$ \\   \bottomrule
\end{tabular}
\caption{Results from single-color and single-channel fits to a mERE with $z=\Mpi$ fixed, for the $SS$ and $AA$ channels. We do not include data for $|\bm{P}|^2=4(2\pi/L)^2$ for the $AA$ channel, due to large observed cutoff effects.  }

\label{tab:SSAAmEREresults}
\end{table}

\begin{figure}[!t]
    \centering
\includegraphics[width=0.47\textwidth]{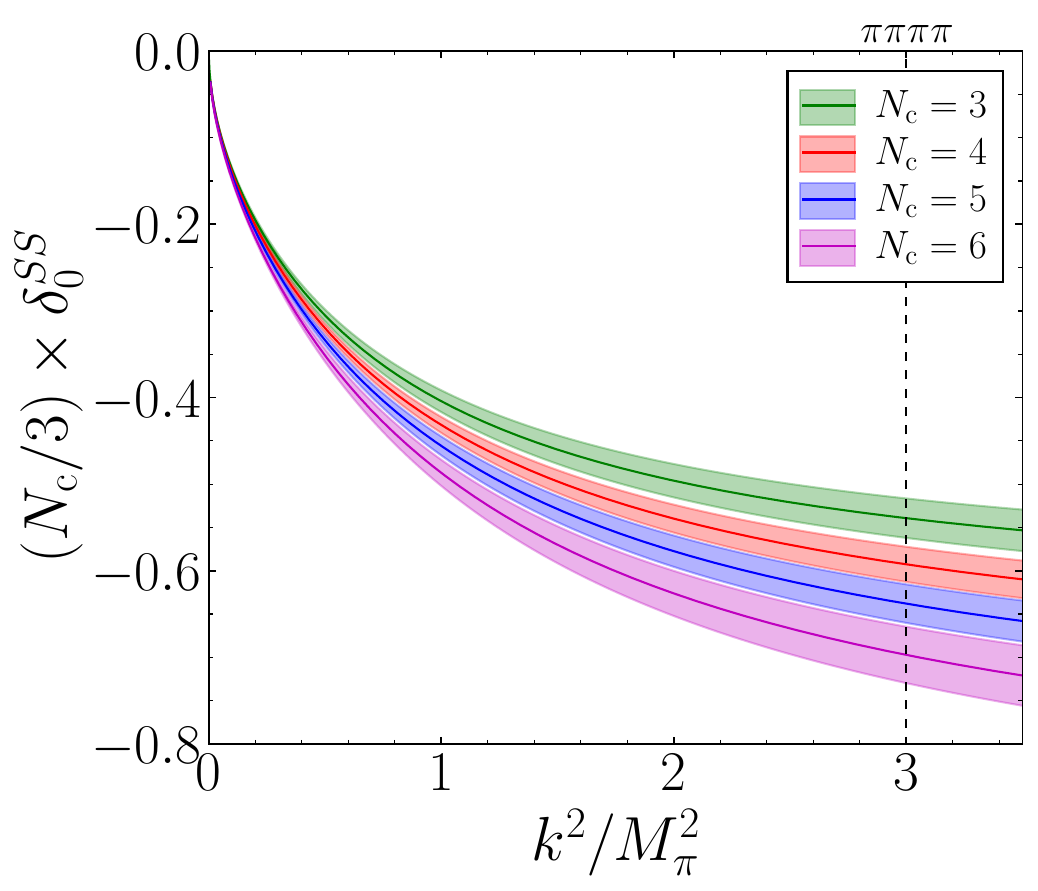} \hspace{0.cm}
\includegraphics[width=0.45\textwidth]{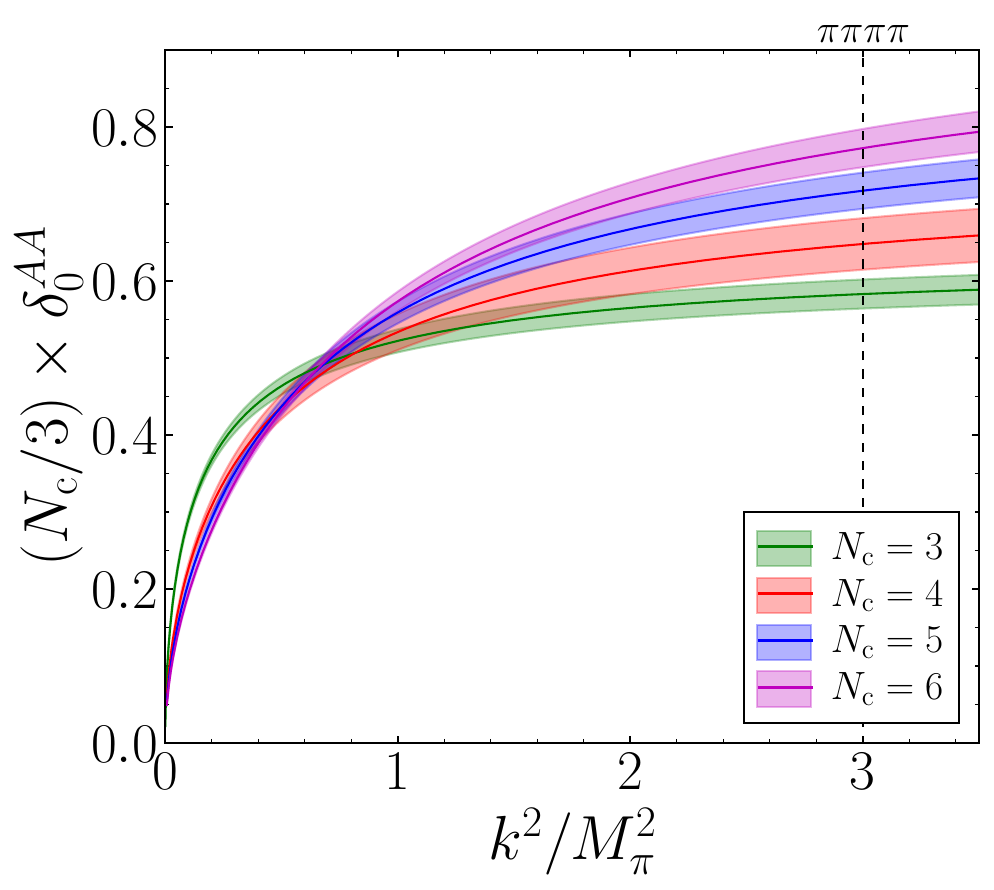}
    \caption{Fit results for the scattering phase shift in the $SS$ and $AA$ channels, obtained from single-$\Nc$ and single-channel fits to a mERE with $z=\Mpi$ fixed. The phase shifts are multiplied by a factor that eliminates the leading $\Nc$ scaling. We only indicate the four-pion inelastic threshold, for a summary of other relevant threshold, see \cref{fig:inelasticthreshold}. }
    \label{fig:collapse}\vspace{0.3cm}

    \centering
    \includegraphics[width=0.88\textwidth]{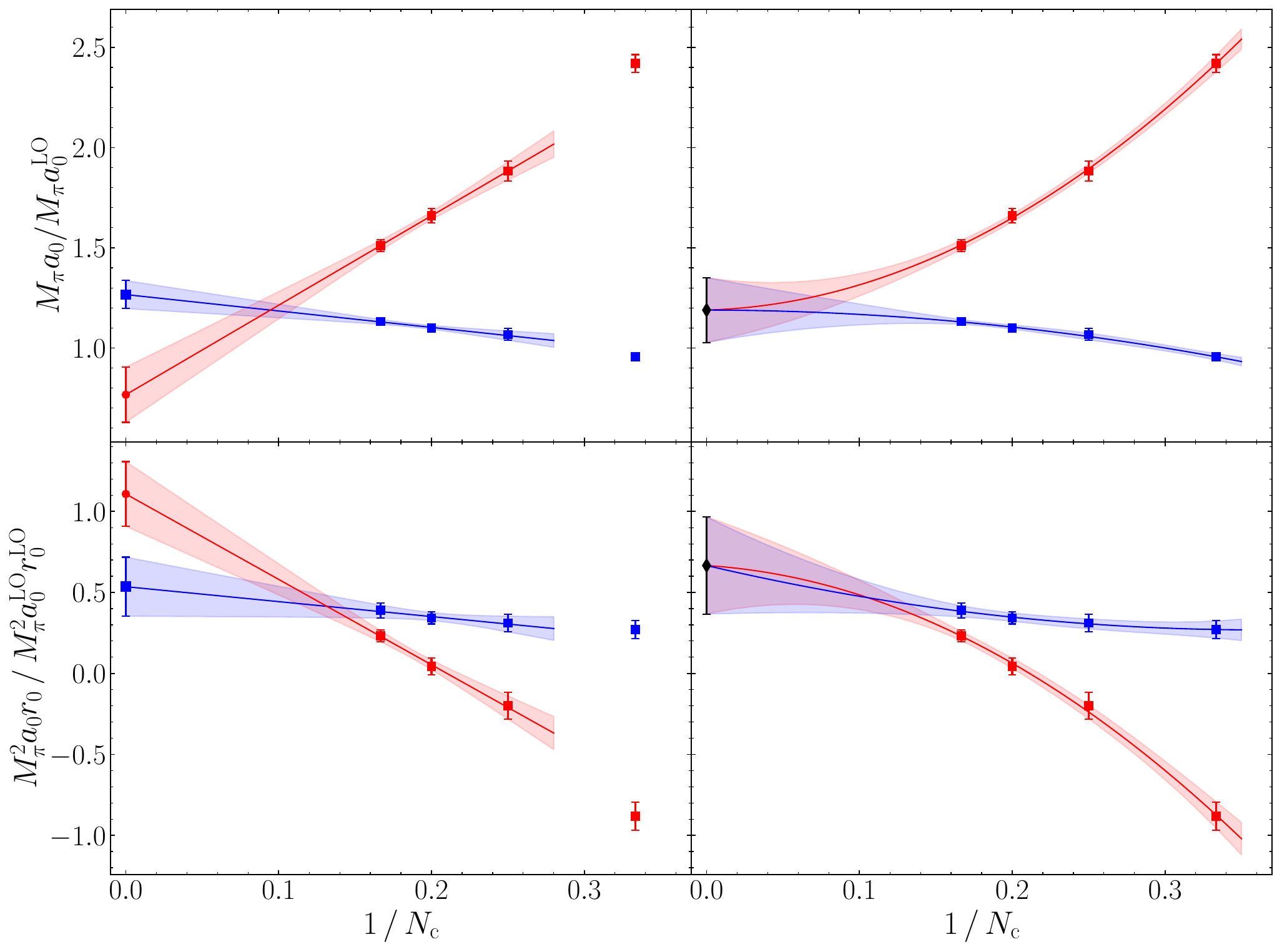}
    \caption{Results for the scattering length and effective length for the $SS$ and $AA$ channel, obtained from single-color and single-channel fits to a mERE---see \cref{eq:modifiedERE}---with fixed $z=\Mpi$. Results are normalized by the LO predictions from ChPT in \cref{eq:ChPTLOpredictiona0r0}, and are extrapolated to the large $\Nc$ limit using a linear relation for $\Nc=4-6$ (left) or a quadratic dependence with constrained large $\Nc$ limit including all $\Nc$ (right).   }\label{fig:ERESSAA}
\end{figure}

The best-fit results for the phase shift are also shown in \cref{fig:collapse}, multiplied by $\Nc/3$ to remove the leading $\Nc$ dependence. We observe that after canceling the leading $\Nc$ scaling, the curves appear to collapse for all $\Nc$, although the results suggest that we are sensitive to subleading effects.

Using \cref{eq:EREandmERErelation}, the scattering length and effective range are determined, which we present in \cref{fig:ERESSAA} as a function of $1/\Nc$, normalized by the LO ChPT prediction,\\
\noindent\begin{equation}\label{eq:ChPTLOpredictiona0r0}
    M_\pi a_0^{SS}=-\Mpi a_0^{AA}=-\frac{\Mpi^2}{16\pi^2\Fpi^2}\,,\quad\quad\quad M_\pi^2a_0^{SS,AA}r_0^{SS,AA}=-3\,.
\end{equation}
We extrapolate our results to the large $\Nc$ limit. We observe that the results for $\Nc=4-6$ are well described by a linear relation on $1/\Nc$, as presented on the left panels of \cref{fig:ERESSAA}. However, the results for both channels do not agree in the large $\Nc$ limit in this case. If we allow for $\cO(\Nc^{-2})$ corrections, it is possible to reproduce the expected common large $\Nc$ limit and also describe the $\Nc=3$ results. This is presented in the right panels of \cref{fig:ERESSAA}, in which the common large $\Nc$ results is imposed. We observe subleading $\Nc$ effects are larger in the $AA$ channel than the $SS$ one. This can be understood from the large $\Nc$ predictions in \cref{eq:scalingNcSSAA}. Subleading corrections are expected to add up in one of the channels and cancel in the other one, depending on the signs of $\tilde{b}$ and $\tilde{c}$ in \cref{eq:scalingNcSSAA}. From our results, we can conclude that subleading corrections add up in the $AA$ channel and cancel in the $SS$ case. 
This can also be understood from ChPT predictions in \cref{app:chptamplitudes}, for example looking at the $q^2$-independent terms in \cref{eq:explicitffactorSSAA,eq:explicitFfactorSSAA}: $\cO(\Nf^{-1})$ terms in the $AA$ channel appear with the same sign as $\cO(\Nf^0)$ and $\cO(\Nf^{-2})$ ones and add up, while in the $SS$ channel they appear with different signs, and so cancel.

\begin{table}[!b]\vspace{0.cm}
    \centering
        \begin{tabular}{ccc}
\toprule
    $\Nc$ &    $\Mpi^3 a_1 \times 10^3$ &  $\chi^2\,/\,\text{dof}$\\  \midrule 

 3 & $11.2(2.8)$  & $19.0/16=1.19$ \\
 4 & $8(4)$ &  $13.6/10=1.36$ \\
 5 & $9(3)$ & $21.7/10=2.17$ \\
 6 & $12.6(29)$ & $18.1/10=1.81$ \\   \bottomrule
\end{tabular}

    \caption{Results for the scattering length in the $AS$ channel, obtained from single-color fits to an ERE---see \cref{eq:ERE}.  }\label{tab:ASEREresults}
\end{table}

\begin{figure}[!b]
    \centering

    \includegraphics[width=0.68\textwidth]{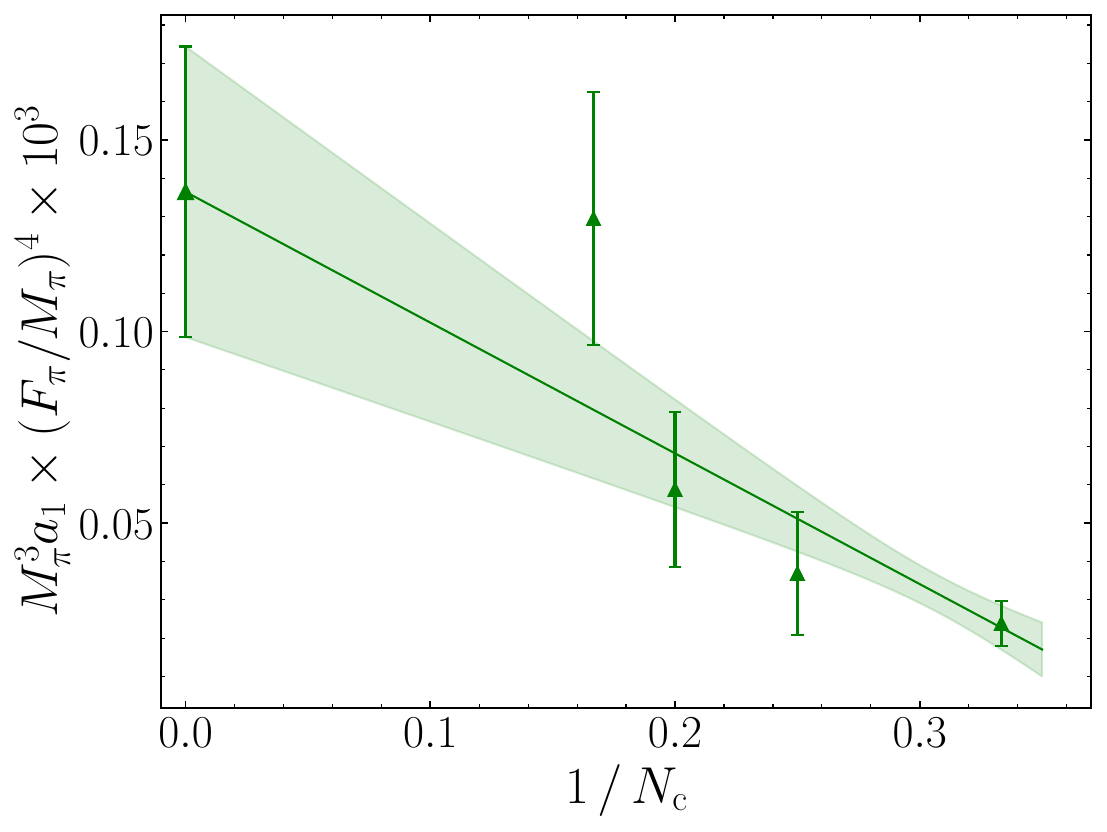}

    \caption{Results for the scattering length in the $AS$ channel, obtained from single-color fits to an ERE---see \cref{eq:ERE}. Results are multiplied by $\Fpi^4/M_\pi^4$ to eliminate leading chiral and $\Nc$ dependencies, and are linearly extrapolated to the large $\Nc$ limit.   }\label{fig:EREAS}
\end{figure}

We also fit the energy results for the AS channel, using an ERE parametrization including a single parameter: the scattering length. Result are presented in \cref{tab:ASEREresults} and \cref{fig:EREAS}, where they are multiplied by a factor $(\Fpi/\Mpi)^4$ to eliminate leading chiral and $\Nc$ dependencies. We observe that the results are reasonably well reproduced by a linear extrapolation to the large $\Nc$ limit. Note that, in this case, the use of a linear extrapolation corresponds to considering $\cO(\Nc^{-3})$ effects in the scattering amplitude, analogously to the quadratic terms used in the constrained extrapolation for the $SS$ and $AA$ channels.

\subsection{ Fits to ChPT at fixed $\Nc$  }

We now match our results to ChPT. We consider both the NLO predictions from the SU(4) theory and the NNLO results from U(4) ChPT, and perform single-color and single-channel fits. Results for the LECs in the $SS$ and $AA$ channels, in which we use the IAM, are presented in \cref{tab:SSAAfitchptSU,tab:SSAAfitchptU} together with the $\chi^2$ of the fits. We observe that the results are nearly identical between the SU(4) and the U(4) theories, with only minor differences in the values of the $L_{AA}$ coefficients. This indicates that, at the pion mass we are working and when using the IAM, loop corrections that include the $\eta'$ are small. We also note that effects related to the higher order $K$ LECs are absorbed into the $L$ LECs, which may also hide the differences between both theories.

\begin{table}[b!]
\centering
\begin{tabular}{cccccc}
\toprule
 Channel, $R$   &    $\Nc$ &    $(L_R/\Nc)\times 10^3$ & $(L_R'/\Nc)\times 10^3$ & $(L_R''/\Nc)\times 10^3$ & $\chi^2\,/\,\text{dof}$\\  \midrule 

\multirow{4}{*}{$SS$} & 3 & $-0.385(28)$  & $-1.4(4)$ & 1.68(28) & $9.1/11=0.83$ \\
& 4 & $-0.39(6)$  & $-1.1(7)$ & 1.45(24) & $5.1/7=0.73$ \\
& 5 & $-0.302(23)$  & $-2.0(3)$ & 1.77(15) & $12.9/7=1.84$ \\
& 6 & $-0.337(22)$  & $-1.1(4)$ & 1.41(22) & $8.4/7=1.20$ \\ \midrule
\multirow{4}{*}{$AA$} & 3 & $0.067(12)$  & $1.58(14)$ & 1.61(12) & $16.1/9=1.79$ \\
& 4 & $-0.02(4)$  & $0.8(5)$ & 1.59(26) & $15.3/6=2.54$ \\
& 5 & $-0.048(21)$  & $0.6(3)$ & 1.46(16) & $19.7/6=3.28$ \\
& 6 & $-0.052(22)$  & $0.4(4)$ & 1.40(20) & $17.5/6=2.92$  \\ \bottomrule
\end{tabular}
\caption{Results from  single-color and single-channel fits to NLO SU(4) ChPT predictions for the $SS$ and $AA$ channels using the IAM. }\vspace{0.3cm}

\label{tab:SSAAfitchptSU}
\end{table}

\begin{table}[b!]
\centering
\begin{tabular}{cccccc}
\toprule
 Channel, $R$   &    $\Nc$ &    $(\tilde{L}_R/\Nc)\times 10^3$ & $(\tilde{L}_R'/\Nc)\times 10^3$ & $(L_R''/\Nc)\times 10^3$ & $\chi^2\,/\,\text{dof}$\\  \midrule 

\multirow{4}{*}{$SS$} & 3 & $-0.365(28)$  & $-1.4(4)$ & 1.68(28) & $9.1/11=0.83$ \\
& 4 & $-0.38(6)$  & $-1.1(7)$ & 1.45(24) & $5.1/7=0.73$ \\
& 5 & $-0.301(23)$  & $-2.0(3)$ & 1.77(15) & $12.9/7=1.84$ \\
& 6 & $-0.338(22)$  & $-1.1(4)$ & 1.41(22) & $8.4/7=1.20$ \\ \midrule
\multirow{4}{*}{$AA$} & 3 & $0.061(12)$  & $1.57(14)$ & 1.61(12) & $16.1/9=1.79$ \\
& 4 & $-0.03(4)$  & $0.9(5)$ & 1.59(27) & $15.3/6=2.54$ \\
& 5 & $-0.060(21)$  & $0.6(3)$ & 1.47(16) & $19.7/6=3.28$ \\
& 6 & $-0.066(22)$  & $0.4(4)$ & 1.40(20) & $17.5/6=2.92$  \\ \bottomrule 
\end{tabular}
\caption{Results from single-color and single-channel fits to NNLO U(4) ChPT predictions for the $SS$ and $AA$ channels, using the IAM. }\vspace{0.3cm}
\label{tab:SSAAfitchptU}
\end{table}

\begin{figure}[!t]
    \centering\vspace{-0.2cm}
\includegraphics[width=0.485\textwidth]{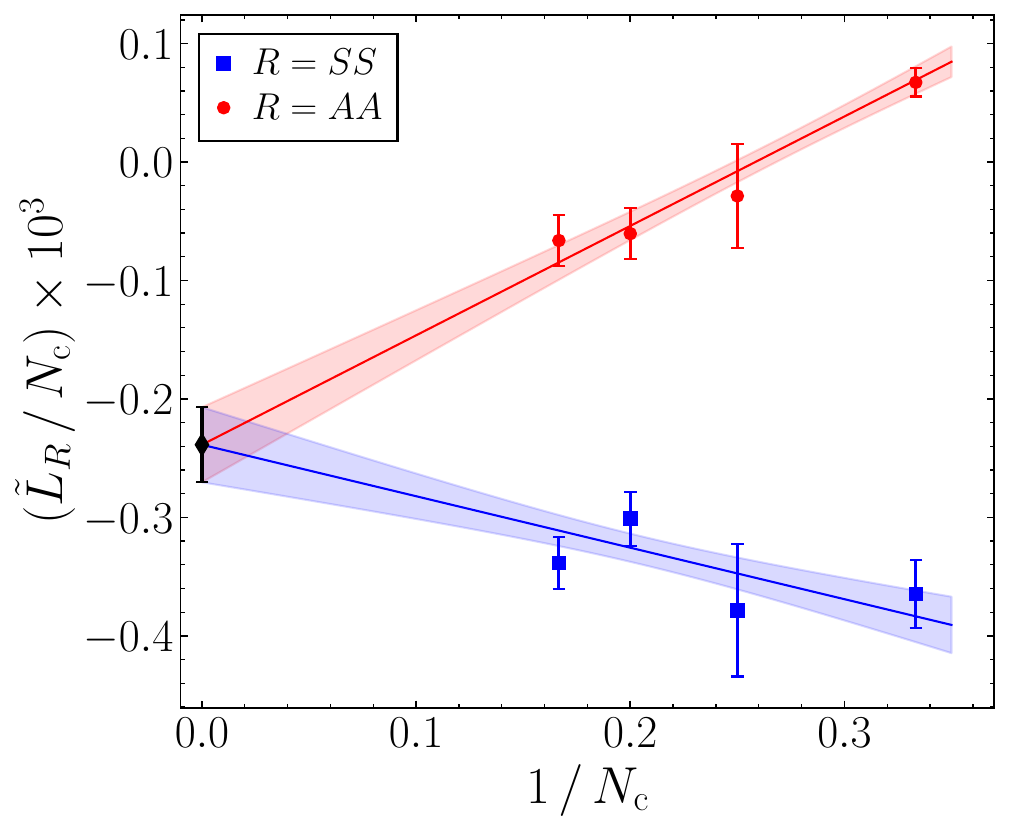} \hspace{0.cm}
\includegraphics[width=0.47\textwidth]{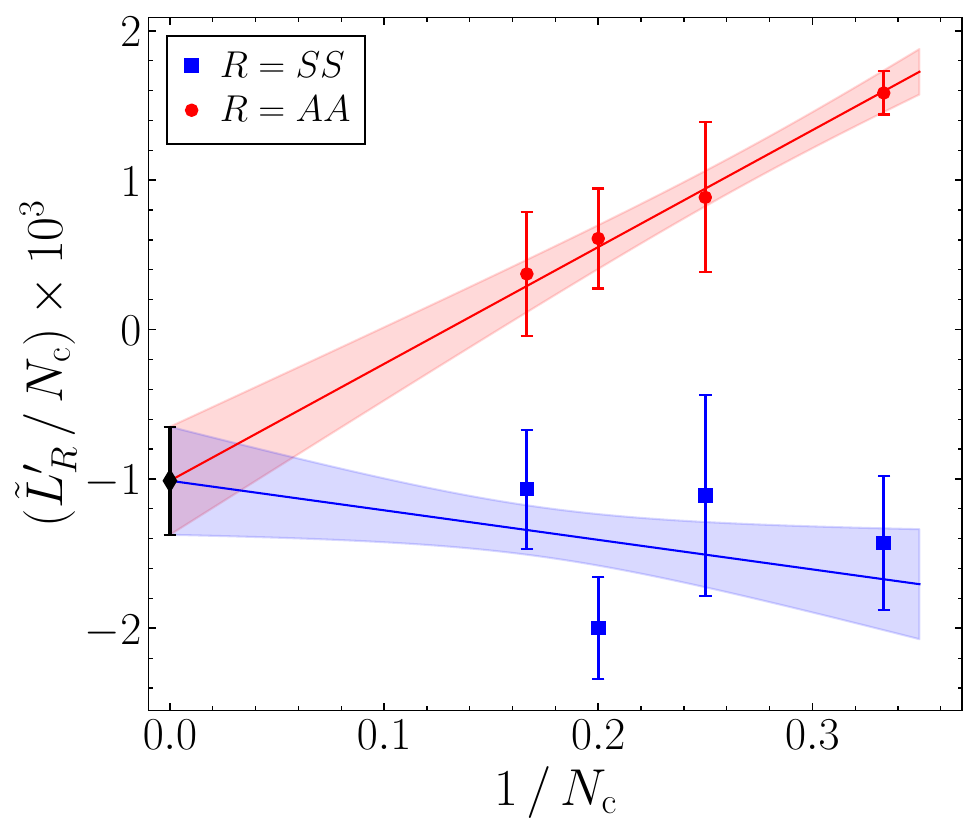} \\[0.cm]
\includegraphics[width=0.47\textwidth]{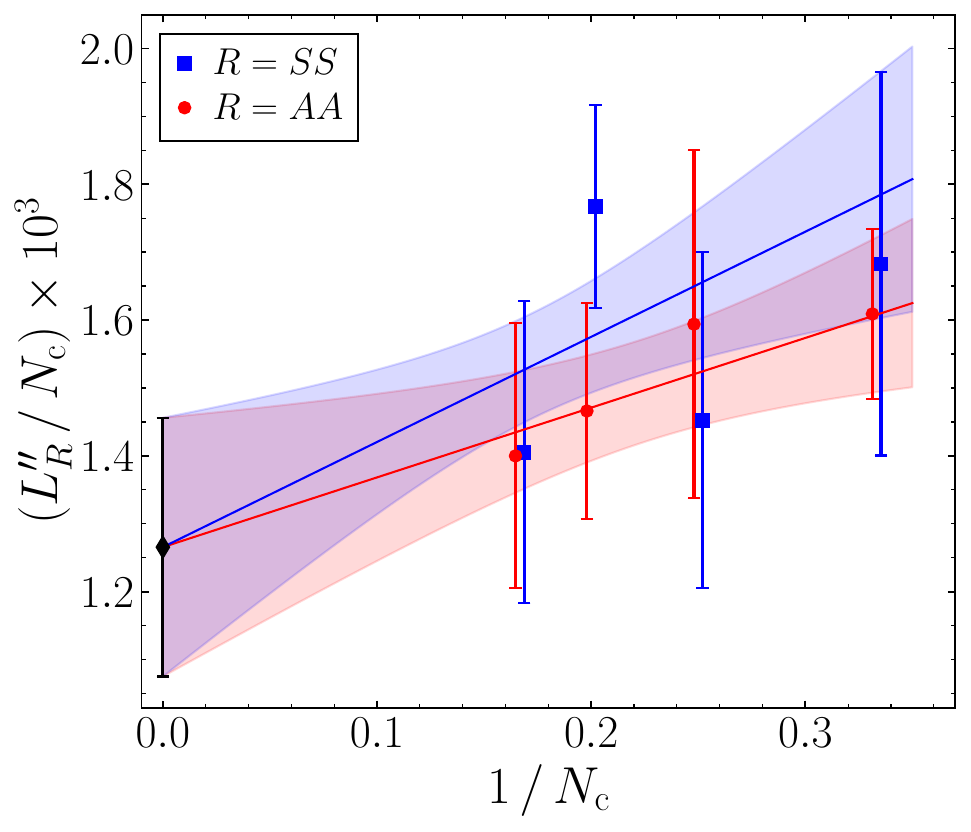} 
    \caption{Results for the U(4) ChPT LECs appearing in the $SS$ and $AA$ channels, obtained from single-color and single-channel fits to U(4) ChPT, together with a constrained linear extrapolation to the large $\Nc$ limit. The LECs in the upper panels are redefined to absorb higher order effects---see \cref{eq:LECsredefinition}. A small horizontal offset is added in the bottom panel for better legibility.}
    \label{fig:LECsSSAA}
\end{figure}

In \cref{fig:LECsSSAA} we present the result for the U(4) LECs for the two channels. We observe our results for all $\Nc$ is well reproduced by a linear $\Nc$ dependence, with a consistent large $\Nc$ limit, and we use them to perform a constrained linear extrapolation  to  large $\Nc$. The fact that our results are well reproduced by a linear $\Nc$ dependence, including a common large $\Nc$ limit, indicates that ChPT predictions capture better the large $\Nc$ scaling of the amplitude than a naive mERE. This is probably related to the use of the physical values of $\Mpi$ and $\Fpi$, which implicitly include $1/\Nc$ effects to all orders, as well as the inclusion of chiral logarithms.

 \begin{table}[t!]
\centering
\begin{tabular}{ccccc}
\toprule
 Fit   &    $\Nc$ &    $L_{AS}\times 10^3$ & $L_{AS}'\times 10^3$ & $\chi^2\,/\,\text{dof}$\\  \midrule 

\multirow{4}{*}{NLO SU(4)} & 3 & $-0.325(27)$  & $0.003(11)$ & $18.7/15=1.25$ \\
& 4 & $-0.28(8)$  & $-0.02(4)$& $12.9/9=1.44$ \\
& 5 & $-0.18(8)$  & $-0.04(4)$& $18.6/9=2.07$ \\
& 6 & $-0.08(12)$  & $0.00(6)$& $16.7/9=1.86$ \\ \midrule
\multirow{4}{*}{NNLO U(4)} & 3 & $-0.324(27)$  & $0.006(11)$ & $18.7/15=1.25$ \\
& 4 & $-0.28(8)$  & $-0.02(4)$& $12.9/9=1.43$ \\
& 5 & $-0.18(8)$  & $-0.04(4)$& $18.6/9=2.07$ \\
& 6 & $-0.08(12)$  & $0.00(6)$& $16.7/9=1.86$ \\ \bottomrule 
\end{tabular}
\caption{Results from single-color and single-channel fits to NLO SU(4) and NNLO U(4) ChPT predictions for the $AS$ channel. }

\label{tab:ASfitchpt}
\end{table}

\begin{figure}[!h]
    \centering
\includegraphics[width=0.455\textwidth]{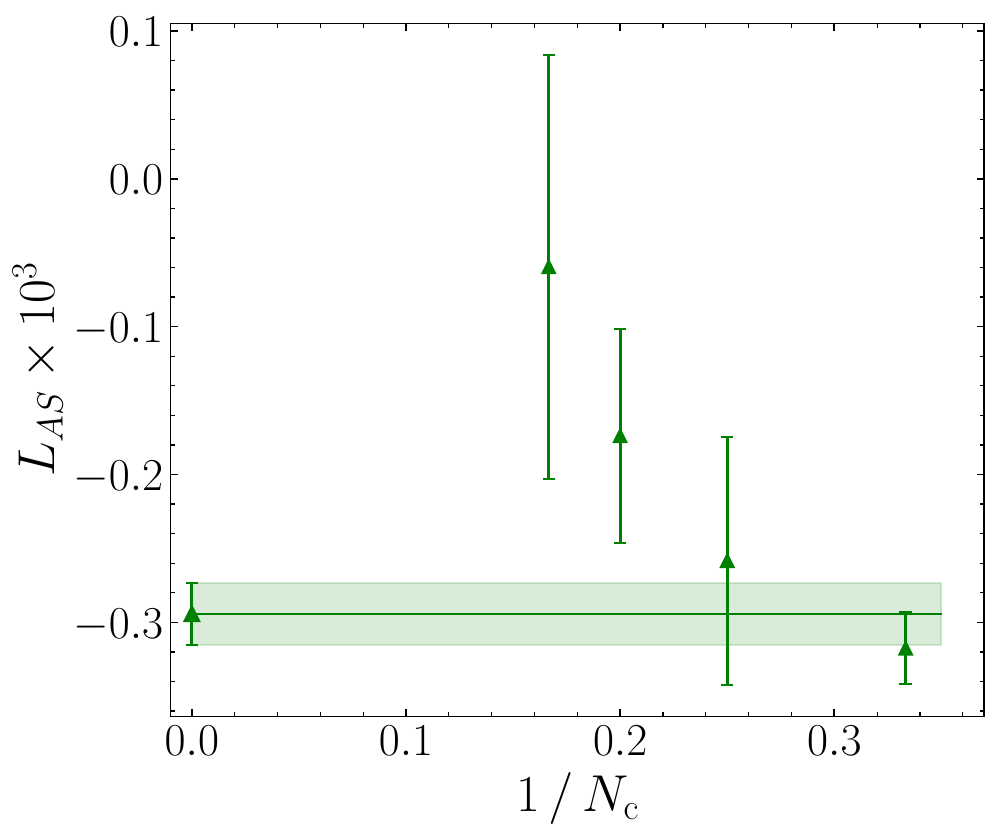} \hspace{0.2cm}
\includegraphics[width=0.47\textwidth]{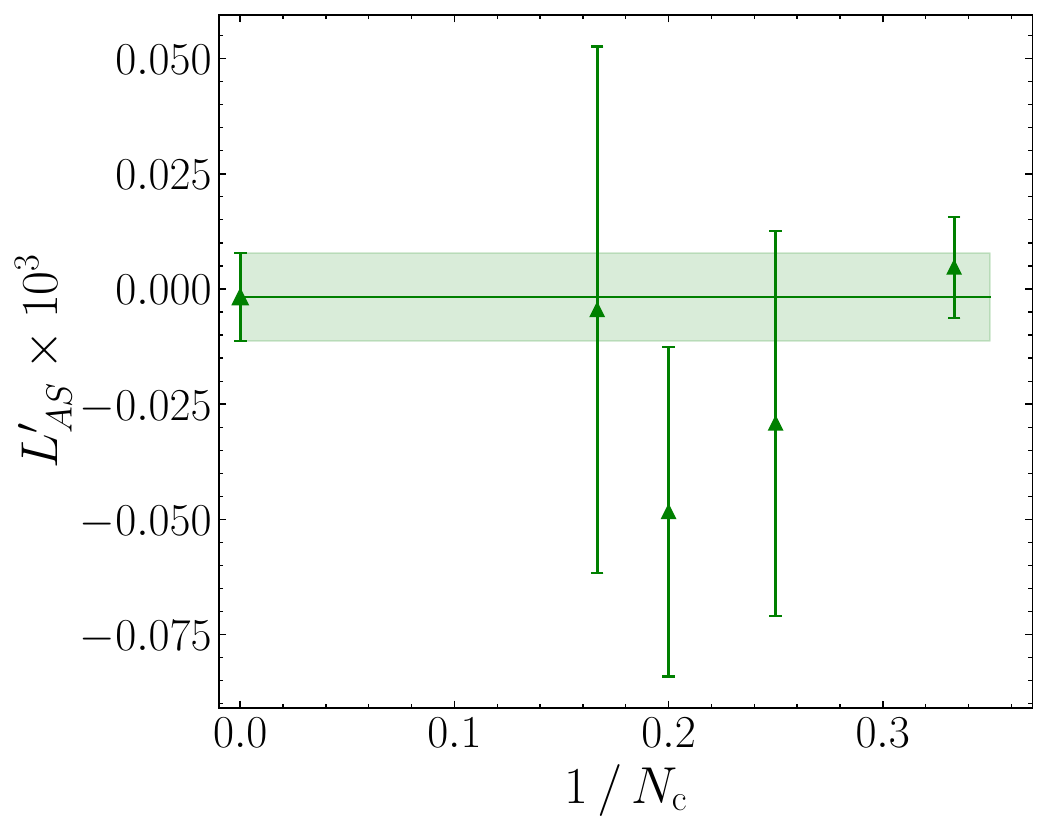} 
    \caption{Results for the LECs in the $AS$ channel, obtained from single-color fits to U(4) ChPT, together with the results of a constant extrapolation to the large $\Nc$ limit.
    }\label{fig:LECsAS}
\end{figure}

Results for the $AS$ channel are presented in \cref{tab:ASfitchpt} and we depict the LECs of U$(4)$ ChPT in \cref{fig:LECsAS}, together with a constant extrapolation to large $\Nc$. Our results are well described by this constant dependence, which is dominated by $\Nc=3$. This is consistent with our results and with the order at which we are working in the chiral expansion---see \cref{eq:NcpredictionsLECsAS}.

\subsection{ Simultaneous $\Nc$ fits to ChPT  }\label{sec:resultsglobalfits}

\begin{figure}[!hp]
    \centering
    \includegraphics[width=0.8\textwidth]{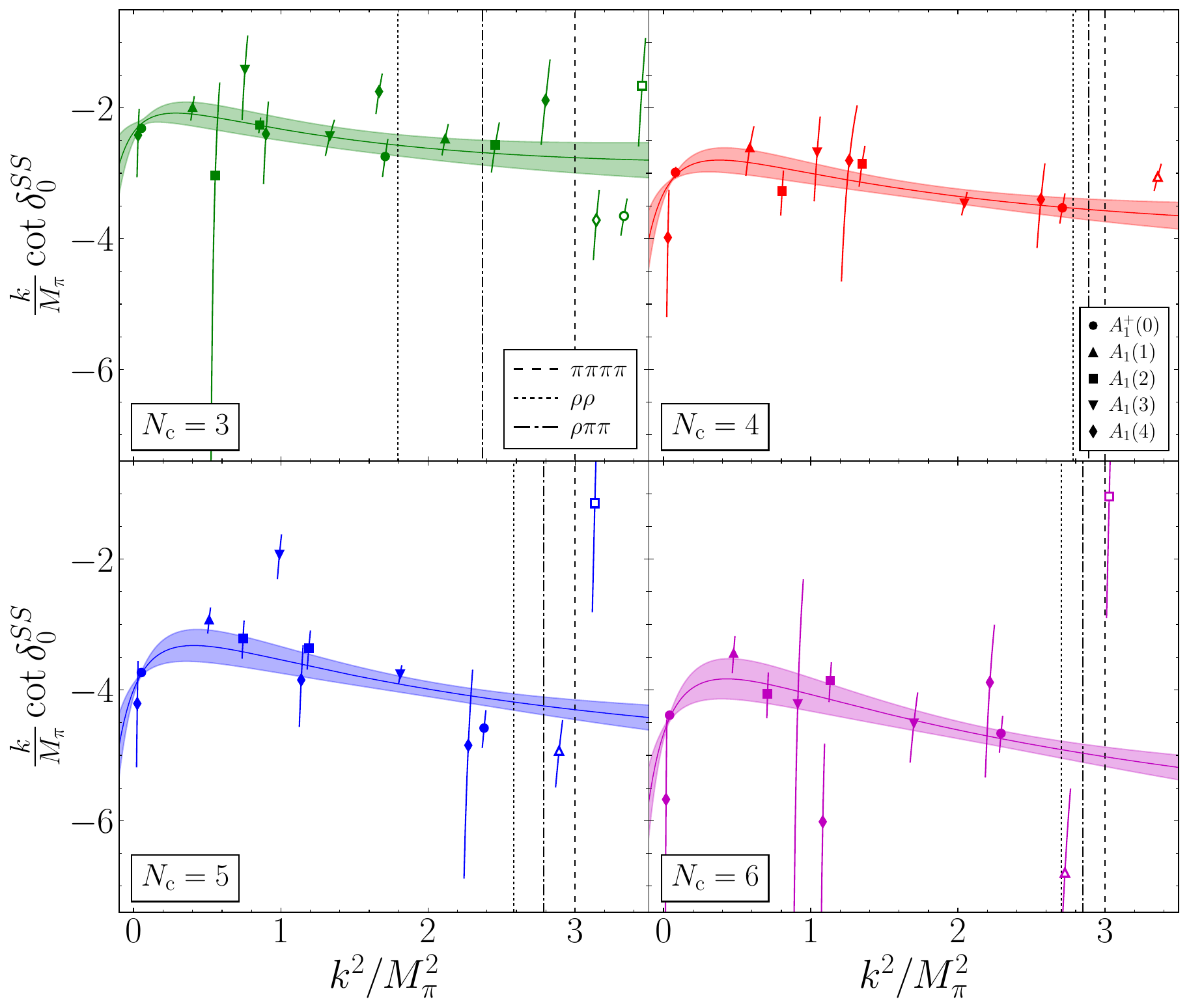}\\
    \hspace{0.22cm}\includegraphics[width=0.783\textwidth]{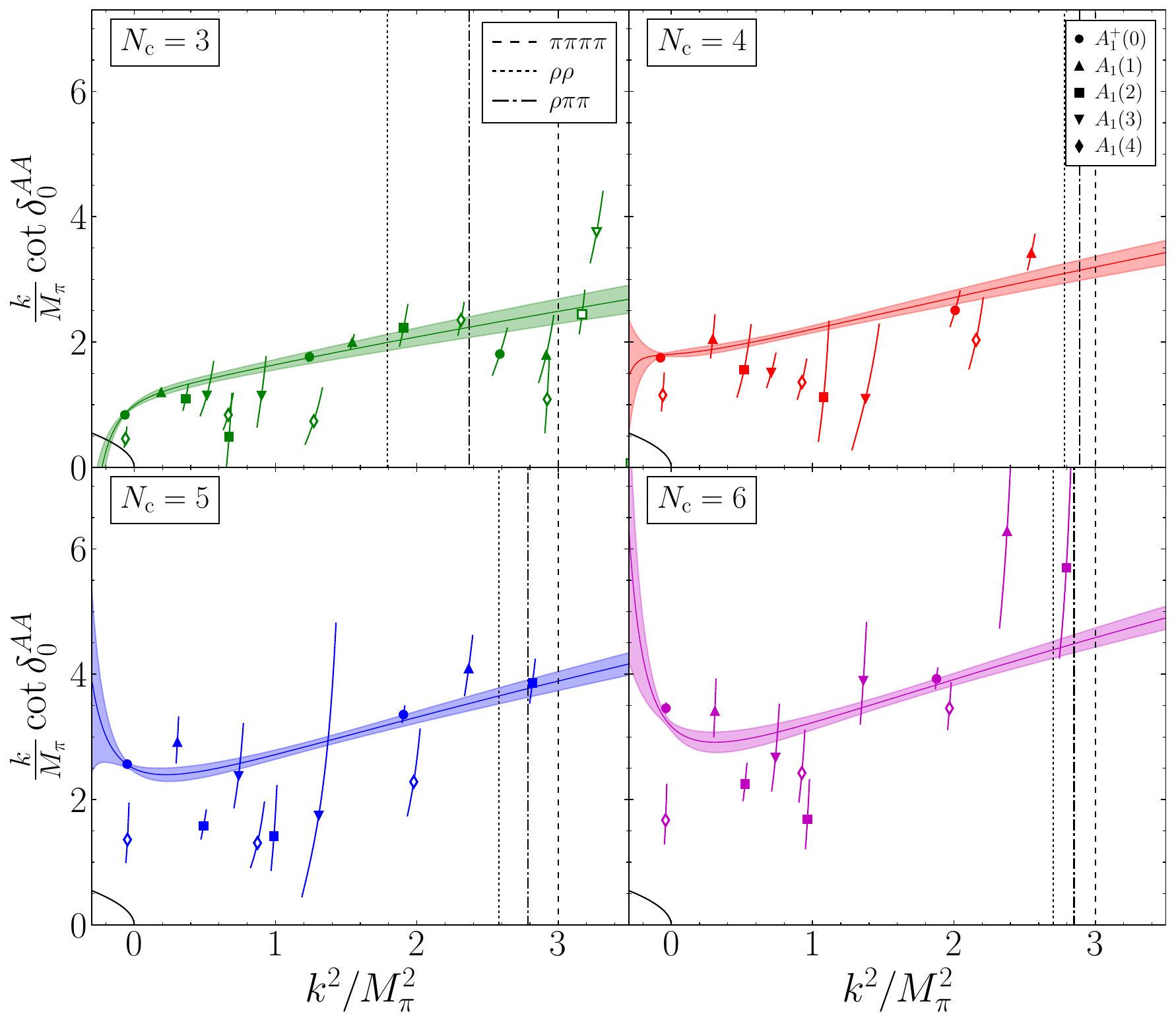}
    \caption{Results for the phase shift for the $SS$ (top) and $AA$ (bottom) channels, together with the best-fit predictions from a combined fit of all $\Nc$ and both channels to NNLO U(4) ChPT using the IAM. Empty points are not considered for the fit. The crossing of the phase shift curves with the solid black line in the bottom left of the $AA$ channel panels signals the presence of a virtual bound state---see \cref{eq:boundstatecondition}.  }
    \label{fig:SSAAkcot}
\end{figure}

We now  perform global fits including all values of $\Nc$ and focus on fitting to NNLO U(4) ChPT. For the $SS$ and $AA$ channels, we parametrize the LECs with a leading and subleading term---as presented in \cref{eq:SUparammetrizationSSAALECs}. This parametrization is also kept for the modified LECs in \cref{eq:LECsredefinition}, where the modification is expected to only affect the leading $\Nc$ part of the LECs at the order we are working, as discussed in \cref{sec:fittingprocedure}. We have performed fits to both channels separately and simultaneously. Results are presented in \cref{tab:SSAAglobalfits} and we show the predictions for the scattering phase shift from the combined-channel fit in \cref{fig:SSAAkcot}, together with results for the scattering phase shift obtained directly from the lattice data using \cref{eq:Luscheralgebraicswave}. In the $AA$ channel, we observe how the predictions for $\Nc=3$ cross the $\sqrt{-k^2}$ curve (solid black), signaling the presence of a virtual bound state. This seems to also be the case for some samples for $\Nc=4$. We discuss this possibility in detail in \cref{sec:boundstate}, but recall for now that the results for the IAM should not be trusted below threshold---see \cref{sec:IAM}.

\begin{table}[t!]
\centering
\begin{tabular}{cccc}
\toprule
    &    $SS$ &    $AA$ & $SS+AA$\\  \midrule 
$\tilde{L}^{(0)}\times10^3$ & $-0.30(5)$ & $$-0.23(4)$$  & $-0.247(28)$ \\
$L_{SS}^{(1)}\times10^3$ & $-0.18(20)$ & --- & $-0.48(23)$ \\
$L_{AA}^{(1)}\times10^3$ & --- & $0.91(14)$  & $0.96(10)$ \\
$\tilde{L}^{\prime(0)}\times10^3$ & $-1.0(8)$ & $-0.9(7)$  & $-0.7(5)$ \\
$L_{SS}^{\prime(1)}\times10^3$ & $-2.0(3.0)$ & --- & $-2.1(2.3)$ \\
$L_{AA}^{\prime(1)}\times10^3$ & --- & $7.6(2.3)$  & $6.8(1.9)$ \\
$L^{\prime\prime(0)}\times10^3$ & $1.2(4)$ & $1.4(4)$  & $1.33(20)$ \\
$L_{SS}^{\prime\prime(1)}\times10^3$ & $1.5(1.9)$ & --- & $0.9(1.1)$ \\
$L_{AA}^{\prime\prime(1)}\times10^3$ & --- &  $0.4(1.3)$ & $0.3(9)$ \\
$\chi^2/\text{dof}$ & $39.9/38=1.05$ &  $72.1/33=2.18$ & $165.3/74=2.23$ \\\bottomrule 
\end{tabular}
\caption{LEC results from all-$\Nc$ fits to NNLO U(4) ChPT using the IAM, for the $SS$ and $AA$ channels. We use a parametrization for the LECs linear in $\Nc$ and consider a constrained large $\Nc$ limit in the combined-channel case. }

\label{tab:SSAAglobalfits}
\end{table}

\begin{figure}[!t]
    \centering
    \includegraphics[width=0.8\textwidth]{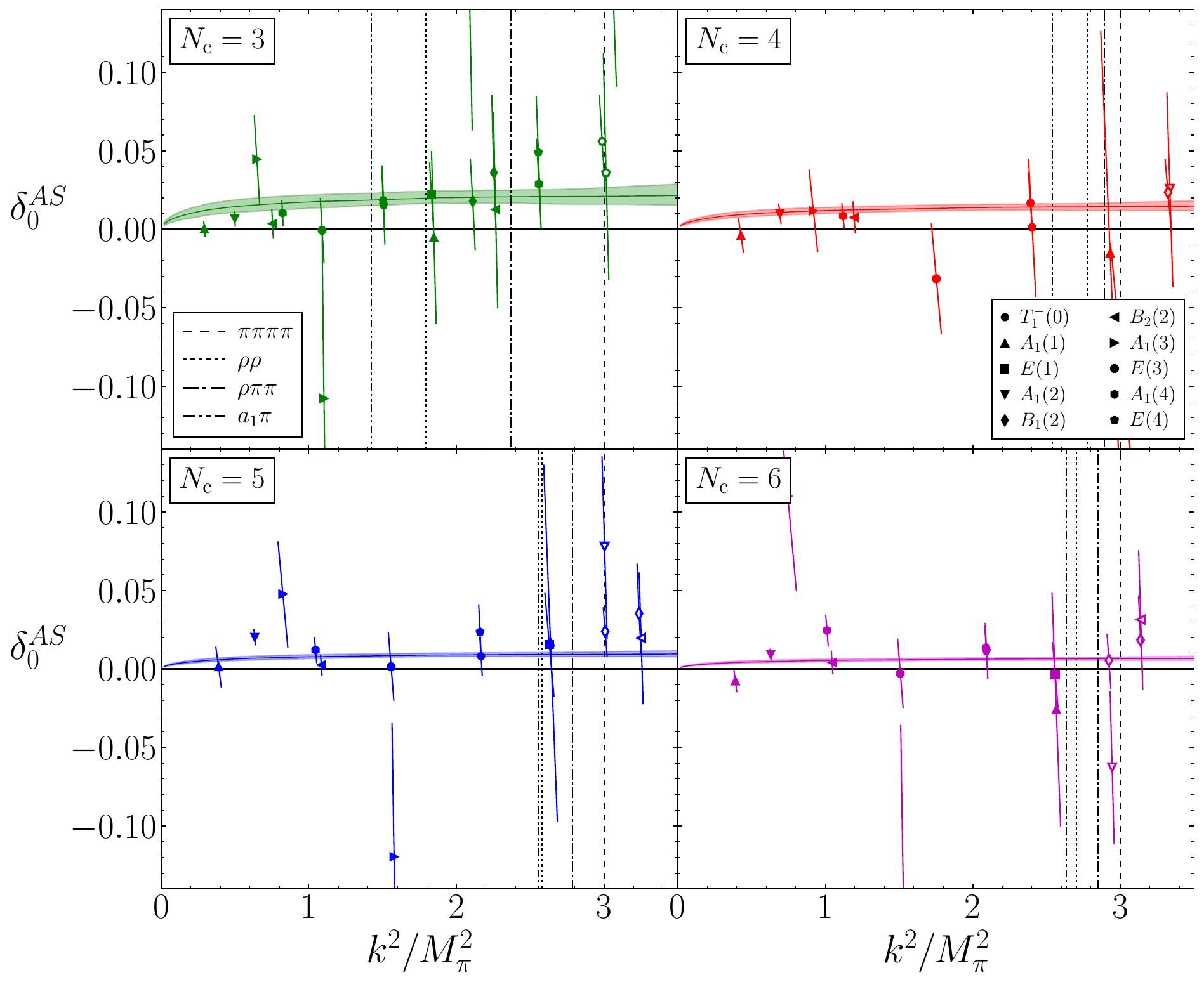}
     \caption{Results for the phase shift for the $AS$ channel, together with the best-fit predictions from a fit of all $\Nc$ to NNLO U(4) ChPT.  }
     \label{fig:ASkcot}
\end{figure}

Concerning the $AS$ channel, we consider a constant parametrization of the LECs, in agreement with \cref{eq:NcpredictionsLECsAS}. Results from a fit are presented in \cref{tab:ASglobalfit}, with the predictions shown in  \cref{fig:ASkcot}, together with the results obtained from the lattice using the single-channel $p$-wave quantization condition. As before, we found that the results follow the expected $\Nc$ scaling.

\begin{table}[t!]
\centering
\begin{tabular}{ccc}
\toprule
  $L_{AS}^{(1)}\times10^3$  &    $L_{AS}^{\prime(1)}\times10^3$ &  $\chi^2/\text{dof}$ \\  \midrule 
$-0.300(27)$ &  $0.001(12)$ &  $78.6/48 = 1.64$ \\
\bottomrule 
\end{tabular}
\caption{LEC results from an all-$\Nc$ fit of the $AS$ channel results to NNLO U(4) ChPT, using a constant parametrization for the LECs.}

\label{tab:ASglobalfit}
\end{table}

\subsection{Virtual bound state in the $AA$ channel}\label{sec:boundstate}

From our results for the scattering amplitudes, we find no evidence of a tetraquark resonance in the pion-pion system above the two-pion threshold and below the four pion threshold, neither in the $AA$ channel, nor in the $AS$ one. However, our fit results indicate the presence of a virtual state pole in the $AA$ channel for $\Nc=3$, on which we focus on the remaining of this section. We do not find a near-threshold virtual or bound state for other $\Nc$. 

A virtual bound state corresponds to a pole on the real $s$-axis of the second Riemann sheet below the two-particle threshold, and are identified as solutions to the condition
\begin{equation}\label{eq:boundstatecondition}
    k\cot\delta_0-\sqrt{-k^2}=0\,.
\end{equation}
From a quantum-mechanical perspective, they are non-normalizable solutions to Schrödinger's equation, and can be smoothly connected to bound states or resonances as the interaction parameters of the potential are varied~\cite{10.1119/1.1987678}.

\begin{figure}[!t]
    \centering
    \includegraphics[width=0.7\textwidth]{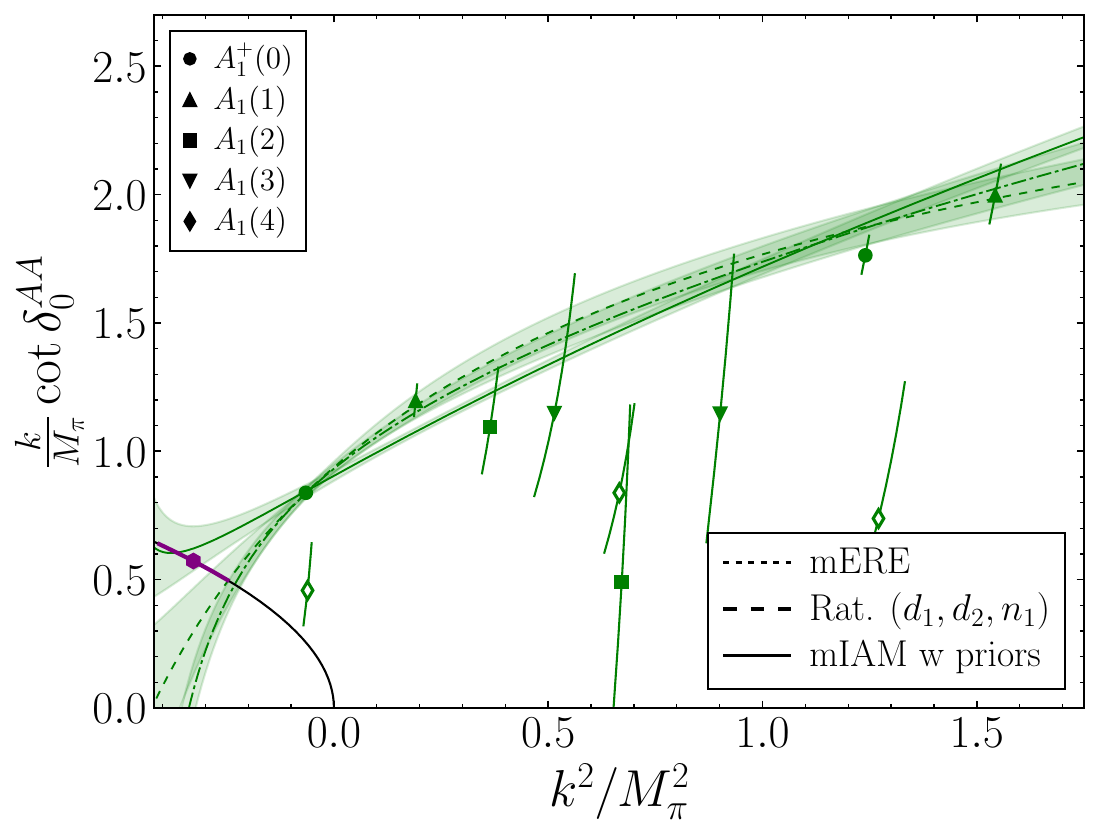}
    \caption{Results for the scattering phase shift of the $AA$ channel with $\Nc=3$ in the near-threshold region, for several parametrizations of the scattering amplitude. We observe that, for all considered cases, the results predict the presence of a virtual bound state. The average location of this state is represented with a purple hexagon.}\label{fig:virtualboundstates}
\end{figure}

The results for a mERE presented in \cref{sec:fitsERE} are consistent with the presence of a virtual bound state for $\Nc=3$ only. 
To further investigate this state, we perform fits using rational parametrizations of the scattering phase shift of the form\\\
\begin{equation}\label{eq:rationalERE}
    k\cot\delta_0^{AA}=\frac{1+n_1 k^2}{d_1 + d_2 k^2 + d_3 k^4}
\end{equation}
with different fit parameters, with the rest set to zero. The results are also consistent with the presence of a virtual bound state, as summarized in \cref{tab:virtualboundstates}. We note that not in all cases the presence of a virtual found state is predicted by all bootstrap samples---the fraction of samples for which a solution to \cref{eq:boundstatecondition} is found is indicated in the second-to-last column of \cref{tab:virtualboundstates}. We also show the prediction for the compositeness parameter introduced in \cref{eq:compositeness}. We observe that in all cases its value is significantly smaller than 0.5, which suggests that the state is dominated by a molecular pion-pion component.  In \cref{fig:virtualboundstates} the predictions for the phase shift in the near-threshold region are shown for the fit to an mERE and the fit to \cref{eq:rationalERE} with $n_1$, $d_1$ and $d_2$ as free parameters, and $d_3=0$. 

\begin{table}[t!]
\centering
\begin{tabular}{ccccc}
\toprule
    Fit &    $a E_\text{virtual}$ & $Z$ &     Fraction of samples & $\chi^2/\text{dof}$ \\  \midrule 
mERE & $1.752(15)$ & 0.335(17) & 100\% & 17.4/10=1.74 \\
Rat. $(d_1,d_2,d_3)$ & $1.458(26)$ & $0.186(8)$ & 100\% & 22.9/9=2.54 \\
Rat. $(d_1,d_2,n_1)$ & $1.73(4)$ & $0.35(3)$ & 100\% & 16.3/9=1.81 \\
Rat. $(d_1,d_2,d_3,n_1)$ & $1.66(15)$ & $0.31(9)$ & 99\% & 16.3/8=2.03   \\
Rat. $(d_1,d_2,d_3,n_1,n_2)$ & $1.5(3)$ & $0.32(8)$ & 93\% & 16.3/7=2.32   \\
mIAM w. priors & $1.63(4)$ & $0.260(14)$ & 53\%  & 39.5/12=3.29 \\\midrule 
Average & $1.63(10)$ & $0.29(5)$ & --- & ---  \\\bottomrule 
\end{tabular}
\caption{Position of the virtual bound state found in the $AA$ channel for $\Nc=3$, for different parametrizations of the phase shift, together with the corresponding compositeness, the fraction of bootstrap samples for which the state is observed and the $\chi^2$ of the fit. For fits to the rational parametrization in \cref{eq:rationalERE}, the fit parameters are indicated in parenthesis, with the remaining ones set to zero. The fit to the mIAM uses the results in \cref{tab:SSAAfitchptU} as Gaussian priors, and the number of prior parameters is added to the dof.}
\label{tab:virtualboundstates}
\end{table}

Obtaining information about poles in the subthreshold region from ChPT amplitudes, on the other hand, is more complicated. The IAM, which we have used to study the $AA$ channel, fails below threshold in this channel as explained in \cref{sec:IAM}. We have found that direct fits to the mIAM---introduced in \cref{eq:mIAM}---are very unstable. To improve the predictive capacity of this model, we use the results from the $AA$ channel fit with $\Nc=3$ in \cref{tab:SSAAfitchptU} as Gaussian priors in a fit to predictions based on the mIAM. The results for this fit are also presented in \cref{fig:virtualboundstates}. In this case, the virtual bound state is only found in around half of the bootstrap samples, at energies compatible with the ones obtained from model-agnostic fits and with similar value of the compositeness parameter. Note, however, that the mIAM provides a worse description of the data, as signaled by the large value of the $\chi^2$. In addition, it is worth mentioning that while a virtual bound state seems to be predicted by some bootstrap samples for $\Nc=4$ when fitting to the IAM, predictions based on the IAM should not be trusted below threshold. A fit to the mIAM, performed in analogy to that for $\Nc=3$, finds no virtual bound state for $\Nc=4$.

An average result for the pole position and the compositeness parameter can be obtained combining the results of all models, finding,
\begin{equation}
    E_\text{virtual}=1.63(10)\Mpi\,,\quad\quad\text{and}\quad\quad Z=0.29(5)\,.
\end{equation}
When moving towards the chiral limit, it is expected that the position of the pole moves smoothly as a function of the pion mass. In particular, virtual bound states usually become resonances at lighter pion masses~\cite{Briceno:2016mjc,Briceno:2017qmb}. We expect that the found state could be related to the experimental tetraquarks when moving from our setup with four degenerate quarks at heavy pion mass towards a more physical scenario, with two light quarks and a heavy charm. 

When then charm quark becomes heavier than the other three flavors, states in the $AA$ channel decompose in general in different irreducible representations of the SU$(3)_\text{f}$ flavor group. In particular, some states such as $D_s\pi$ or $DK$ would transform as a sextet, while others like $D^-D_s^+$ or $\pi^-K^+$ are part of an octet. We note that the SU$(3)_\text{f}$ irreps are not in one-to-one relation with those of SU$(4)_\text{f}$. 

Finally, we note that our results are qualitatively consistent with previous lattice studies. Meson scattering in the SU$(3)_\text{f}$ sextet was studied in \rcite{Yeo:2024chk} for $\Mpi\approx 700$ MeV, also finding evidence of a virtual bound state of molecular nature. 

\subsection{Comparison to previous literature}

We have shown how our results follow the expected $\Nc$ scaling, with sensitivity to subleading $\Nc$ corrections. Now, we present a quantitative comparison  to previous literature. Combining the results of the global fits in \cref{sec:resultsglobalfits}, it is possible to obtain information about the scaling of several combinations of LECs. The results from the fit to all $\Nc$ and both  $SS$ and $AA$ channels can be used, following \cref{eq:structureLECsAASSAS}, to find\\
\noindent\begin{equation}\label{eq:resultsSSAA}
    \begin{array}{rl}
         (\tilde{L}_0 + \tilde{L}_3 - \tilde{L}_5 + \tilde{L}_8)\times 10^{-3} & = -0.247(28)\Nc  + 0.06(4)\Nf +\cO(\Nc^{-1})\,,  \\ [5pt]
         (4\tilde{L}_0 +2 \tilde{L}_3 - \tilde{L}_5)\times 10^{-3} & = -0.7(5)\Nc  + 0.6(4)\Nf +\cO(\Nc^{-1})\,,  \\ [5pt]
         (3L_0 - L_3)\times 10^{-3} & = 1.33(20)\Nc + 0.16(22)\Nf +\cO(\Nc^{-1})\,,  \\ [5pt]
         ( L_1 +  L_2 -  L_4 +  L_6)\times 10^{-3} & = -0.36(4)+\cO(\Nc^{-1})\,,  \\ [5pt]
         (2 L_1 + 3 L_2 -  L_4)\times 10^{-3} & = -2.2(6)+\cO(\Nc^{-1})\,,  \\ [5pt]
         (L_1 + 2 L_2)\times 10^{-3} & = 0.14(25)+\cO(\Nc^{-1})\,,
    \end{array}
\end{equation}
where the tildes indicate that we have reabsorbed the leading $\Nc$ effects of higher-order LECs---see \cref{eq:LECsredefinition}. We recall that this redefinition only affects the $\cO(\Nc)$ terms at the order we are working in the chiral expansion---see the discussion below \cref{eq:LECsredefinition}. From the fit to the $AS$ channel, on the other hand, we obtain,
\begin{equation}
    \begin{array}{rl}
    (-2L_1+L_2+L_4)\times 10^{-3} & = -0.300(27)+\cO(\Nc^{-1})\,,\\[5pt]
    (-2L_1 + L_2)\times 10^{-6} & = 1(12)+\cO(\Nc^{-1})\,.
    \end{array}
\end{equation}

We can compare our results to those we obtained in \rcite{Baeza-Ballesteros:2022azb}, using the same setup and predictions from standard ChPT. In particular, we focus our comparison on subleading $\Nc$ terms, which are not affected by the redefinition in \cref{eq:LECsredefinition} at the order we are working,
\begin{equation}
\begin{array}{rl}
    ( L_1 +  L_2 -  L_4 +  L_6)\times 10^{-3} & = -0.88(10)+\cO(\Nc^{-1})\,,
\end{array} \quad\quad \text{Ref.~\cite{Baeza-Ballesteros:2022azb}}\,.
\end{equation}
We observe a significant difference, which could originate from higher-order effects arising in the IAM, but also be related to other differences in the analysis, such as having neglected discretization effects, which were studied in better detail in \rcite{Baeza-Ballesteros:2022azb}, or of higher-order terms in the chiral expansion, which could be sizable at the pion masses we have considered and we cannot properly distinguish when working at fixed pion mass. We do not attempt to estimate these effects here, and leave their study for a future work. 

We also compare our results to those available in the literature for $\Nf=3$. This comparison is limited due to our inability to separate the higher-order effects from the $K$ LECs, which affect the leading $\Nc$ corrections, and also for the appearance of the $L_0$ LEC in this limit, which is implicitly absorbed into $L_1$, $L_2$ and $L_3$ at $\Nf=3$. We compare results for $L_{SS}''$ and $L_4$, which are not be affected by these limitations.

This comparison requires us to evaluate the results for $\Nf=3$ and $\Nc=3$ and to change the renormalization scale~\cite{Bijnens:2009qm} from the used $\mu=1.54(3)$ GeV (averaged over all ensembles) to the standard $\mu=770$ MeV scale at the physical mass of the $\rho$ resonance. The results for this comparison are presented in \cref{tab:LECcomparison}, together with phenomenological results from \rcite{Pelaez:2003ip}, in which the IAM method was used to fit to scattering data, and \rcite{Bijnens:2014lea}, and lattice results from \rcite{MILC:2010hzw}.

\begin{table}[b!]
\centering
\begin{tabular}{ccccc}
\toprule
    &    This work &    \rcite{Pelaez:2003ip} & \text{ref.~\cite{Bijnens:2014lea}, table 1, col. 2}  & \rcite{MILC:2010hzw}\\  \midrule 
$L_{SS}''\times 10^{-3}$ & $8.9(8)_\text{stat}(2)_\mu$ & $3.0(5)$  & $4.4(5)$ & --- \\
$L_4\times 10^{-3}$ & $0.25(4)_\text{stat}(2)_\mu$ & $0.200(4)$  & --- & $-0.02(56)$ \\
\bottomrule 
\end{tabular}
\caption{Predictions for $\Nf=3$ for different LEC combinations at a scale $\mu=770$ MeV, together with results from the literature.  }

\label{tab:LECcomparison}
\end{table}

We observe good agreement for the result of the $L_4$ LEC, specially when compared to results of \cite{Pelaez:2003ip}. However, a  discrepancy exists for $L_{SS}''$. As before, this may me related to unaccounted systematics, such as cutoff effects and higher orders in the chiral expansion. A precise investigation of these potential systematics would require the use of finer lattice and lighter pion masses, and is left for future work.

Finally, we compare our results to analytical predictions for the LECs based on the resonant chiral theory (RChT)~\cite{Ecker:1988te,Pich:2002xy,Ledwig:2014cla}. This model assumes that the LECs at large $\Nc$ are saturated by the contribution obtained from integrating our the low-lying resonances. For the $L_{SS,AA}''$ relevant in this work, RChT predicts
\begin{equation}
\begin{array}{rl}
    (3L_0 - L_3)\times 10^{-3} & =1.57 \Nc +\cO(1)\,.
    \end{array}\quad\quad \text{Ref.~\cite{Ledwig:2014cla}, table 1, col. 2}
\end{equation}
This results can be compared to that in the third row in \cref{eq:resultsSSAA}, observing a good agreement with the large $\Nc$ result. Other LECs cannot be directly compared due to the redefinition in \cref{eq:LECsredefinition}.

\section{Conclusions} \label{sec:conclusions}

In this work, we have studied the scaling of meson-meson scattering amplitudes towards the large $\Nc$ limit of QCD with $\Nf=4$ degenerate quark flavors. We have focused on three scattering channels: the $SS$ channel, which is analogous to the isospin-two channel of two-flavor QCD, the $AA$ channel, that only exists for $\Nf\geq 4$ and presents attractive interactions, and the $AS$ channel, which contains odd partial waves. The last two channels, in particular, are candidates to contain a tetraquark state, which could be connected to recently-found exotic states at LHCb~\cite{LHCb:2020bls,LHCb:2020pxc,LHCb:2022sfr,LHCb:2022lzp}. Using the large $\Nc$ limit and ChPT, we have constrained the scaling of the scattering amplitudes for all these channels, including \mbox{one-loop} effects in large $\Nc$ ChPT. In particular, we have shown that interactions in the $AS$ channel are suppressed as $\cO(\Nc^{-2})$, compared to the generic $\cO(\Nc^{-1})$ of the other two channels.

We have performed lattice simulations at $\Nc=3-6$, using the HiRep code~\cite{DelDebbio:2008zf,DelDebbio:2009fd} and ensembles with fixed values of the pion mass, $\Mpi\approx 560$ MeV, and the lattice spacing, $a\approx 0.075$ fm. We have accurately determined the finite-volume energies below the four-pion threshold for different values of the total momentum and all cubic-group irreps that contain two-pion states in the lowest partial wave.  We have used a large set of interpolating operators, consisting of two-pion, two-vector-meson and local tetraquark operators, finding a small impact of the last two types on the determination of pion-pion finite-volume energies. 

In the $SS$ and $AA$ channels, we have fitted our data to  a model-agnostic parametrization, based on a mERE, as well as to one-loop predictions from ChPT combined within the IAM. We find that both fits provide similar descriptions of our data, and are sensitive to subleading $\Nc$ corrections. However, the results for the $AA$ channel show a large dispersion, with worse $\chi^2$ values, possibly affected by significant discretization effects, as found in \rcite{Baeza-Ballesteros:2022azb}. 
Concerning the $AS$ channel, we find interactions to be very weak, confirming the expectations from the large $\Nc$ limit and ChPT. As before, the results are well described by an ERE and ChPT predictions to one loop.

While our results for the $AA$ and $AS$ channels show no evidence of a resonance in the elastic region, they are consistent with the presence of a virtual bound state in the $AA$ channel for $\Nc=3$, at an energy $E_\text{virtual}=1.63(10)\Mpi$, which is not found for other values of $\Nc$. The compositeness parameter is found to be $Z=0.29(5)$, consistent with the state being dominated by a molecular component. These results are in agreement with previous lattice studies performed with a heavy charm~\cite{Yeo:2024chk}. The position of the pole is expected be a smooth function of the quark masses. In particular, virtual bound states of two pions are expected to become resonances as the pion mass is reduced, so the found state could be related to some of the tetraquark states found in LHCb---the $T_{cs0}^0(2900)$, $T_{c\bar{s}0}^{++}(2900)$ and $T_{c\bar{s}0}^0(2900)$. Quantitatively establishing this connection, however, would require a dedicated study with varying pion mass and non-degenerate quarks, and is left for future work. 

From matching to large $\Nc$ ChPT, we have  constrained the value of certain combinations of LECs and compared them with other phenomenological determinations. With more values of the pion mass, it should be possible to constrain the values of all LECs appearing in the chiral Lagrangian up to $\cO(p^6)$, and up to subleading $\Nc$ corrections. We leave for future work a global fit of the results in this work together with those from \rrcite{Hernandez:2019qed,Baeza-Ballesteros:2022azb}, which include results at lighter pion mass and other lattice spacings, that should allow to better control discretization and higher order chiral effects.

As demonstrated by this work, varying the number of colors in QCD remains a powerful tool to probe its nonperturbative features, especially in the realm of multi-hadron dynamics. 
\acknowledgments

The authors would like to thank Claudio Bonanno, Thomas DeGrand, Jeremy R. Green and Andrew D. Hanlon for useful discussions. PH acknowledges support from the EU H2020 research and innovation programme under the MSC grant agreement No 860881-HIDDeN, and the Staff Exchange grant agreement No-101086085-ASYMMETRY, from projects PID2020-113644GB-I00 and PID2023-148162NB-C21 from the Spanish Ministerio de Ciencia e Innovaci\'on and Agencia Estatal de Investigaci\'on (MICIU/AEI) and the European Regional Development Funds (FEDER), and by Generalitat Valenciana through the grant CIPROM/2022/69. The work of JBB has been partially supported by the aforementioned funding and the Spanish grant FPU19/04326 of Ministerio de Universidades. The work of FRL is supported in part by Simons Foundation grant 994314 (Simons Collaboration on Confinement and QCD Strings) and the Mauricio and Carlota Botton Fellowship.

We thank Mare Nostrum 4 (BSC), Finis Terrae II (CESGA), Tirant 3 (UV), Lluis Vives (Servei d’Inform\`atica UV) and XULA (CIEMAT) for the computing time provided.

\newpage
\appendix

\section{Properties of the correlation functions}\label{app:technicaldetailscorrelator}

In this appendix, we review some properties of the quark contractions contributing to the correlation functions of the channels of interest---see \cref{sec:largeNcamplitudes}. We work in the infinite-volume theory, although the same conclusions also apply for lattice determinations. Also, we focus on two-meson external states, as the results are straightforward to generalize to local tetraquarks---see the discussion in \cref{sec:computationcorrelationfunctions}. 

To determine the correlation functions, we use some representative states of each channel. For the purpose of this appendix, we choose
\begin{equation}
    \begin{array}{rl}
         \begin{array}{rl}\label{eq:generaloperators}
O^{SS}_{\Gamma_1,\Gamma_2}(\bm{p}_1,\bm{p}_2)  = \frac{1}{2} & \left[(\bar{d}\Gamma_1 u)(\bm{p}_1)(\bar{s}\Gamma_2 c)(\bm{p}_2)+(\bar{s}\Gamma_1 u)(\bm{p}_1)(\bar{d}\Gamma_2 c)(\bm{p}_2) + (\bm{p}_1\leftrightarrow \bm{p}_2, \Gamma_1\leftrightarrow\Gamma_2)\right]\,,\\[10pt]
 O^{AA}_{\Gamma_1,\Gamma_2}(\bm{p}_1,\bm{p}_2)  = \frac{1}{2} & \left[(\bar{d}\Gamma_1 u)(\bm{p}_1)(\bar{s}\Gamma_2 c)(\bm{p}_2)-
(\bar{s}\Gamma_1 u)(\bm{p}_1)(\bar{d}\Gamma_2 c)(\bm{p}_2) + (\bm{p}_1\leftrightarrow \bm{p}_2, \Gamma_1\leftrightarrow\Gamma_2)\right]\,,\\[10pt]
O^{AS}_{\Gamma_1,\Gamma_2}(\bm{p}_1,\bm{p}_2)  = \frac{1}{2} & \left[(\bar{d}\Gamma_1 u)(\bm{p}_1)(\bar{s}\Gamma_2 c)(\bm{p}_2)+(\bar{s}\Gamma_1 u)(\bm{p}_1)(\bar{d}\Gamma_2 c)(\bm{p}_2) - (\bm{p}_1\leftrightarrow \bm{p}_2, \Gamma_1\leftrightarrow\Gamma_2)\right]\,,\\[10pt]
O^{SA}_{\Gamma_1,\Gamma_2}(\bm{p}_1,\bm{p}_2)  = \frac{1}{2} & \left[(\bar{u}\Gamma_1 d)(\bm{p}_1)(\bar{c}\Gamma_2 s)(\bm{p}_2)-(\bar{u}\Gamma_1 s)(\bm{p}_1)(\bar{c}\Gamma_2 d)(\bm{p}_2) - (\bm{p}_1\leftrightarrow \bm{p}_2, \Gamma_1\leftrightarrow\Gamma_2)\right]\,,
\end{array}
    \end{array}
\end{equation}
where $u$, $d$, $s$ and $c$ refer to the up, down, strange and charm quarks, in this same order, and $\Gamma_1$ and $\Gamma_2$ are combinations of Dirac matrices---see \cref{eq:Diracstructuresummary,tab:summaryCPcomplex} for a summary of the combinations used in this work. Using these operators, the correlation functions can be naively computed,
\begin{equation}\label{eq:generalizednaivecontractions}
\begin{array}{rl}
	C_{SS}(t) & = D(\bm{p}_1,\Gamma_1',\bm{p}_2,\Gamma_2';\bm{k}_1,\Gamma_1,\bm{k}_2,\Gamma_2;t)-C(\bm{p}_1,\Gamma_1',\bm{p}_2,\Gamma_2';\bm{k}_1,\Gamma_1,\bm{k}_2,\Gamma_2;t)\\ &+(\bm{p}\leftrightarrow -\bm{p}, \Gamma_1\leftrightarrow \Gamma_2)\,,\\[5pt]
    C_{AA}(t) & = D(\bm{p}_1,\Gamma_1',\bm{p}_2,\Gamma_2';\bm{k}_1,\Gamma_1,\bm{k}_2,\Gamma_2;t)+C(\bm{p}_1,\Gamma_1',\bm{p}_2,\Gamma_2';\bm{k}_1,\Gamma_1,\bm{k}_2,\Gamma_2;t)\\ &+(\bm{p}\leftrightarrow -\bm{p}, \Gamma_1\leftrightarrow \Gamma_2)\,,\\[5pt]
    C_{AS}(t) & = D(\bm{p}_1,\Gamma_1',\bm{p}_2,\Gamma_2';\bm{k}_1,\Gamma_1,\bm{k}_2,\Gamma_2;t)-C(\bm{p}_1,\Gamma_1',\bm{p}_2,\Gamma_2';\bm{k}_1,\Gamma_1,\bm{k}_2,\Gamma_2;t)\\ &-(\bm{p}\leftrightarrow -\bm{p}, \Gamma_1\leftrightarrow \Gamma_2)\,, \\[5pt]
    C_{SA}(t) & = D(\bm{p}_1,\Gamma_1',\bm{p}_2,\Gamma_2';\bm{k}_1,\Gamma_1,\bm{k}_2,\Gamma_2;t)+C(\bm{p}_1,\Gamma_1',\bm{p}_2,\Gamma_2';\bm{k}_1,\Gamma_1,\bm{k}_2,\Gamma_2;t)\\ &-(\bm{p}\leftrightarrow -\bm{p}, \Gamma_1\leftrightarrow \Gamma_2)\,,
\end{array}
\end{equation}
where $\{\bm{k}_1,\bm{k}_2\}$ and $\{\bm{p}_1,\bm{p}_2\}$ indicate the three-momenta of the particles in the initial and final state, respectively, and similarly for the Dirac structures, $\{\Gamma_1,\Gamma_2\}$ and $\{\Gamma_1',\Gamma_2'\}$. Also, we have defined generalized versions of the disconnected and connected contractions in \cref{eq:Ddiagram,eq:Cdiagram}
\begin{multline}\label{eq:Ddiagram2}
D(\bm{p}_1,\Gamma_1',\bm{p}_2,\Gamma_2';\bm{k}_1,\Gamma_1,\bm{k}_2,\Gamma_2;t)=\int \text{d}\bm{x}_1\text{d}\bm{x}_2\text{d}\bm{y}_1\text{d}\bm{y}_2\,\text{e}^{-i(\bm{p}_1\bm{y}_1+\bm{p}_2\bm{y}_2)}\text{e}^{i(\bm{k}_1\bm{x}_1+\bm{k}_2\bm{x}_2)}\\
\times\left\langle D(x_3,\Gamma_3,x_4,\Gamma_4;x_1,\Gamma_1,x_2,\Gamma_2)\right\rangle\,,
\end{multline}
\begin{multline}\label{eq:Cdiagram2}
C(\bm{p}_1,\Gamma_1',\bm{p}_2,\Gamma_2';\bm{k}_1,\Gamma_1,\bm{k}_2,\Gamma_2;t)=\int \text{d}\bm{x}_1\text{d}\bm{x}_2\text{d}\bm{y}_1\text{d}\bm{y}_2\,\text{e}^{-i(\bm{p}_1\bm{y}_1+\bm{p}_2\bm{y}_2)}\text{e}^{i(\bm{k}_1\bm{x}_1+\bm{k}_2\bm{x}_2)}\\
\times\left\langle C(x_3,\Gamma_3,x_4,\Gamma_4;x_1,\Gamma_1,x_2,\Gamma_2)\right\rangle\,.
\end{multline}
Here $x_i=(0,\bm{x}_i)$ and $y_i=(t,\bm{x}_i)$, and we have introduced the position-space contractions
\begin{multline}\label{eq:Ddiagramposition}
D(y_1,\Gamma_1',y_2,\Gamma_2';x_1,\Gamma_1,x_2,\Gamma_2)=\\ \Tr\left[\Gamma_1^{\prime\text{f}}S(y_1,x_1)\Gamma_1^\text{i}S^\dagger(y_1,x_1)\right]\Tr\left[\Gamma_2^{\prime\text{f}}S(y_2,x_2)\Gamma_2^\text{i}S^\dagger(y_2,x_2)\right]\,,
\end{multline}
\begin{multline}\label{eq:Cdiagramposition}
C(y_1,\Gamma_1',y_2,\Gamma_2';x_1,\Gamma_1,x_2,\Gamma_2)=\\ \Tr\left[\Gamma_2^{\prime\text{f}}S(y_2,x_1)\Gamma_1^\text{i}S^\dagger(y_1,x_1)\Gamma_1^{\prime\text{f}}S(y_1,x_2)\Gamma_2^\text{i}S^\dagger(y_2,x_2)\right]\,,
\end{multline}
with $\Gamma^\text{i}=\gamma_0\Gamma^\dagger\gamma_0\gamma_5$ and $\Gamma^\text{f}=\gamma_5\Gamma$, where the $\gamma_5$ factors originate from using the \mbox{one-end-trick}. 

For the $SA$ and $AS$ channels, the correlation functions can be simplified. It is straightforward to note that the associated operators in \cref{eq:generaloperators} are related by a charge-conjugation transformation, which changes quarks to antiquarks, and viceversa,
\begin{equation}
    O^{AS}_{\Gamma_1,\Gamma_2}(\bm{p}_1,\bm{p}_2)\xrightarrow{C} \pm O^{SA}_{\Gamma_1,\Gamma_2}(\bm{p}_1,\bm{p}_2)
\end{equation}
where the sign depends on the particular choice of Gamma structures. For the choices in this work it is always a plus sign---see \cref{tab:summaryCPcomplex}. Since charge conjugation is a symmetry of QCD, this implies that, for all the operators considered, the correlation functions for the $AS$ and $SA$ channels have to be equal, $C_{AS}(t)=C_{SA}(t)$. From \cref{eq:generalizednaivecontractions}, this implies
\begin{equation}
       C(\bm{p}_1,\Gamma_1',\bm{p}_2,\Gamma_2';\bm{k}_1,\Gamma_1,\bm{k}_2,\Gamma_2;t) = C(\bm{p}_1,\Gamma_1',\bm{p}_2,\Gamma_2';\bm{k}_2,\Gamma_2,\bm{k}_1,\Gamma_1;t)\,.
\end{equation}
Thus, the correlation functions for the $AS$ and $SA$ channels simplify to
\begin{equation}
    C_{AS}(t) = C_{SA}(t)  = D(\bm{p}_1,\Gamma_1',\bm{p}_2,\Gamma_2';\bm{k}_1,\Gamma_1,\bm{k}_2,\Gamma_2;t)-(\bm{p}\leftrightarrow -\bm{p}, \Gamma_1\leftrightarrow \Gamma_2)\,.
\end{equation}

\begin{table}[t]
\renewcommand{\arraystretch}{1.1}
\centering
\begin{tabular}{cccccc}
\toprule
 $O_{\Gamma_1,\Gamma_2}$ & $\Gamma_1$ & $\Gamma_2$ &  $\eta^{C}$ &  $\eta^{P}$ & $\xi$  \\ \midrule 
 $\pi\pi$ & $\gamma_5$ & $\gamma_5$ & $1$ & $1$ & 1 \\ 
 $\rho\rho$ & $\gamma_i$ & $\gamma_j$ & 1 & $1$ & 1  \\ \midrule
 $T_{P_a P_a}$ & $\gamma_5$ & $\gamma_5$ & 1 & $1$ & 1  \\ 
 $T_{P_b P_b}$ & $\gamma_5\gamma_0$ & $\gamma_5\gamma_0$ & 1 & $1$ & 1  \\ 
 $T_{S S}$ & $\mathbbm{1}$ & $\mathbbm{1}$ & 1 & $1$ & 1  \\ 
 $T_{V V}^{ij}$ & $\gamma_i$ & $\gamma_j$ & 1 & $1$ & 1  \\ 
 $T_{AA}^{ij}$ & $\gamma_5\gamma_i$ & $\gamma_5\gamma_j$ & 1 & 1 & 1  \\ \midrule
 $T_{P_a A}^i$ & $i\gamma_5$ & $\gamma_5\gamma_i$ & $-1$ & 1 & $-1$ \\ 
 $T_{P_b A}^i$ & $\gamma_5\gamma_0$ & $\gamma_5\gamma_i$ & $-1$ & 1 & $1$  \\ \bottomrule
\end{tabular}
\caption{Transformation phases for the different operators used in this work under charge conjugation, parity and hermitian conjugation.
} 
\label{tab:summaryCPcomplex}
\end{table}

Next, we can study whether the correlation functions for all the considered channels are real or imaginary. We can perform a parity and charge conjugation transformation to the disconnected and connected quark contractions, and then complex conjugate them to find
\begin{equation}
\begin{array}{rl}
D(\bm{p}_1,\Gamma_1',\bm{p}_2,\Gamma_2';\bm{k}_1,\Gamma_1,\bm{k}_2,\Gamma_2;t) & = \eta_C\eta_P D(-\bm{p}_1,\Gamma_1',-\bm{p}_2,\Gamma_2';-\bm{k}_1,\Gamma_1,-\bm{k}_2,\Gamma_2;t)\\ 
& = \eta_C\eta_P\xi D(\bm{p}_1,\Gamma_1',\bm{p}_2,\Gamma_2';\bm{k}_1,\Gamma_1,\bm{k}_2,\Gamma_2;t)
\end{array}
\end{equation}
\begin{equation}
\begin{array}{rl}
C(\bm{p}_1,\Gamma_1',\bm{p}_2,\Gamma_2';\bm{k}_1,\Gamma_1,\bm{k}_2,\Gamma_2;t)& = \eta_C\eta_P C(-\bm{p}_1,\Gamma_1',-\bm{p}_2,\Gamma_2';-\bm{k}_2,\Gamma_2,-\bm{k}_1,\Gamma_1;t)  \\
     & = \eta_C\eta_P\xi C(\bm{p}_1,\Gamma_1',\bm{p}_2,\Gamma_2';\bm{k}_1,\Gamma_1,\bm{k}_2,\Gamma_2;t) 
\end{array}
\end{equation}
where $\eta_P=\pm 1$ and $\eta_C=\pm 1$ are the signs arising from transforming all Dirac structures under parity and charge conjugation, respectively, and $\xi=\pm 1$ is the sign originating from complex conjugating. For all the choices of $\Gamma$ used in this work, these are summarized in \cref{tab:summaryCPcomplex}. We stress that these transformation do not require the states considered to be symmetric under parity or charge conjugation. In fact, none of the states is $C$-invariant, and the same is the case for non-zero total momentum. We note that the combination of a parity transformation and complex conjugation ensures the total momentum is not transformed.

From here, we conclude that the correlation functions are either purely real or purely imaginary, depending on the choice of Dirac structures. In the case of two-pion states, we have $\eta^{P}=\eta^{C}=\xi=1$, and so the correlation functions are purely real. This same conclusion holds in general as long as the two bilinears in the initial and final states have the same type of gamma structure, which may be different between the initial and the final operators. This is the case, for example, of all the correlation function considered in this work involving a combination of two-pion, two-vector-meson or scalar-tetraquark operators. The only exceptions are correlation functions involving vector tetraquarks, which are $CP$-odd. Different cases need to be considered then:
\begin{itemize}
    \item If the initial- and final-state operators are the same local vector  tetraquarks, the correlation functions are real.
    \item If the one of the initial- or final-state operators is a two-particle state (either two pions or two vector mesons) and the other one is $T_{P_a A}$, the correlation functions are also real.
    \item For the correlation function between a two-particle state and the $T_{P_b A}$ tetraquark, the correlation function is imaginary.
    \item Finally, correlation functions between the two types of vector tetraquark operators are also purely imaginary.
\end{itemize}
Overall, this implies that the correlation functions computed in this work are expected to be real in general, and so we can drop the imaginary parts that arise due to the use of stochastic sources and is observed to be compatible with zero. The only exception are correlation functions in the $AS$ channel that involve one $T_{P_b A}$ operator, for which the correlation function is imaginary, and so we drop the real part. 

\section{Summary of ChPT predictions}\label{app:chptamplitudes}

This appendix summarizes ChPT results for the scattering amplitudes of the $SS$, $AA$ and $AS$ channels, both in the SU($\Nf$) and U($\Nf$) theories, which are matched to lattice results to constrain the values and large $\Nc$ scaling of the LECs. We focus on studying single-channel processes at the lowest partial wave, for which Lüscher's formalism~\cite{Luscher:1986pf,Luscher:1990ux} allows to directly determine the scattering phase shift from the lattice data. Two-pion scattering amplitudes, $\cM_2$, are first projected to definite partial wave $\ell$,
\begin{equation}\label{eq:partialwaveprojeciton}
    \cM_{2,\ell}=\frac{2\ell+1}{2}\int_{-1}^1  \cM_2(s,t,u) P_\ell (\cos\theta)\text{d}\cos\theta\,,
\end{equation}
where $P_\ell(x)$ are Legendre's polynomials and $\theta$ is the scattering angle. From the partial-wave projected amplitude, the scattering phase shift is determined as
\begin{equation}
    k\cot\delta_\ell = 16\pi\sqrt{s}\,\text{Re}\left(\cM_{2,\ell}^{-1}\right)\,.
    \label{eq:kcot}
\end{equation}
In this work, when matching our lattice results for the $SS$ and $AA$ channels to ChPT, we use the IAM redefinition of the scattering amplitude to compute the scattering phase shift---see \cref{eq:IAMSU}.

Predictions of the scattering amplitude in SU($\Nf$) ChPT are known up to NNLO~\cite{Bijnens:2011fm}. Results at LO are presented in \cref{eq:chptLOpredictions}, while NLO predictions are summarized in \cref{eq:chptNLOpredictionsSSAA,eq:chptNLOpredictionsAS}, projected to the lowest non-zero partial waves. An explicit expression of the linear combinations of the LECs is given in \cref{eq:chptNLOpredictionsSSAALECs,eq:chptNLOpredictionsASLEcs}, respectively, and we present here the expression for the remaining parts of the amplitudes. For the $SS$ and $AA$ channels, we have
\begin{equation}\label{eq:explicitffactorSSAA}
    \begin{array}{rl}
        f_{SS,AA}(\Mpi^2,q^2,\mu,\Nf) & =   \displaystyle \frac{1}{4\pi^2}\left[-1-\frac{1}{\Nf^2}\pm\frac{1}{\Nf}+q^2\left(-3\mp\frac{\Nf}{18}\right)+q^4\left(-\frac{10}{3}\mp\frac{11\Nf}{27}\right)\right]\\[15pt]
\displaystyle &+\frac{1}{4\pi^2}\left[-1-\frac{1}{\Nf^2}\pm\frac{1}{\Nf}+q^2\left(-3\mp\frac{\Nf}{6}\right)+q^4\left(-\frac{10}{3}\mp\frac{5N_\text{f}}{9}\right)\right]\ln{\frac{M_\pi^2}{\mu^2}} \\[10pt]
\displaystyle &+ \left(2+8q^2+8q^4\right)\bar{J}(s) + F_{SS,AA}(q^2)\,,
    \end{array}
\end{equation}
where the upper (lower) signs correspond to the $SS$ ($AA$) channel, and we have defined
\begin{equation}\label{eq:explicitFfactorSSAA}
    \begin{array}{rl}
F_{SS,AA}(q^2)=&\displaystyle\int_{-1}^1\left[1+\frac{2}{\Nf^2}\mp\frac{2}{\Nf}+\frac{2\Nf}{3}+q^2\left(2\mp\frac{4N_\text{f}}{3}-2x\right) \right.\\[10pt]
&\displaystyle\left. + q^4 \left(2\pm\Nf-4x\mp\frac{4\Nf x}{3}+2x^2\mp\frac{\Nf x^2}{3}\right)\right] \bar{J}[-2q^2(1-x)]\,\text{d} x,
\end{array}
\end{equation}
For the $AS$ channel,
\begin{equation}
    \begin{array}{rl}
        f_{AS}(\Mpi^2,q^2,\mu,\Nf) = &  -\frac{1}{\pi^2}\left(1+\ln{\frac{M_\pi^2}{\mu^2}}\right) + F_{AS}(q^2)\,,
    \end{array}
\end{equation}
where
\begin{equation}
    \begin{array}{rl}
    \displaystyle F_{AS}(q^2)=\frac{3}{4}\int_{-1}^1&\displaystyle\left\{\left[\frac{1}{\Nf^2}-1-2q^2(1-x)\right]\bar{J}[-2q^2(1-x)]\right.\\[10pt]
    & \displaystyle\left.-\left[\frac{1}{\Nf^2}-1-2q^2(x+1)\right]\bar{J}[-2q^2(x+1)]\right\}\text{d}x\,.
    \end{array}
\end{equation}

One-loop predictions for the scattering amplitudes in U($\Nf$) ChPT take the form presented in \cref{eq:UchptNLOpredictionsSSAA,eq:UchptNLOpredictionsAS}. For the $SS$ and $AA$ channels, the LO contribution is the same as in the SU($\Nf$) theory, the NLO part contains tree-level diagrams with $\cO(\Nc)$ LECS, 
\begin{equation}
    \frac{\Fpi^4}{\Mpi^4}\left.\cM_{2,s}^{SS,AA,\NLO}\right|_{\text{U}(\Nf)}  = \pm32 \Nc L^{(0)} \pm 32 q^2 \Nc L^{\prime(0)} \pm \frac{128}{3}q^4 \Nc L^{\prime\prime(0)}\,, 
\end{equation}
and the NNLO part of the amplitude includes loop diagrams and tree level diagrams with single insertions of $\cO(1)$ LECs or two insertions of $\cO(\Nc)$ LECs,
\begin{equation}
\begin{array}{rl}
    \frac{\Fpi^4}{\Mpi^4}\left.\cM_{2,s}^{SS,AA,\NNLO}\right|_{\text{U}(\Nf)}  = & \pm32 L_{SS,AA}^{(1)} \pm 32 q^2 L_{SS,AA}^{\prime(1)} \pm \frac{128}{3}q^4 L_{SS,AA}^{\prime\prime(1)}+f_{SS,AA}(\Mpi^2, q^2, \mu, \Nf)\\[10pt]
    & -\frac{\Mpi^2}{\Fpi^2}(K_{SS,AA}+q^2 K_{SS,AA}')+\Delta \cM_2^{AA}(\Mpi^2, q^2, \mu, \Nf)\,,  \,, 
    \end{array}
\end{equation}
where the upper (lower) signs, refer to the $SS$ and $AA$ channel, respectively. Also, the leading and subleading parts of the LECs are introduced in \cref{eq:SUparammetrizationSSAALECs}, while the $K$ LECs are defined in \cref{eq:largeNcscalingK} and are equal for both channels.
The contribution from loop diagrams including the $\eta'$ take the form
\begin{equation}
    \begin{array}{rl}
        \Delta \cM_2^{SS,AA}(\Mpi^2,\tilde{t},\tilde{u},\mu,\Nf) & = \mp \frac{2}{\Nf}\left(1\mp\frac{2}{\Nf}\right)B_1(\tilde{t})-\frac{2}{\Nf^2}B_2(\tilde{t})+(\tilde{t}\leftrightarrow \tilde{u})\,,\end{array}
\end{equation}
where we define two kinematic functions,\footnote{We note there was a typo in the definition of $B_2(z)$ in eq.~(A.13) of the published version of \rcite{Baeza-Ballesteros:2022azb}, where the $\bar{J}$ function had the incorrect normalization and sign. }\begin{equation}\label{eq:chptB1}
\begin{array}{rl}
B_1(z)=&\displaystyle\frac{1}{(4\pi)^2}\left\{\frac{1}{M_{\eta'}^2-M_\pi^2}\left(M_{\eta'}^2\log{\frac{M_{\eta'}^2}{\mu^2}}-M_\pi^2\log{\frac{M_\pi^2}{\mu^2}}\right)\right.\\[15pt]
&\displaystyle\left.+\int_0^1 \log\left[\frac{M_\pi^2x+M_{\eta'}^2(1-x)-x(1-x)z}{M_\pi^2x+M_{\eta'}^2(1-x)}\right]\right\}\text{d}x\,,
\end{array}
\end{equation}
\begin{equation}\label{eq:loopint2}
\displaystyle B_2(z)=\frac{1}{(4\pi)^2}+\frac{1}{(4\pi)^2}\log{\frac{M_{\eta'}^2}{\mu^2}}-\bar{J}\left(z\frac{M_\pi}{M_{\eta'}^2}\right)\,.
\end{equation}
This can then be projected to the relevant partial wave using \cref{eq:partialwaveprojeciton}. On the other hand, the $AS$ channel starts at NNLO in the U($Nf$) theory, and the correction from loops including the $\eta'$ is given in \cref{eq:ASUchptcorrection}. 

\section{Summary of $\pi\pi$ finite-volume energies}\label{app:summaryenergies}

In this appendix, we summarize the lattice results for the pion-pion finite-volume energies obtained in this work, which are used in \cref{sec:scatteringfits} to constrain the $\Nc$ scaling of the scattering amplitude. Results for the $SS$, $AA$ and $AS$ channels are presented in \cref{tab:SSenergies,tab:AAenergies,tab:ASenergies}, in this same order.

 \begin{table}[h!]
\centering
\begin{tabular}{cccccc}
\toprule
   &     &   \multicolumn{4}{c}{$E_{SS}^*/\Mpi$} \\
Irrep  & Level  & $\Nc=3$ & $\Nc=4$ &    $\Nc=5$ & $\Nc=6$\\  \midrule 

\multirow{2}{*}{$A_1^+(0)$} & $n=0 $ & $2.052(6)$ & $2.079(7)$ & $2.052(4)$ & $2.041(4)$ \\
 & $n=1 $ & $3.292(14)$ & $3.852(5)$ & $3.677(5)$ & $3.629(3)$ \\\midrule
\multirow{2}{*}{$A_1(1)$} & $n=0 $ & $2.366(11)$ & $2.519(23)$ & $2.460(8)$ & $2.430(8)$ \\
 & $n=1 $ & $3.530(11)$ & --- & --- & --- \\\midrule
\multirow{4}{*}{$A_1(2)$} & $n=0 $ & $2.493(26)$ & $2.687(9)$ & $2.640(6)$ & $2.613(6)$ \\
 & $n=1 $ & $2.725(6)$ & $3.065(15)$ & $2.959(6)$ & $2.920(6)$ \\
 & $n=2 $ & $3.613(23)$ & --- & --- & ---\\
 & $n=3 $ & $3.719(16)$ & --- & --- & --- \\\midrule
\multirow{2}{*}{$A_1(3)$} & $n=0 $ & $2.650(16)$ & $2.859(15)$ & $2.822(11)$ & $2.765(29)$ \\
 & $n=1 $ & $3.055(19)$ & $3.489(11)$ & $3.353(5)$ & $3.287(15)$ \\\midrule
\multirow{3}{*}{$A_1(4)$} & $n=0 $ & $2.030(10)$ & $2.029(10)$ & $2.024(6)$ & $2.017(6)$ \\
 & $n=1 $ & $2.756(15)$ & $3.01(4)$ & $2.925(9)$ & $2.886(8)$ \\
 & $n=2 $ & $3.266(12)$ & --- & --- & --- \\
\bottomrule 
\end{tabular}
\caption{Results for the finite-volume pion-pion energies in the CM frame for the $SS$ channel. Only those energies  used for the analysis of scattering amplitudes in \cref{sec:scatteringfits} are listed. The first column indicates the cubic-group irrep, with the number in parenthesis referring to $\bm{P}^2/(2\pi/L)^2$. }

\label{tab:SSenergies}
\end{table}

 \begin{table}[h!]
\centering
\begin{tabular}{cccccc}
\toprule
   &     &   \multicolumn{4}{c}{$E_{AA}^*/\Mpi$} \\
Irrep  & Level  & $\Nc=3$ & $\Nc=4$ &    $\Nc=5$ & $\Nc=6$\\  \midrule 

\multirow{4}{*}{$A_1^+(0)$} & $n=0 $ & $1.933(6)$ & $1.923(6)$ & $1.950(4)$ & $1.962(4)$ \\
 & $n=1 $ & $2.993(7)$ & $3.469(18)$ & $3.409(4)$ & $3.393(3)$ \\
 & $n=2 $ & $3.788(29)$ & --- & --- & --- \\\midrule
\multirow{3}{*}{$A_1(1)$} & $n=0 $ & $2.182(7)$ & $2.275(16)$ & $2.285(8)$ & $2.288(8)$ \\
 & $n=1 $ & $3.189(8)$ & $3.767(12)$ & $3.670(15)$ & $3.68(3)$ \\
 & $n=2 $ & $3.958(25)$ & --- & --- & --- \\\midrule
\multirow{4}{*}{$A_1(2)$} & $n=0 $ & $2.336(18)$ & $2.42(4)$ & $2.442(16)$ & $2.468(14)$ \\
 & $n=1 $ & $2.586(16)$ & $2.884(25)$ & $2.821(15)$ & $2.803(11)$ \\
 & $n=2 $ & $3.410(15)$ & $3.989(20)$ & $3.908(9)$ & $3.897(25)$ \\
 & $n=3 $ & $3.552(15)$ & --- & --- & --- \\\midrule
\multirow{2}{*}{$A_1(3)$} & $n=0 $ & $2.46(4)$ & $2.614(26)$ & $2.637(26)$ & $2.637(21)$ \\
 & $n=1 $ & $2.758(24)$ & $3.08(6)$ & $3.04(8)$ & $3.072(15)$ \\\midrule
\multirow{4}{*}{$A_1(4)$} & $n=0 $ & $1.937(13)$ & $1.940(13)$ & $1.952(11)$ & $1.960(9)$ \\
 & $n=1 $ & $2.58(3)$ & $2.775(23)$ & $2.74(4)$ & $2.775(15)$ \\
 & $n=2 $ & $3.01(4)$ & $3.554(29)$ & $3.451(26)$ & $3.446(8)$ \\
 & $n=3 $ & $3.640(11)$ & --- & --- & --- \\\bottomrule 
\end{tabular}
\caption{Same as \cref{tab:SSenergies} for the $AA$ channel. Energies for the $A_1(4)$ irrep are not used for the final fits, due to large observed systematic deviations of the corresponding scattering phase shift---see \cref{sec:fittingprocedure}. }

\label{tab:AAenergies}
\end{table}

 \begin{table}[h!]
\centering
\begin{tabular}{cccccc}
\toprule
   &     &   \multicolumn{4}{c}{$E_{AS}^*/\Mpi$} \\
Irrep  & Level  & $\Nc=3$ & $\Nc=4$ &    $\Nc=5$ & $\Nc=6$\\  \midrule 

\multirow{1}{*}{$T_1^-(0)$} & $n=0 $ & $3.164(4)$ & $3.682(4)$ & $3.559(4)$ & $3.515(3)$ \\\midrule
\multirow{2}{*}{$A_1(1)$} & $n=0 $ & $2.273(8)$ & $2.390(17)$ & $2.357(16)$ & $2.358(8)$ \\
  & $n=1 $ & $3.375(9)$ & $3.965(32)$ & $3.814(24)$ & $3.776(14)$ \\\midrule
\multirow{1}{*}{$E(1)$} & $n=0 $ & $3.365(8)$ & $4.00(4)$ & $3.811(17)$ & $3.773(9)$ \\\midrule
\multirow{2}{*}{$A_1(2)$} & $n=0 $ & $2.448(8)$ & $2.602(9)$ & $2.556(6)$ & $2.553(6)$ \\
  & $n=1 $ & $3.514(12)$ & $4.164(5)$ & $4.002(8)$ & $3.972(6)$ \\\midrule
\multirow{2}{*}{$B_1(2)$} & $n=0 $ & $3.528(7)$ & $4.158(5)$ & $4.004(4)$ & $3.961(4)$ \\
  & $n=1 $ & $3.608(4)$ & $4.277(5)$ & $4.117(3)$ & $4.068(5)$ \\\midrule
\multirow{2}{*}{$B_2(2)$} & $n=0 $ & $2.649(5)$ & $2.963(6)$ & $2.887(4)$ & $2.861(4)$ \\
  & $n=1 $ & $3.614(5)$ & $4.283(6)$ & $4.124(3)$ & $4.068(4)$ \\\midrule
\multirow{2}{*}{$A_1(3)$} & $n=0 $ & $2.567(16)$ & $2.771(24)$ & $2.701(26)$ & $2.64(5)$ \\
  & $n=1 $ & $2.900(9)$ & $3.335(15)$ & $3.211(9)$ & $3.185(14)$ \\\midrule
\multirow{1}{*}{$E(3)$} & $n=0 $ & $2.891(8)$ & $3.318(21)$ & $3.199(11)$ & $3.168(12)$ \\\midrule
\multirow{2}{*}{$A_1(4)$} & $n=0 $ & $2.699(8)$ & $2.912(9)$ & $2.860(7)$ & $2.836(8)$ \\
  & $n=1 $ & $3.773(6)$ & $4.475(8)$ & $4.307(6)$ & $4.256(3)$ \\\midrule
\multirow{1}{*}{$E(4)$} & $n=0 $ & $3.166(5)$ & $3.689(12)$ & $3.555(4)$ & $3.517(4)$ \\\bottomrule
\end{tabular}
\caption{Same as \cref{tab:SSenergies} for the $AS$ channel. }

\label{tab:ASenergies}
\end{table}

\clearpage

\bibliographystyle{JHEP}      
\bibliography{bibtexref.bib,manualbib.bib}

\end{document}

%% file: definitions.tex
\usepackage{cleveref}

\newcommand{\Nf}[0]{N_\text{f}}
\newcommand{\Nc}[0]{N_\text{c}}

\newcommand{\Mpi}[0]{M_\pi}
\newcommand{\Metap}[0]{M_{\eta'}}
\newcommand{\Fpi}[0]{F_\pi}

\newcommand{\cG}[0]{\mathcal{G}}
\newcommand{\cK}[0]{\mathcal{K}}
\newcommand{\cM}[0]{\mathcal{M}}
\newcommand{\cO}[0]{\mathcal{O}}
\newcommand{\cZ}[0]{\mathcal{Z}}

\newcommand{\LO}[0]{\mathrm{LO}}
\newcommand{\NLO}[0]{\mathrm{NLO}}
\newcommand{\NNLO}[0]{\mathrm{NNLO}}

\newcommand{\rcite}[1]{ref.~\cite{#1}}
\newcommand{\rrcite}[1]{refs.~\cite{#1}}

\newcommand{\mathbbm}[1]{\text{\usefont{U}{bbm}{m}{n}#1}}